\documentclass{article}
\pdfoutput=1
\usepackage[utf8]{inputenc}
\usepackage[T1]{fontenc}
\usepackage{amsthm}
\usepackage{amssymb}
\usepackage{mathrsfs}
\usepackage{amsmath}
\usepackage{amsfonts,amssymb}
\usepackage{mathtools}
\usepackage{fullpage}
\usepackage{hyperref}
\usepackage{multirow}
\usepackage{relsize}
\usepackage{lmodern}
\usepackage{import}
\usepackage{bm}
\usepackage{braket}
\usepackage{blkarray}

\mathtoolsset{showonlyrefs}

\usepackage[usenames, dvipsnames]{xcolor}
\usepackage{tikz}
\usepackage{tikz-3dplot}
\usetikzlibrary{arrows,decorations,matrix,calc,decorations.markings,arrows.meta}
\usepackage{ifthen}

\newtheorem{Theo}{Theorem}
\newtheorem*{Theo*}{Theorem}
\newtheorem{Prop}[Theo]{Proposition}
\newtheorem*{Prop*}{Proposition}
\newtheorem{Cor}[Theo]{Corollary}
\newtheorem{Le}[Theo]{Lemma}
\newtheorem{Ex}[Theo]{Example}
\theoremstyle{definition}
\newtheorem{Def}[Theo]{Definition}
\newtheorem*{Defs*}{Definitions}
\newtheorem{Rk}[Theo]{Remark}

\DeclareMathOperator{\Pf}{\mathrm{Pf}}

\DeclareMathOperator{\am}{\mathrm{am}}
\DeclareMathOperator{\sn}{\mathrm{sn}}
\DeclareMathOperator{\cn}{\mathrm{cn}}
\DeclareMathOperator{\dn}{\mathrm{dn}}
\DeclareMathOperator{\dc}{\mathrm{dc}}
\DeclareMathOperator{\nc}{\mathrm{nc}}
\DeclareMathOperator{\nd}{\mathrm{nd}}
\renewcommand{\sc}{\mathrm{sc}}

\newcommand{\R}{\mathbb{R}}
\newcommand{\C}{\mathbb{C}}

\renewcommand{\Pr}{\mathbb{P}}

\newcommand{\Z}{\mathbb{Z}}
\newcommand{\T}{\mathbb{T}}
\newcommand{\ZZ}{\mathcal{Z}}

\newcommand{\QQ}{\mathcal{Q}}
\newcommand{\VV}{\mathcal{V}}
\newcommand{\EE}{\mathcal{E}}
\newcommand{\FF}{\mathcal{F}}
\newcommand{\BB}{\mathcal{B}}
\newcommand{\WW}{\mathcal{W}}

\newcommand{\s}{\bm{\sigma}}
\renewcommand{\a}{\bm{\alpha}}
\renewcommand{\t}{\bm{\tau}}
\renewcommand{\H}{\bm{H}}

\newcommand{\1}{\bm{1}}
\newcommand{\0}{\bm{0}}

\renewcommand{\u}{\mathrm{u}}
\renewcommand{\v}{\mathrm{v}}
\newcommand{\x}{\mathrm{x}}
\newcommand{\y}{\mathrm{y}}
\newcommand{\w}{\mathrm{w}}
\renewcommand{\b}{\mathrm{b}}

\title{The free-fermion eight-vertex model: couplings, bipartite
  dimers and $Z$-invariance}
\author{Paul Melotti\thanks{Université de Fribourg. Email: \texttt{paul.melotti@unifr.ch}}}
\date{\today}

\begin{document}

\maketitle

\begin{abstract}
  We study the eight-vertex model at its free-fermion point. We
  express a new ``switching'' symmetry of the model in several forms:
  partition functions, order-disorder variables, couplings, Kasteleyn
  matrices. This symmetry can be used to relate free-fermion 8V-models
  to free-fermion 6V-models, or bipartite dimers. We also define new
  solution of the Yang-Baxter equations in a ``checkerboard'' setting,
  and a corresponding $Z$-invariant model. Using the bipartite dimers
  of Boutillier, de Tilière and Raschel
  \cite{BoutillierDeTiliereRaschel}, we give exact local formulas for
  edge correlations in the $Z$-invariant free-fermion 8V-model on
  lozenge graphs, and we deduce the construction of an ergodic Gibbs
  measure.
\end{abstract}

\tableofcontents

\section{Introduction}
\label{sec:intro}

\begin{figure}[!ht]
  \centering
  \begin{tikzpicture}
  \newcommand\nord[2]{
    \draw [ultra thick] (#1,#2) -- (#1,#2+1);
  }
  \newcommand\est[2]{
    \draw [ultra thick] (#1,#2) -- (#1+1,#2);
  }
  \newcommand\haut[2]{
    \draw [decoration={markings,
    mark=at position 0.5 with {\arrow{>}}},postaction={decorate}] (#1,#2) -- (#1,#2+1);
  }
  \newcommand\bas[2]{
    \draw [decoration={markings,
    mark=at position 0.5 with {\arrow{>}}},postaction={decorate}] (#1,#2) -- (#1,#2-1);
  }
  \newcommand\gauche[2]{
    \draw [decoration={markings,
    mark=at position 0.5 with {\arrow{>}}},postaction={decorate}] (#1,#2) -- (#1-1,#2);
  }
  \newcommand\droite[2]{
    \draw [decoration={markings,
    mark=at position 0.5 with {\arrow{>}}},postaction={decorate}] (#1,#2) -- (#1+1,#2);
  }

  \begin{scope}[xshift=6.5cm]
  \clip (-0.7,-0.7) rectangle (4.7,4.7);
  \foreach \x in {-1,...,4} \foreach \y in {-1,...,4}
  {
    \pgfmathparse{mod(\x+\y,2) ? "gray!50" : "white"}
    \edef\colour{\pgfmathresult}
    \path[fill=\colour] (\x,\y) rectangle (\x+1,\y+1);
  }
  \nord{0}{-1};
  \nord{1}{-1};
  \nord{3}{-1};
  \nord{4}{-1};
  \nord{1}{0};
  \nord{2}{0};
  \nord{3}{0};
  \nord{2}{1};
  \nord{3}{1};
  \nord{0}{2};
  \nord{2}{2};
  \nord{3}{2};
  \nord{4}{2};
  \nord{0}{3};
  \nord{2}{3};
  \nord{3}{3};
  \nord{4}{3};
  \nord{0}{4};
  \nord{3}{4};
  \nord{4}{4};
  \est{-1}{0};
  \est{2}{0};
  \est{3}{0};
  \est{1}{1};
  \est{2}{1};
  \est{3}{1};
  \est{4}{1};
  \est{1}{2};
  \est{2}{2};
  \est{3}{2};
  \est{0}{2};
  \est{-1}{4};
  \est{0}{4};
  \est{1}{4};
  \end{scope}

  \begin{scope}[xshift=0cm]
    \clip (-0.7,-0.7) rectangle (4.7,4.7);
    \foreach \x in {-1,...,4} \foreach \y in {-1,...,4}
    {
      \pgfmathparse{mod(\x+\y,2) ? "gray!50" : "white"}
      \edef\colour{\pgfmathresult}
      \path[fill=\colour] (\x,\y) rectangle (\x+1,\y+1);
    }
    \haut{0}{-1};
    \haut{4}{-1};
    
    \gauche{0}{0};
    \haut{0}{0};
    \gauche{1}{0};
    \haut{1}{0};
    \droite{1}{0};
    \bas{1}{0};
    \droite{2}{0};
    \bas{2}{0};
    \haut{3}{0};
    \bas{3}{0};
    \gauche{4}{0};
    \haut{4}{0};
    \gauche{5}{0};
    
    \gauche{0}{1};
    \droite{0}{1};
    \haut{1}{1};
    \droite{1}{1};
    \haut{2}{1};
    \bas{2}{1};
    \gauche{3}{1};
    \droite{3}{1};
    \gauche{5}{1};

    \droite{-1}{2};
    \droite{0}{2};
    \bas{0}{2};
    \gauche{2}{2};
    \droite{2}{2};
    \haut{3}{2};
    \bas{3}{2};
    \gauche{4}{2};
    \bas{4}{2};
    \gauche{5}{2};
    
    \gauche{0}{3};
    \haut{0}{3};
    \droite{0}{3};
    \bas{0}{3};
    \haut{1}{3};
    \bas{1}{3};
    \gauche{2}{3};
    \haut{2}{3};
    \droite{2}{3};
    \bas{2}{3};
    \gauche{4}{3};
    \haut{4}{3};
    \droite{4}{3};
    \bas{4}{3};

    \gauche{0}{4};
    \droite{0}{4};
    \gauche{2}{4};
    \haut{2}{4};
    \gauche{3}{4};
    \haut{3}{4};
    \droite{3}{4};
    \bas{3}{4};
    \gauche{5}{4};

    \bas{0}{5};
    \bas{1}{5};
    \bas{4}{5};
  \end{scope}
  
\end{tikzpicture}

  \caption{Two equivalent representations of an eight-vertex
    configuration on $\Z^2$.}
  \label{fig:8v_z2}
\end{figure}
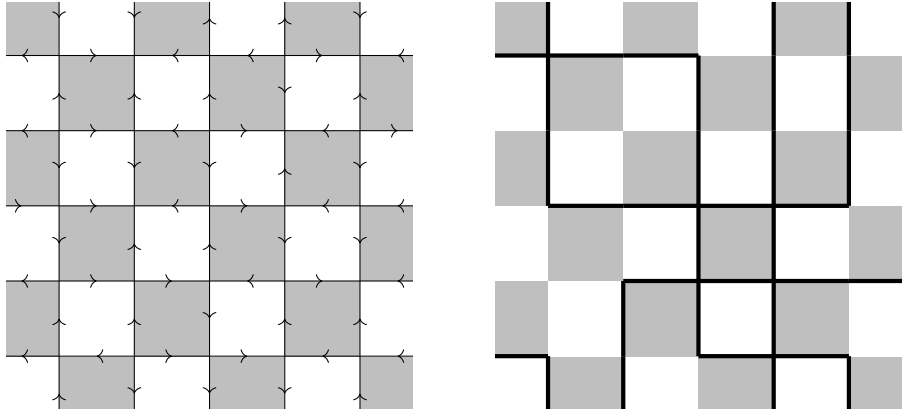

The eight-vertex model, or 8V-model for short, was introduced by
Sutherland \cite{Sutherland} and Fan and Wu \cite{FanWu} as a
generalization of the 6V-model, which itself finds its origins in the
study of square ice \cite{Slater, Lieb}. The configurations of the
8V-model are orientations of the edges of $\Z^2$ such that every
vertex has an even number of incoming edges, like in
Figure~\ref{fig:8v_z2}. There are eight possible local configurations,
hence the name. In the most classical setting, the model is
characterized by four local weights $a,b,c,d$ such that opposite local
configurations are given the same weight, see \cite{Baxter:exactly}.

This model attracted attention during the 70s and 80s, and was
famously solved on the square lattice and a few other regular lattices
using transfer matrices methods, see again \cite{Baxter:exactly} and
references therein. It exhibits phenomena that were surprising at the
time; for instance, it has been predicted that some critical exponents
depend \emph{continuously} on $a,b,c,d$. This is the case of the
exponent $\nu$ for the correlation length: in a generic
\emph{disordered} phase, with inverse temperature $\beta$, the
edge-edge correlations between faraway edges $e,e'$ should decay as
$\exp\left( -\frac{|e-e'|}{\xi(\beta)} \right)$. As the system becomes
critical one should have $\xi(\beta)\sim(\beta-\beta_c)^{-\nu}$. It
seems that these predictions and the computation of $\nu$ still
require mathematical investigation. In what follows, we focus on the
special \emph{free-fermion} case $a^2+b^2=c^2+d^2$ on more generic
lattices. A consequence of the techniques developed here is a
construction of a Gibbs measure on these graphs and a proof that
$\nu=1$ for free-fermion models, in accordance with Baxter's
computation.

\begin{figure}[!ht]
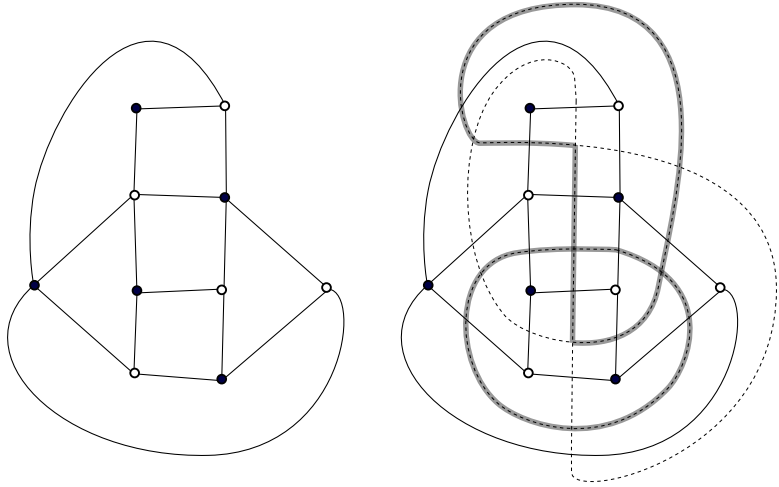

  \centering
  \def\svgwidth{5cm}
  \input{ex8v_prim.tex}
  \def\svgwidth{5cm}
  \input{ex8v_dual.tex}
  \caption{Left: a quadrangulation $\QQ$ in the spherical case. Right:
    the same quadrangulation with its dual $\QQ^*$ (dashed) and an
    eight-vertex configuration.}
  \label{fig:globconf}
\end{figure}

Let us now describe our setting. A configuration can be represented
equivalently as a polygon, by choosing a checkerboard coloring of the
faces of $\Z^2$ and drawing in bold the edges oriented with, say, a
white face on their left.
In this paper we extend definitions to graphs that are dual of a planar
quadrangulation $\QQ$, whose set of vertices (resp. edges, faces) we
denote by $\VV$ (resp. $\EE,\FF$); an example is displayed in
Figure~\ref{fig:globconf}.
More precisely, the Boltzmann weight of a configuration is the product
of local weights associated to local configurations at a face $f$ of
the quadrangulation $\QQ$, as in Figure~\ref{fig:locconf}, that are
denoted $A(f),B(f),C(f),D(f)$ (we use uppercase notations to emphasize
that these are functions of the faces). The case of a 6V-model corresponds to
$D(f)=0$ at every face. Notice that complementary configurations have
the same weight, which means that we are in a ``zero field'' case. To
make these weights well-defined, notice also that we fixed a bipartite
coloring of $\QQ$. This is sometimes referred to as a
\emph{checkerboard model} \cite{Baxter:checkerboard,BaxterPerkAuYang},
or a \emph{staggered} model in the case of the square lattice
\cite{HsueLinWu}. Checkerboard (or alternating, or staggered) 8V
models have attracted some attention, in particular for their relation
with the Ashkin-Teller model \cite{Wegner,Baxter:exactly}, but little
is known about them in general. In this paper we investigate
checkerboard 8V-models that satisfy the \emph{free-fermion} condition:
\begin{equation}
  \label{eq:56}
  A^2 + B^2 = C^2 + D^2.
\end{equation}
Under this condition, the theory of transfer matrices can be adapted,
making computation easier than for the complete 8V-model
\cite{BazhanovStroganov1, BazhanovStroganov2, BazhanovStroganov3,
  Felderhof, Felderhof2, Felderhof3}. There exists a different method
using a correspondence with \emph{dimers} on a non-bipartite decorated
graph, leading to the computation of Pfaffians
\cite{LinWang,FanWu,HsueLinWu,Lin,Lin2}. The dimer model is an object
of interest of its own in the mathematical community, an important
example being the celebrated description of Gibbs measures and phase
diagram of bipartite dimers on periodic graphs
\cite{ckp,kos}. Adapting these results to non-bipartite settings is an
important open problem in the community, making the previous
correspondence of 8V configurations to non-bipartite dimers of little
help for the description of, say, Gibbs measures and correlations. In
what follows we provide a way to transform these non-bipartite dimers
into bipartite ones. From the point of view of the dimer model,
unexpectedly, this provides a family of examples of non-bipartite
dimers whose statistics can be related to bipartite ones, and whose
spectral curve is reducible, as shown in formula \eqref{eq:p8vprod}.

Our method is based on a
``switching'' result, that we introduce now.
If we perform a \emph{gauge transformation} by multiplying all weights $A(f),B(f),C(f),D(f)$ at a face $f$ by the
same constant, the relative weight of different 8V-configurations are
unchanged. Thus an 8V
model with weights satisfying the free-fermion condition \eqref{eq:56}
can be effectively represented by two free parameters per face, say
$\alpha(f),\beta(f) \in \R/2\pi\Z$; see \eqref{eq:pds86} for the exact
parametrization. Our parametrization is such that when
$\alpha = \beta$, the model becomes a 6V one. We denote by
$X_{\alpha,\beta}$ the whole set of weights corresponding to
$\alpha,\beta$, and by $\ZZ_{8V}(\QQ,X_{\alpha,\beta})$ the partition
function; when $\alpha=\beta$, we denote it by
$\ZZ_{6V}(\QQ,X_{\alpha,\alpha})$ to emphasize that it becomes a
6V-model. The choice of parameters $\alpha,\beta$ is such that we have
the following ``switching'' relation, see Theorem~\ref{theo:8vswitch}
for a generalized statement and \eqref{eq:defc} for the value of the
constant $c_{\alpha,\beta}$:
\begin{Theo}
  \label{theo:z_intro}
  Let $\QQ$ be a quadrangulation of the sphere. For any
  $\alpha,\beta,\alpha',\beta':\FF \to \left( 0,\frac{\pi}{2} \right)$,
  \begin{equation}
    \frac{\ZZ_{8V}(\QQ,X_{\alpha,\beta})}{\sqrt{c_{\alpha,\beta}}} \frac{\ZZ_{8V}(\QQ,X_{\alpha',\beta'})}{\sqrt{c_{\alpha',\beta'}}} =
    \frac{\ZZ_{8V}(\QQ,X_{\alpha,\beta'})}{\sqrt{c_{\alpha,\beta'}}} \frac{\ZZ_{8V}(\QQ,X_{\alpha',\beta})}{\sqrt{c_{\alpha',\beta}}}.
  \end{equation}
\end{Theo}
In particular, for $(\alpha',\beta')=(\beta,\alpha)$, we recover
6V-models on the right-hand side. This gives a new relation between
free-fermion 8V-models and 6V ones:
\begin{Cor}
  Let $\QQ$ be a quadrangulation of the sphere. For any
  $\alpha,\beta:\FF \to \left( 0,\frac{\pi}{2} \right)$,
  \begin{equation}
    \left(\ZZ_{8V}(\QQ,X_{\alpha,\beta}) \right)^2 =
    \ \frac{c_{\alpha,\beta}}{\sqrt{c_{\alpha,\alpha} c_{\beta,\beta}}} \ \ZZ_{6V}(\QQ,X_{\alpha,\alpha}) \ \ZZ_{6V}(\QQ,X_{\beta,\beta}).
  \end{equation}
\end{Cor}

This illustrates how the switching identity of
Theorem~\ref{theo:z_intro} can turn useful, as free-fermion 6V models
are related to bipartite dimers. Theorem~\ref{theo:z_intro} also
suggests that other hidden features of free-fermion 8V-models might
exist. We identify several of them.

\medskip

First, it hints at a possible coupling of pairs of
8V-configurations. If $\tau,\tau'$ are two 8V-configurations, seen as
subsets of the dual edges of $\QQ$, their XOR is still an
8V-configuration; we denote it by $\tau \oplus \tau'$. To ensure that
the 8V-weight actually define a probability measure, we impose a
sufficient condition \eqref{eq:61} on the parametrization. We prove
the following:

\begin{Theo}
  \label{theo:xorcoupl}
  Let $\QQ$ be a quadrangulation of the sphere, and let
  $\alpha,\beta,\alpha',\beta':\FF \to (0,\frac{\pi}{2})$ be such that
  $(\alpha,\beta), (\alpha',\beta'), (\alpha,\beta'), (\alpha',\beta)$
  all satisfy \eqref{eq:61}. Let
  $\tau_{\alpha,\beta}, \tau_{\alpha',\beta'}, \tau_{\alpha,\beta'},
  \tau_{\alpha',\beta}$ be independent 8V-configurations with the
  corresponding Boltzmann distributions. Then
  $\tau_{\alpha,\beta} \oplus \tau_{\alpha',\beta'}$ and
  $\tau_{\alpha,\beta'} \oplus \tau_{\alpha',\beta}$ are equal in
  distribution.
\end{Theo}

Theorem \ref{theo:xorcoupl} is proved via the formalism of
order-disorder variables \cite{KadanoffCeva,Dubedat}, see
Theorem~\ref{theo:8vswitch} for a generalized statement. By specifying
again to
$(\alpha',\beta')=(\beta,\alpha)$ it implies that the XOR of two
independent 8V-configurations (with the same distribution) is
distributed as the XOR of two independent 6V-configurations (with
different distributions). This may be worth investigating from the point of
view of conformal objects in the limit: the XOR of free-fermion
6V-models are effectively double-dimer loops for two independent,
differently distributed (bipartite) dimers configurations. In the identically
distributed case, these are conjectured to converge to CLE(4)
\cite{Kenyon:CLE4,Dubedat:CLE4,BasokChelkak}, but the
differently-distributed case seems more mysterious at the moment. In
any case, the potential occurrence of such conformal ensembles in the
context of 8V-models is surprising.

Although being unexpected, this XOR property is reminiscent of the
coupling identities of \cite{BoutillierDeTiliere:xor,Dubedat}. However
these previous identities involved two independent Ising models, while
our results are naturally associated with four Ising models (see
Corollary~\ref{cor:8vff_is}) and cannot be deduced immediately from
their work.

\medskip

\begin{figure}[h]
  \centering
  \includegraphics[width=3.5cm]{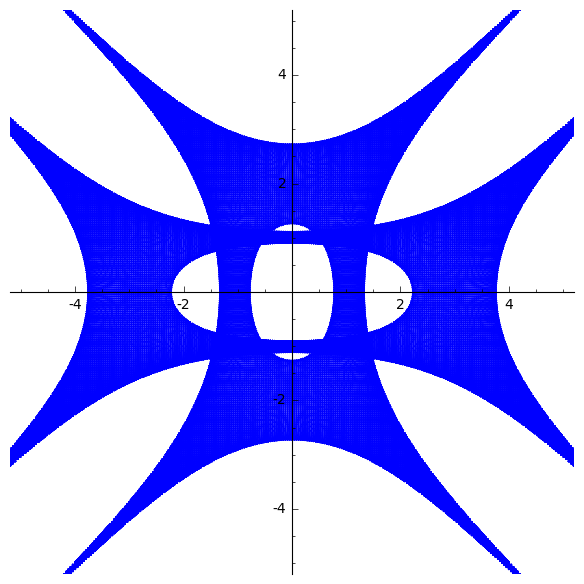}
  \hspace{0.2cm}
  \includegraphics[width=3.5cm]{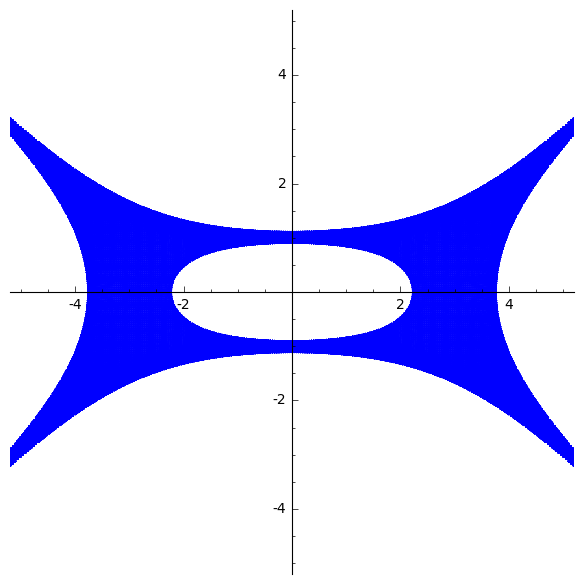}
  \hspace{0.2cm}
  \includegraphics[width=3.5cm]{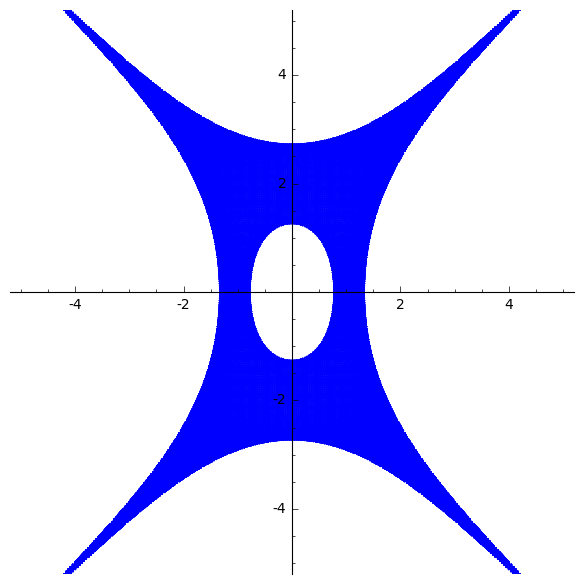}
  \caption{Amoebas of the curves defined by $P^{8V}_{\alpha,\beta}$,
    $P^{6V}_{\alpha}$ and $P^{6V}_{\beta}$ for the square
    lattice. The code for computing and plotting these amoebas is
    written in Sage for the Jupyter notebook and can be found in the
    sources of this paper (\texttt{Amoeba.ipynb}).
  }
  \label{fig:amoebas}
\end{figure}

Second, it is natural to wonder what happens if $\QQ$ is a
quadrangulation of the torus. This is useful in particular to
understand periodic boundary conditions and construct infinite
measures on the full plane, as we explain later. In the toric case,
the 8V-weights $X_{\alpha,\beta}$ are naturally associated with a
characteristic (Laurent) polynomial of two complex variables, denoted
$P^{8V}_{\alpha,\beta}(z,w)$, and defined in \eqref{eq:defP} just like
in the case of the dimer model \cite{kos}. The analogous statement of
Theorem~\ref{theo:z_intro} is the following, which works under milder
hypothesis \eqref{eq:16} (see Theorem~\ref{theo:pol8vswitch}
for a complete statement):
\begin{Theo}
  \label{theo:pswitch_intro}
  Let $\QQ$ be a quadrangulation of the torus. Let
  $\alpha,\beta,\alpha',\beta':\FF\to \left[ 0,2\pi \right)$ be such
  that $(\alpha,\beta)$ and $(\alpha',\beta')$ satisfy
  \eqref{eq:16}. Then
  \begin{equation}
    c_{\alpha,\beta} c_{\alpha',\beta'} \ \ P^{8V}_{\alpha,\beta}(z,w)
    P^{8V}_{\alpha',\beta'}(z,w) \ = \
    c_{\alpha,\beta'} c_{\alpha',\beta} \ \
    P^{8V}_{\alpha,\beta'}(z,w) P^{8V}_{\alpha',\beta}(z,w).
  \end{equation}
\end{Theo}
In particular (see Corollary~\ref{cor:pol8v6v}),
\begin{equation}
  \label{eq:p8vprod}
  P^{8V}_{\alpha,\beta}(z,w) = \tilde{c} \ P^{6V}_{\alpha}(z,w) P^{6V}_{\beta}(z,w).
\end{equation}
The polynomials $P^{6V}_{\alpha}$ and $P^{6V}_{\beta}$ correspond to
bipartite dimers \cite{WuLin, Nienhuis, Dubedat,
  BoutillierDeTiliere:xor}. The curves defined by their zero locus in
$\C^2$ are Harnack curves \cite{kos}. Thus the zero locus of
$P^{8V}_{\alpha,\beta}$ is the union of two Harnack curves. This can
be observed in the amoebae of Figure~\ref{fig:amoebas}, (the amoeba is
the image of the zero locus under the map
$(z,w) \mapsto (\log|z|,\log|w|)$ ).

\medskip

Third, the 8V-models at the free-fermion point corresponds to
non-bipartite dimers, for which we can define a version of a Kasteleyn
matrix $K_{\alpha,\beta}$, see Section~\ref{sec:skew-herm}. The
elements of the inverse of $K_{\alpha,\beta}$ are related to the
correlations of the 8V-model (see Proposition~\ref{prop:corrkast}),
and thus encode the physical behaviour of the model. It
is possible to get a relation between those inverse matrices; precise
statements are given in Theorem~\ref{theo:km1sphere} on the sphere and
in Theorem~\ref{theo:km1torus} on the torus, and the matrix $T$ is
defined by \eqref{eq:8}.
\begin{Theo}
  \label{theo:km1_intro}
  \begin{equation}
    \label{eq:58}
    K_{\alpha,\beta}^{-1} = \frac12 \left( (I+T)
      K_{\alpha,\beta'}^{-1} + (I-T) K_{\alpha',\beta}^{-1}\right).
  \end{equation}
\end{Theo}
This has the remarkable property of holding for all $\alpha',\beta'$,
even though this is not apparent in the right-hand-side. Again we can
set $(\alpha',\beta')=(\beta,\alpha)$, so that this formula relates
8V-correlations to 6V ones, \textit{i.e.} to bipartite dimers
\cite{WuLin,Nienhuis,Dubedat,BoutillierDeTiliere:xor}. We give several
consequences of this identity with the solution of the 8V-model in the
$Z$-invariant regime on lozenge graphs.

\medskip

A model is said to be \emph{$Z$-invariant} when it satisfies a form of
the Yang-Baxter equations, or a star-triangle transformation (see
Figure~\ref{fig:startri}). In the approach via transfer matrices, this
property is often seen as a sufficient condition for the commutativity
of transfer matrices \cite{Baxter:exactly}, see also \cite{Belavin,
  Dutyshev}. However it has also been shown by Baxter that the
star-triangle move is enough information to guess the behavior of the
model on very generic lattices \cite{Baxter8}, without relying on
transfer matrices methods. In particular, it should imply a form of
\textit{locality} for the model, which means that the two-point
correlations depend only on a path (any path) between the two
points. This property is surprising, since in general correlations are
expected to depend on the geometry of the whole graph. We prove that
it holds for $Z$-invariant free-fermion models on lozenge graphs,
which we introduce now.

\begin{figure}[ht]
  \centering \begin{tikzpicture}[scale=0.7]

\begin{scope}[yshift=0cm] 
  \begin{scope}[xshift=9cm]
  \coordinate (hh) at (0 ,2.31) ;
  \coordinate (bd) at (1.633,-1.633) ;
  \coordinate (bg) at (-2.08,-1.002) ;
  \coordinate (hd) at ($(hh)+(bd)$) ;
  \coordinate (hg) at ($(hh)+(bg)$) ;
  \coordinate (bb) at ($(bd)+(bg)$) ;
  \coordinate (cc) at ($(hd) + (bb)- (bd)$);
  \draw [gray] (hh) -- (hd) -- (bd) -- (bb) -- (bg) -- (hg) -- cycle;
  \draw [gray] (hd) -- (cc);
  \draw [gray] (hg) -- (cc);
  \draw [gray] (bb) -- (cc);
  \node [draw=black, fill=black,thick,circle,inner sep=0pt,minimum size=4pt] at (hh) {};
  \node [draw=black, fill=white,circle,thick,inner sep=0pt,minimum size=4pt] at (hd) {};
  \node [draw=black, fill=white,thick,circle,inner sep=0pt,minimum size=4pt] at (hg) {};
  \node [draw=black, fill=black,thick,circle,inner sep=0pt,minimum size=4pt] at (bd) {};
  \node [draw=black, fill=black,thick,circle,inner sep=0pt,minimum size=4pt] at (bg) {};
  \node [draw=black, fill=white,thick,circle,inner sep=0pt,minimum size=4pt] at (bb) {};
  \node [draw=black, fill=black,thick,circle,inner sep=0pt,minimum size=4pt] at (cc) {};
  \draw [dashed] (hh) -- (cc);
  \draw [dashed] (bg) -- (cc);
  \draw [dashed] (bd) -- (cc);
  \draw ($(cc)+(1.5,0)$) node [] {$\frac{\pi}{2} - \theta_1$};
  \draw ($(cc)+(-0.6,1.1)$) node [] {$\frac{\pi}{2} - \theta_2$};
  \draw ($(cc)+(-0.6,-1)$) node [] {$\frac{\pi}{2} - \theta_3$};
  \draw ($(cc) + 0.25*(hh) - 0.25*(cc)$) arc (80:133:0.7cm);
  \draw ($(cc) + 0.25*(bd) - 0.25*(cc)$) arc (-45:30:0.5cm);
  \draw ($(cc) + 0.25*(bb) - 0.25*(cc)$) arc (-90:-155:0.6cm);
\end{scope}
\begin{scope}[xshift=3.5cm, yshift=-0.2cm]
  \draw[>=stealth,<->, line width = 1pt] (0,0) -- (2,0);
\end{scope}
\begin{scope}[xshift=0cm]
  \coordinate (hh) at (0 ,2.31) ;
  \coordinate (bd) at (1.633,-1.633) ;
  \coordinate (bg) at (-2.08,-1.002) ;
  \coordinate (hd) at ($(hh)+(bd)$) ;
  \coordinate (hg) at ($(hh)+(bg)$) ;
  \coordinate (bb) at ($(bd)+(bg)$) ;
  \draw [gray] (hh) -- (hd) -- (bd) -- (bb) -- (bg) -- (hg) -- cycle;
  \draw [gray] (bd) -- (0,0);
  \draw [gray] (bg) -- (0,0);
  \draw [gray] (hh) -- (0,0);
  \node [draw=black, fill=black,thick,circle,inner sep=0pt,minimum size=4pt] at (hh) {};
  \node [draw=black, fill=white,thick,circle,inner sep=0pt,minimum size=4pt] at (hd) {};
  \node [draw=black, fill=white,thick,circle,inner sep=0pt,minimum size=4pt] at (hg) {};
  \node [draw=black, fill=black,thick,circle,inner sep=0pt,minimum size=4pt] at (bd) {};
  \node [draw=black, fill=black,thick,circle,inner sep=0pt,minimum size=4pt] at (bg) {};
  \node [draw=black, fill=white,thick,circle,inner sep=0pt,minimum size=4pt] at (bb) {};
  \node [draw=black, fill=white,thick,circle,inner sep=0pt,minimum size=4pt] at (0,0) {};
  \draw [dashed] (hh) -- (bg) -- (bd) -- (hh);
  \draw ($(hh)+(-0.3,-1.2)$) node [] {$\theta_1$};
  \draw ($(bg)+(1.3,0.2)$) node [] {$\theta_2$};
  \draw ($(bd)+(-0.8,1.3)$) node [] {$\theta_3$};
  \draw ($(hh) + 0.22*(bg) - 0.22*(hh)$) arc (-115:-80:0.8cm);
  \draw ($(bg) + 0.22*(bd) - 0.22*(bg)$) arc (-25:20:0.7cm);
  \draw ($(bd) + 0.22*(hh) - 0.22*(bd)$) arc (110:133:0.9cm);
\end{scope}
\end{scope}

\end{tikzpicture}
  \caption{Star-triangle move on a lozenge graph. The angles satisfy
    $\theta_1 + \theta_2 + \theta_3 =
    \frac{\pi}{2}$.}
  \label{fig:st_loz}
\end{figure}
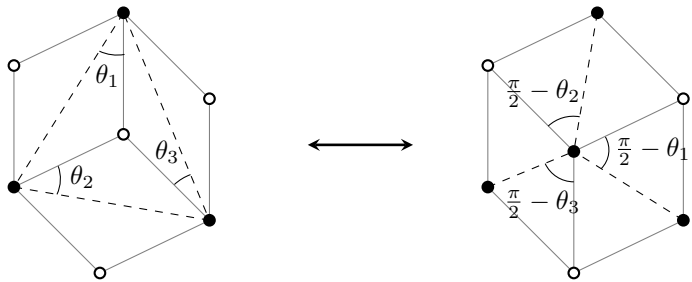

One way to interpret $Z$-invariance geometrically is to use
\textit{isoradial} graphs. For these graphs, the faces of the
quadrangulation $\QQ$ are supposed to be rhombi with the same
edge-length (we call $\QQ$ a \emph{lozenge graph}), and the weights
$A(f),B(f),C(f),D(f)$ at a face $f$ are supposed to depend on the
half-angle $\theta$ of the rhombus $f$ at the black vertices. The
angles $\theta$ satisfy some relations under star-triangle
transformation (see Figure~\ref{fig:st_loz}), and the goal is to
define Boltzmann weights in terms of $\theta$ so as to transform these
relations into the Yang-Baxter equations. Several $Z$-invariant models
have been studied on lozenge graphs, including the bipartite dimer
model \cite{Kenyon:lap}, Ising model
\cite{BoutillierDeTiliere:loc,BoutillierDeTiliereRaschel}, Laplacian
(or spanning forest model)
\cite{Kenyon:lap,BoutillierDeTiliereRaschel:lap}, random cluster model
\cite{DCLiManolescu}. The results of \cite{KenyonSchlenker} also imply
that we do not lose anything by considering the Yang-Baxter equations
on a lozenge graph rather than on a pseudoline arrangement, like that
of \cite{Baxter8}.

The $Z$-invariant weights of the 8V model in the non-checkerboard case
have been parameterized by Baxter \cite{Baxter:8partition, Baxter8,
  Baxter:exactly} and Zamolodchikov \cite{Zamolodchikov}. Other
techniques have appeared since to classify these solutions \cite{SUAW,
  GalleasMartins, KS:YBE, Vieira}. By considering checkerboard
Yang-Baxter equations, more solutions can appear, as noted for
instance in \cite{PerkAuYang}, although no complete parametrization is
known. Here we introduce what seems to be the
first set of checkerboard $Z$-invariant weights for the 8V model, all
included in the free-fermion case.

Let $k,l$ be complex numbers such that $k^2,l^2 \in (-\infty,1)$. For
any real number $x$ let $x_k = \frac{2\mathrm{K}(k)}{\pi} x$, where
$\mathrm{K}(k)$ is the complete elliptic integral of the first kind,
and similarly for $x_l$. Then the following 8V-weights of lozenge
graphs, expressed in terms of Jacobi's elliptic functions (see
\cite{AbramowitzStegun,Lawden}) at a face $f$ with half-angle
$\theta$, satisfy the Yang-Baxter equations:
\begin{equation}
  \label{eq:62}
  \begin{split}
    &A(f) = \sn\left(\theta_k|k\right) +
    \sn\left(\theta_l|l\right) \\
    &B(f) = \cn\left(\theta_k|k\right) +
    \cn\left(\theta_l|l\right) \\
    &C(f) = 1+ \sn\left(\theta_k|k\right)
    \sn\left(\theta_l|l\right) + \cn\left(\theta_k|k\right)
    \cn\left(\theta_l|l\right) \\
    &D(f) = \cn\left(\theta_k|k\right)
    \sn\left(\theta_l|l\right) - \sn\left(\theta_k|k\right)
    \cn\left(\theta_l|l\right)
  \end{split}.
\end{equation}
We prove this in Proposition~\ref{prop:pdszinv}.  When
$(1-k^2)(1-l^2)=1$ (or $k^*=l$ in the notations of
\cite{BoutillierDeTiliereRaschel}) the weights no longer depend on the
bipartite coloring of $\QQ$ (\textit{i.e} if a face $f$ has a
half-angle $\theta$ and $g$ has a half-angle $\frac{\pi}{2}-\theta$,
then $A(f)=B(g)$, etc.), and we recover Baxter's solution in the
free-fermion case. When $k=l$ we get a $Z$-invariant 6V model whose
corresponding dimer model can be found in
\cite{BoutillierDeTiliereRaschel}. At this point it is unknown if
this new parametrization of checkerboard Yang-Baxter equations could
be extended outside of the free-fermion manifold.

We prove that this choice of local weights indeed provides a Gibbs
measure with the locality property, confirming in that case the
prediction of Baxter. To be more precise, every edge of the lozenge
graph $e\in \EE$ is associated with two vertices
$\mathrm{b}_e,\mathrm{w}_e$ that correspond to the ``leg'' of a
decorated graph $G^T$, see Figure~\ref{fig:weightshwl}. In Section
\ref{sec:zinv} we introduce a local operator $K_{k,l}^{-1}$ on the
vertices of $G^T$. It satisfies the following:
\begin{Theo}
  \label{theo:gibbs}
  Let $\QQ$ be a planar lozenge graph. For any $0\leq k < l < 1$,
  there exists a unique probability measure $\mathcal{P}_{8V}$ on the
  space of 8V-configurations equipped with the $\sigma$-field
  generated by cylinders. It is such that for any
  $e_1,\dots,e_p \in \EE$, let $\mathrm{V}$ be the associated vertices
  of $G^T$:
  $\mathrm{V} =
  \{\mathrm{b}_1,\mathrm{w}_1,\dots,\mathrm{b}_p,\mathrm{w}_p\}$, then
  \begin{equation}
    \label{eq:42}
    \mathcal{P}_{8V}(e_1,\dots,e_m \in \tau) = \sqrt{\det \left[\left(K^{-1}_{k,l}\right)_{V}\right]}.
  \end{equation}
  Moreover,
  $\mathcal{P}$ is a translation invariant ergodic Gibbs measure.
\end{Theo}

The construction is based on the switching property in operator form from
Theorem~\ref{theo:km1_intro} and on the work on
bipartite dimers of \cite{BoutillierDeTiliereRaschel}.
When $\QQ$ is $\Z^2$-periodic, it yields the free energy
\eqref{eq:freee}. It also allows for the computation of asymptotics of
coefficients, using results from
\cite{BoutillierDeTiliereRaschel}: under some technical hypothesis, we
show that when $0<k<l<1$ the coefficients of the inverse Kasteleyn
matrix between points at distance $r$ decays as
$r^{-\frac12}e^{-r/\zeta_k}$ (see Corollary~\ref{cor:asymp}). Notice
that the effect of $l$ vanishes in the asymptotics. When $k=0$ the
decay is polynomial, corresponding to a critical model. When
$k \to 0$, we prove that the quantity $\zeta_k$ is a $\Theta(k^{-2})$
in Proposition~\ref{prop:exp_crit}. As $k^2$ plays the role of
$(\beta-\beta_c)$ in usual statistical mechanics terms, this critical
exponent is compatible with that of the correlation length, $\nu=1$,
in the universality class of the Ising model, and with previous predictions
\cite{Baxter:exactly}.

\subsection*{Outline of the paper}

In Section~\ref{sec:def} we properly define the 8V, Ising and dimer
models in spherical, toric and planar settings. We also introduce the
formalism of order-disorder correlators.

In Section~\ref{sec:coupl} we restrict ourselves to the spherical
case. We compute the correlators of free-fermion 8V models and relate
them to Ising ones, generalizing results of \cite{Dubedat}, see
Corollary~\ref{cor:8vff_is}. We prove the coupling result of
Theorem~\ref{theo:xorcoupl}, first in correlators terms
(Theorem~\ref{theo:8vswitch}), then in probabilistic terms; the latter
is deduced from the former by using a discrete Fourier transformation
described in Appendix~\ref{sec:1form}. The results of
Section~\ref{sec:coupl} imply
Theorem~\ref{theo:z_intro}. Sections~\ref{sec:dim}~and~\ref{sec:zinv}
are independent of Section~\ref{sec:coupl}.

In Section~\ref{sec:dim} we define the dimer model associated to the
8V-model, and appropriate versions of Kasteleyn matrices, one being
skew-symmetric and the other being skew-hermitian; they are related by
a diagonal conjugation, see Lemma~\ref{le:diagconj}. We show that the
edge correlations can be expressed as minors of the inverse Kasteleyn
matrices, see Proposition~\ref{prop:corrkast}. We prove the relation
of inverses of Theorem~\ref{theo:km1_intro}. This gives an alternative
proof of Theorem~\ref{theo:z_intro}, as well as its toric counterpart,
Theorem~\ref{theo:pswitch_intro}.

In Section~\ref{sec:zinv} we prove that the weights \eqref{eq:62}
satisfy the Yang-Baxter equations. Using the relation to 6V models
coming from Section~\ref{sec:dim}, we give a local formula for the
inverse Kasteleyn matrix $K_{k,l}^{-1}$ in the full plane in
Section~\ref{sec:local}. We prove the mentioned asymptotics and
critical exponent in Corollary~\ref{cor:asymp} and
Proposition~\ref{prop:exp_crit}. Finally we construct an ergodic Gibbs
measure in the $Z$-invariant case in Section~\ref{sec:gibbs} and prove
Theorem~\ref{theo:gibbs}.

\subsection*{Acknowledgements}
I am very grateful to Cédric Boutillier and Béatrice de Tilière for
the motivation of this paper, as well as their help and supervision. I
also thank Béatrice de Tilière for her critical help in proving
Theorem~\ref{theo:pol8vswitch}. I am grateful to Yacine Ikhlef for
useful discussions.
The author acknowledges support from the Agence Nationale de la Recherche,
Grant Number ANR-18-CE40-0033 (ANR DIMERS).

\section{Definitions}
\label{sec:def}

Let $\QQ$ be a \textit{quadrangulation} of a surface $\mathcal{S}$, that is
a finite connected graph $\QQ=(\VV,\EE)$, without multi-edges and
self-loops, embedded on $\mathcal{S}$ so that edges do
not intersect, and so that the \textit{faces} of $\QQ$ are homeomorphic to
disks and have degree $4$. We denote by $\FF$ its set of faces. We
will focus on three cases:
\begin{itemize}
\item \textit{the spherical case} where $\mathcal{S}$ is the
  two-dimensional sphere and $\QQ$ is finite;
\item \textit{the planar case} where $\mathcal{S}$ is $\R^2$ and $\QQ$
  is infinite and covers the whole plane;
\item \textit{the toric case} where $\mathcal{S}$ is the
  two-dimensional torus and $\QQ$ is finite and bipartite.
\end{itemize}

Since $\QQ$ is bipartite in all these cases, we can fix a partition of
$\VV$ into a set of black vertices $\BB$ and white vertices $\WW$,
such that edges only connect black and white vertices
together. We also set $G^{\BB}$ (resp. $G^{\WW}$) to be the graph
formed by black (resp. white) vertices, joined \textit{iff} they form
the diagonal of a face of $\QQ$. Finally, in the toric case, we
suppose that there are two simple cycles $\gamma^{\BB}_x$ and
$\gamma^{\BB}_y$ on $G^{\BB}$ that wind once, respectively horizontally
and vertically on the torus; see Figure~\ref{fig:torus}.

The dual of $\QQ$, denoted by $\QQ^*$, is the embedded graph whose set
of vertices is $\FF$ and which has edges connecting elements of $\FF$ that
are adjacent in $\QQ$. We denote by $\EE^*$ the set of edges of $\QQ^*$;
for an edge $e\in \EE$, we denote by $e^*$ its corresponding dual edge.

\subsection{Eight-vertex-model}
\begin{figure}[!h]
  \centering \begin{tikzpicture}[scale=0.4]
   
  \draw [color=gray] (-2,0) -- (0,-1) -- (2,0) -- (0,1) -- cycle;
  \draw [ultra thick] (-1.5,0.75) -- (0,0) -- (1.5,0.75);
  \node [draw=black, fill=black,thick,circle,inner sep=0pt,minimum size=4pt] at (-2,0) {};
  \node [draw=black, fill=white,thick,circle,inner sep=0pt,minimum size=4pt] at (0,1) {};
  \node [draw=black, fill=black,thick,circle,inner sep=0pt,minimum size=4pt] at (2,0) {};
  \node [draw=black, fill=white,thick,circle,inner sep=0pt,minimum size=4pt] at (0,-1) {};
  \begin{scope}[xshift = 3.3cm]
    \node [] at (0,0) {$A(f)$};
  \end{scope}
  \begin{scope}[xshift=0cm, yshift=-3.5cm]
    \draw [color=gray] (-2,0) -- (0,-1) -- (2,0) -- (0,1) -- cycle;
    \draw [ultra thick] (-1.5,-0.75) -- (0,0) -- (1.5,-0.75);
    \node [draw=black, fill=black,thick,circle,inner sep=0pt,minimum size=4pt] at (-2,0) {};
    \node [draw=black, fill=white,thick,circle,inner sep=0pt,minimum size=4pt] at (0,1) {};
    \node [draw=black, fill=black,thick,circle,inner sep=0pt,minimum size=4pt] at (2,0) {}; 
    \node [draw=black, fill=white,thick,circle,inner sep=0pt,minimum size=4pt] at (0,-1){};
    \begin{scope}[xshift = 3.3cm]
      \node [] at (0,0) {$A(f)$};
    \end{scope}
  \end{scope}

\begin{scope}[xshift=8cm]
  \draw [color=gray] (-2,0) -- (0,-1) -- (2,0) -- (0,1) -- cycle;
  \draw [ultra thick] (-1.5,0.75) -- (0,0) -- (-1.5,-0.75);
  \node [draw=black, fill=black,thick,circle,inner sep=0pt,minimum size=4pt] at (-2,0) {};
  \node [draw=black, fill=white,thick,circle,inner sep=0pt,minimum size=4pt] at (0,1) {};
  \node [draw=black, fill=black,thick,circle,inner sep=0pt,minimum size=4pt] at (2,0) {};
  \node [draw=black, fill=white,thick,circle,inner sep=0pt,minimum size=4pt] at (0,-1) {};
  \begin{scope}[xshift = 3.3cm]
    \node [] at (0,0) {$B(f)$};
  \end{scope}
  \begin{scope}[xshift=0cm, yshift=-3.5cm]
    \draw [color=gray] (-2,0) -- (0,-1) -- (2,0) -- (0,1) -- cycle;
    \draw [ultra thick] (1.5,-0.75) -- (0,0) -- (1.5,0.75);
    \node [draw=black, fill=black,thick,circle,inner sep=0pt,minimum size=4pt] at (-2,0) {};
    \node [draw=black, fill=white,thick,circle,inner sep=0pt,minimum size=4pt] at (0,1) {};
    \node [draw=black, fill=black,thick,circle,inner sep=0pt,minimum size=4pt] at (2,0) {}; 
    \node [draw=black, fill=white,thick,circle,inner sep=0pt,minimum size=4pt] at (0,-1){};
    \begin{scope}[xshift = 3.3cm]
      \node [] at (0,0) {$B(f)$};
    \end{scope}
  \end{scope}
\end{scope}

\begin{scope}[xshift=16cm]
  \draw [color=gray] (-2,0) -- (0,-1) -- (2,0) -- (0,1) -- cycle;
  \draw [ultra thick] (-1.5,0.75) -- (1.5,-0.75);
  \draw [ultra thick] (1.5,0.75) -- (-1.5,-0.75);
  \node [draw=black, fill=black,thick,circle,inner sep=0pt,minimum size=4pt] at (-2,0) {};
  \node [draw=black, fill=white,thick,circle,inner sep=0pt,minimum size=4pt] at (0,1) {};
  \node [draw=black, fill=black,thick,circle,inner sep=0pt,minimum size=4pt] at (2,0) {};
  \node [draw=black, fill=white,thick,circle,inner sep=0pt,minimum size=4pt] at (0,-1) {};
  \begin{scope}[xshift = 3.3cm]
    \node [] at (0,0) {$C(f)$};
  \end{scope}
  \begin{scope}[xshift=0cm, yshift=-3.5cm]
    \draw [color=gray] (-2,0) -- (0,-1) -- (2,0) -- (0,1) -- cycle;
    \node [draw=black, fill=black,thick,circle,inner sep=0pt,minimum size=4pt] at (-2,0) {};
    \node [draw=black, fill=white,thick,circle,inner sep=0pt,minimum size=4pt] at (0,1) {};
    \node [draw=black, fill=black,thick,circle,inner sep=0pt,minimum size=4pt] at (2,0) {}; 
    \node [draw=black, fill=white,thick,circle,inner sep=0pt,minimum size=4pt] at (0,-1){};
    \begin{scope}[xshift = 3.3cm]
      \node [] at (0,0) {$C(f)$};
    \end{scope}
  \end{scope}
\end{scope}

\begin{scope}[xshift=24cm]
  \draw [color=gray] (-2,0) -- (0,-1) -- (2,0) -- (0,1) -- cycle;
  \draw [ultra thick] (-1.5,0.75) -- (1.5,-0.75);
  \node [draw=black, fill=black,thick,circle,inner sep=0pt,minimum size=4pt] at (-2,0) {};
  \node [draw=black, fill=white,thick,circle,inner sep=0pt,minimum size=4pt] at (0,1) {};
  \node [draw=black, fill=black,thick,circle,inner sep=0pt,minimum size=4pt] at (2,0) {};
  \node [draw=black, fill=white,thick,circle,inner sep=0pt,minimum size=4pt] at (0,-1) {};
  \begin{scope}[xshift = 3.3cm]
    \node [] at (0,0) {$D(f)$};
  \end{scope}
  \begin{scope}[xshift=0cm, yshift=-3.5cm]
    \draw [color=gray] (-2,0) -- (0,-1) -- (2,0) -- (0,1) -- cycle;
    \draw [ultra thick] (1.5,0.75) -- (-1.5,-0.75);
    \node [draw=black, fill=black,thick,circle,inner sep=0pt,minimum size=4pt] at (-2,0) {};
    \node [draw=black, fill=white,thick,circle,inner sep=0pt,minimum size=4pt] at (0,1) {};
    \node [draw=black, fill=black,thick,circle,inner sep=0pt,minimum size=4pt] at (2,0) {}; 
    \node [draw=black, fill=white,thick,circle,inner sep=0pt,minimum size=4pt] at (0,-1){};
    \begin{scope}[xshift = 3.3cm]
      \node [] at (0,0) {$D(f)$};
    \end{scope}
  \end{scope}
\end{scope}

\end{tikzpicture}
  \caption{The eight possible configurations for $\tau$ at a face $f\in \FF$ and
    their local weight $w_f(\tau)$.}
  \label{fig:locconf}
\end{figure}
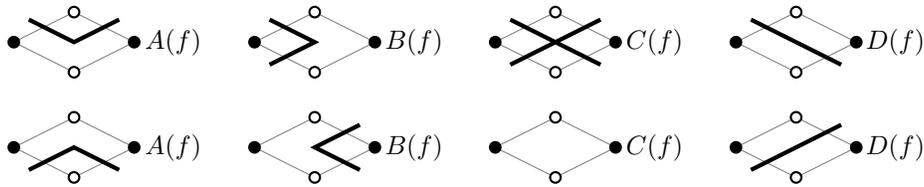

An 8V-configuration is a subset $\tau \subset \EE^*$
such that, at each face $f\in \FF$, an even number of dual edges that
belong to $\tau$ meet at $f$. Thus at any face $f\in \FF$, $\tau$ has
to be one of the eight types shown in Figure~\ref{fig:locconf}. Let
$\Omega(\QQ)$ be the set of all 8V-configurations on $\QQ$.

Let $A,B,C,D$ be four functions from $\FF$ to $\R$. We associate to
$f$ a local weight function $w_f$, such that $w_f(\tau)$ is either
$A(f),B(f),C(f)$ or $D(f)$ depending on the local configuration, as in
Figure~\ref{fig:locconf}. In the spherical and toric cases, the global
weight of $\tau$ is defined as
\begin{equation}
  \label{eq:1}
  w_{8V}(\tau) = \prod_{f \in \FF}w_f(\tau).
\end{equation}
It is more difficult to make sense of formula~\eqref{eq:1} in the
planar case, where the product becomes infinite, requiring the
construction of an appropriate Gibbs measure. We discuss this
construction in the case of a $Z$-invariant, free-fermion model in
Section~\ref{sec:gibbs}.

For the remainder of this section, we only deal with the spherical and
toric cases.
Let $X = (A,B,C,D)$ denote the whole set of weights. The partition
function of the model is
\begin{equation}
  \label{eq:2}
  \ZZ_{8V} = \ZZ_{8V}(\QQ,X) = \sum_{\tau\in\Omega(\QQ)}w_{8V}(\tau).
\end{equation}
When the weights $X$ take values in positive real numbers, the
Boltzmann measure associated to $X$ is the probability measure on
$\Omega(\QQ)$ defined by
\begin{equation}
  \label{eq:pboltzmann}
  \Pr_{8V}(\tau) = \frac{w(\tau)}{\ZZ_{8V}}.
\end{equation}

Even if we
are only interested in the positive real values \textit{in fine}, it is
convenient to let the weights $X$ take any real values. In this case,
$w(\tau)$ and $\ZZ_{8V}$ are still well-defined but \eqref{eq:pboltzmann} does
not define a probability measure.

A \emph{gauge transformation} at some face $f\in\FF$ consists in
multiplying the weights $A(f),B(f),C(f),D(f)$ by the same constant
$\lambda\neq 0$. This has the effect of multiplying all weights
$w_{8V}(\tau)$ by $\lambda$, and $\ZZ_{8V}$ is also multiplied by
$\lambda$. Thus the Boltzmann measure is unchanged.

The weights $X=(A,B,C,D)$ are said to be \emph{free-fermion} if
\begin{equation}
  \label{eq:64}
  \forall f \in \FF,\  A(f)^2 + B(f)^2 = C(f)^2 + D(f)^2.
\end{equation}
They are said to be \emph{standard} if
\begin{equation}
  \label{eq:65}
  \forall f \in \FF,\  C(f) \neq 0.
\end{equation}

\medskip

\begin{Le}
  \label{le:ff}
  Let $a,b,c,d$ be real numbers such that $c \neq 0$ and
  \begin{equation}
    \label{eq:25}
    a^2 + b^2 = c^2 + d^2.
  \end{equation}
  Then there exists a couple $(\alpha,\beta) \in (\R/2\pi\Z)^2$
  such that in homogeneous coordinates,
  \begin{align}
    \label{eq:weight_form1}
    \begin{bmatrix}
      a:b:c:d
    \end{bmatrix}
    & =
    \begin{bmatrix}
      \sin \alpha  + \sin \beta :\\
      \cos \alpha + \cos \beta :\\
      1 + \sin \alpha \sin \beta + \cos \alpha \cos \beta :\\
      \cos \alpha \sin \beta - \sin \alpha \cos \beta
    \end{bmatrix}
    \\
    \label{eq:weight_form2}
    & =
    \begin{bmatrix}
      \sin\left(\frac{\alpha + \beta}{2}\right) :
      \cos\left(\frac{\alpha + \beta}{2}\right) :
      \cos\left(\frac{\alpha - \beta}{2}\right) :
      \sin\left(\frac{-\alpha + \beta}{2}\right)
    \end{bmatrix}.
  \end{align}
\end{Le}

\begin{proof}
  We can rewrite \eqref{eq:25} as
  \begin{equation}
    \label{eq:3}
    \left(\frac{a}{\lambda}\right)^2 +
    \left(\frac{b}{\lambda}\right)^2 =
    \left(\frac{c}{\lambda}\right)^2 +
    \left(\frac{d}{\lambda}\right)^2 = 1
  \end{equation}
  for some constant $\lambda > 0$. Thus there exists $u,v\in\R/2\pi\Z$
  such that
  \begin{equation}
    \label{eq:12}
    [a:b:c:d] = [\sin u : \cos u : \cos v : - \sin v].
  \end{equation}
  Then we define $\alpha=u+v, \beta = u-v$, which gives the form
  \eqref{eq:weight_form2} of the homogeneous coordinates. The form
  \eqref{eq:weight_form1} is obtained by multiplying all weights of
  \eqref{eq:weight_form2} by
  $2 \cos\left(\frac{\alpha-\beta}{2}\right)$, which is non
  zero because $c\neq 0$, and performing simple trigonometric
  computations.
\end{proof}

For that reason, we fix two functions $\alpha, \beta : \FF \to
\R/2\pi\Z$ and we define the associated free-fermion weights by the
following formula, implicitly evaluated at any $f\in\FF$:
\begin{equation}
  \label{eq:pds86}
  X_{\alpha,\beta} =
  \begin{pmatrix}
    A_{\alpha,\beta}\\
    B_{\alpha,\beta}\\
    C_{\alpha,\beta}\\
    D_{\alpha,\beta}
  \end{pmatrix}
  =
  \begin{pmatrix}
    \sin \alpha  + \sin \beta \\
    \cos \alpha + \cos \beta \\
    1 + \sin \alpha \sin \beta + \cos \alpha \cos \beta \\
    \cos \alpha \sin \beta - \sin \alpha \cos \beta
  \end{pmatrix}.
\end{equation}
By Lemma~\ref{le:ff}, any standard free-fermion 8V-model can be
written in this way, after proper gauge transformations.

\begin{Rk}\leavevmode
  \begin{itemize}
  \item The weights $X_{\alpha,\beta}$ satisfy
    \begin{equation}
      \label{eq:relc}
      2C_{\alpha,\beta} =
      A^2_{\alpha,\beta} +
      B_{\alpha,\beta}^2
      = C_{\alpha,\beta}^2 +
      D_{\alpha,\beta}^2.
    \end{equation}
    As a result, given standard free-fermion 8V-weights $X=(A,B,C,D)$,
    one gets to the weights $X_{\alpha,\beta}$ by applying
    gauge transformations at each face with parameter $\lambda(f) =
    \frac{2C(f)}{A(f)^2+B(f)^2}$.

    \item The weight $X_{\alpha,\beta}$ are standard \emph{iff}
    \begin{equation}
      \label{eq:16}
      \forall f \in \FF, \ \alpha(f)  - \beta(f) \neq \pi [2\pi].
    \end{equation}
    We also say that $\alpha,\beta$ are \emph{standard} when
    \eqref{eq:16} is satisfied. Note that if this is not the case at
    some face $f\in\FF$, then all the weights $A(f),B(f),C(f),D(f)$ vanish.
  \item If $\alpha,\beta$ lie in the range
    \begin{equation}
      \label{eq:61}
      \forall f \in \FF, \ 0 < \alpha(f) \leq \beta(f) < \frac{\pi}{2}
    \end{equation}
    then the weights
    $A_{\alpha,\beta}, B_{\alpha,\beta}, C_{\alpha,\beta}$ are
    positive and $D_{\alpha,\beta}$ is non-negative. As a result, the
    Boltzmann measure is a probability measure.
  \item If $\alpha=\beta$, then the weights
    $D_{\alpha,\beta}$ vanish and we are left with a 6V
    model. We simply denote $X_{\alpha}$ the weights in that
    case, and $\ZZ_{6V}(\QQ,X_{\alpha})$ for the partition
    function. We have
    \begin{equation}
      \label{eq:63}
      [A_{\alpha} : B_{\alpha} :
      C_{\alpha}] = [\sin \alpha : \cos \alpha : 1].
    \end{equation}
  \item Switching $\alpha$ and $\beta$ has the effect of multiplying
    the weights $D$ by $-1$. Since any 8V-configuration contains an
    even number of $D$ faces (see \cite{Dubedat}), in both the
    spherical and toric cases,
    \begin{equation}
      \label{eq:switch_a_b}
      \ZZ_{8V}(\QQ,X_{\alpha,\beta}) = \ZZ_{8V}(\QQ,X_{\beta,\alpha}).
    \end{equation}
  \end{itemize}
\end{Rk}

\subsection{Ising model}
\label{sec:is}

Let $\alpha,\beta :\FF \to (0,\frac{\pi}{2})$. Let
$J^{\alpha}_{\BB},J^{\beta}_{\WW}: \FF \to \R$ be the following
\emph{coupling constants}:
\begin{equation}
  \label{eq:couplis}
  \forall f \in \FF, \ \ \ \
  J^{\alpha}_{\BB}(f) = \frac12 \ln\left(\frac{1+\sin \alpha(f)}{\cos
  \alpha(f)} \right), \ \ \ \ \ \ \ \
  J^{\beta}_{\WW}(f) = \frac12 \ln\left(\frac{1+\cos \beta(f)}{\sin
  \beta(f)} \right).
\end{equation}

An \emph{spin configuration} on $G^{\BB}$ (resp. $G^{\WW}$)
is an application from $\BB$ (resp. $\WW$) to $\{\pm 1\}$. The weight
of such a configuration $\sigma_{\BB}$ (resp. $\sigma_{\WW}$) is
defined as
\begin{equation}
  \label{eq:pdsis1}
    w_{\BB}(\sigma_{\BB}) = \prod_{f\in
      \FF}e^{J^{\alpha}_{\BB}(f) \ \sigma_u \sigma_v},  \ \ \ \ \ \ \ \
    w_{\WW}(\sigma_{\WW}) = \prod_{f\in
      \FF}e^{J^{\beta}_{\WW}(f) \ \sigma_x
      \sigma_y},
\end{equation}
where $u,v$ are the black vertices of $f$, and $x,y$ its white vertices.

The partition functions are:
\begin{equation}
    \ZZ_{\text{Ising }}^{\BB}  = \ZZ_{\text{Ising }}^{\BB} (J^{\alpha}_{\BB}) =
    \sum_{\sigma_\BB}w_{\BB}(\sigma_{\BB}),  \ \ \ \ \ \ \ \
    \ZZ_{\text{Ising }}^{\WW} = \ZZ_{\text{Ising }}^{\WW} (J^{\beta}_{\WW}) =
    \sum_{\sigma_{\WW}}w_{\WW}(\sigma_{\WW}),
\end{equation}
where the sums are over spin configurations.
Again, the associated Boltzmann measure is
\begin{equation}
  \label{eq:pboltzmannIs}
    \Pr_{\text{Ising}}^{\BB}(\sigma_{\BB}) =
    \frac{w_{\BB}(\sigma_{\BB})}{ \ZZ_{\text{Ising }}^{\BB} },  \ \ \ \ \ \ \ \
    \Pr_{\text{Ising}}^{\WW}(\sigma_{\WW}) =
    \frac{w_{\WW}(\sigma_{\WW})}{ \ZZ_{\text{Ising }}^{\WW} }.
\end{equation}

\subsection{Dimer model}

Let $G=(V,E)$ be a finite graph, equipped with real weights on the edges
$(\nu_e)_{e\in E}$. A \emph{dimer configuration}, or \emph{perfect matching},
is a subset of edges $m\subset E$ such that every vertex of $G$ is
adjacent to exactly one edge of $m$. We denote by $\mathcal{M}(G)$ the
set of all dimer configurations on $G$.

The Boltzmann measure on $\mathcal{M}(G)$ is defined by
\begin{equation}
  \label{eq:46}
  \Pr_{\text{dim}}(m) = \frac{\prod_{e\in m} \nu_e}{\ZZ_{\text{dim}}(G,\nu)}
\end{equation}
where $\ZZ_{\text{dim}}(G,\nu)$ is the partition function:
\begin{equation}
  \label{eq:48}
  \ZZ_{\text{dim}}(G,\nu) = \sum_{m\in\mathcal{M}(G)}\prod_{e\in m} \nu_e.
\end{equation}

\subsection{Order and disorder variables}

The notions of order and disorder variables were defined by Kadanoff
and Ceva for the Ising model \cite{KadanoffCeva} and play a central
role in the study of spinor and fermionic observables; see for
instance \cite{ChelkakCimasoniKassel} for a unifying review. The
definition for the Ising model is classical; for the 8V model, we
adapt definitions of Dubédat \cite{Dubedat}.

For these definitions $\QQ$ can be a quadrangulation in the spherical
or toric case.

\subsubsection{Ising correlators}
\label{sec:ising-model}

Let $B_0\subset \BB$ and $W_1\subset \WW$ be two subset of black and
white vertices of $\QQ$, of even cardinality. Let $\gamma_{B_0}$ be
the union of disjoint simple paths on $G^{\BB}$ that connect the
vertices of $B_0$ pairwise (these are called \textit{order lines});
$\gamma_{B_0}$ can be alternatively seen as a subset of $\FF$. We
similarly define $\gamma_{W_1}$ as a union of disjoint simple paths on
$G^{\WW}$ that connect the $W_1$ pairwise (also called
\textit{disorder lines}).

Let $\alpha : \FF \to \left(0,\frac{\pi}{2}\right)$. We modify the
coupling constants $J^{\alpha}_{\BB}$ by adding $i\frac{\pi}{2}$ to
the coupling constant at $f$ when $f\in \gamma_{B_0}$, and afterwards
multiplying the coupling constant by $-1$ when $f\in \gamma_{W_1}$
(the order is important). Let $J'$ be these new coupling
constants. Then the \emph{mixed correlator} of
Kadanoff and Ceva is defined as
\begin{equation}
  \label{eq:is_cor_b}
  \left<\prod_{b\in B_0}\sigma_b \prod_{w\in W_1}\mu_w
  \right>^{\text{Ising}
    \BB}_{\alpha,\gamma_{B_0},\gamma_{W_1}}
  = \left<\sigma(B_0) \mu(W_1) \right>^{\text{Ising}
    \BB}_{\alpha,\gamma_{B_0},\gamma_{W_1}}
  := \ZZ^{\BB}_{\text{Ising}}(J').
\end{equation}
This depends on the choice of paths and the order of operations on
the coupling constants, but only up to a global sign. The \emph{order}
variables are simply the spins $\sigma_b$, with $\left< \cdot \right>$
representing an unnormalized expectation under the Boltzmann
measure. The \emph{disorder} variables $\mu_w$ represent defects in
the configuration, and are conjugated with order variables under
Kramers-Wannier duality \cite{KramersWannier}. Again we refer to
\cite{KadanoffCeva}.

Similarly for the Ising model on $G^{\WW}$, if $W_0\subset \WW$ and
$B_1\subset\BB$ are even subsets, we chose paths
$\gamma_{W_0},\gamma_{B_1}$. Then for
$\beta : \FF \to \left(0,\frac{\pi}{2}\right)$, we add
$i\frac{\pi}{2}$ to the constants $J^{\WW}_{\beta}$ on $\gamma_{W_0}$,
then multiply the constants by $-1$ on $\gamma_{B_1}$, and we name the
new constants $J''$. The mixed correlator is
\begin{equation}
  \label{eq:is_cor_w}
  \left<\prod_{w\in W_0}\sigma_w \prod_{b\in B_1}\mu_b
  \right>^{\text{Ising}
    \WW}_{\beta,\gamma_{W_0},\gamma_{B_1}}
  = \left<\sigma(W_0) \mu(B_1) \right>^{\text{Ising}
    \WW}_{\beta,\gamma_{W_0},\gamma_{B_1}}
  := \ZZ^{\WW}_{\text{Ising}}(J'').
\end{equation}

\subsubsection{8V correlators}
\label{sec:8v-corr}

Order and disorder variables for the 8V-model are defined in
\cite{Dubedat}. The following definition is original but it is easy to
check that it is equivalent to that of \cite{Dubedat}. In the 8V case,
order and disorder variables can be located on either black or white
vertices of $\QQ$.

\begin{Def}
  \label{def:corr8}
  Let $\QQ$ be a quadrangulation in the spherical case. Let
  $X=(A,B,C,D):\FF\to\R^4$ be a family of weights. For every
  $f\in \FF$, we define operators
  $\nu^{\BB}_f,\nu^{\WW}_f,\xi^{\BB}_f,\xi^{\WW}_f$ that act on $X$ by
  transforming $X(f)$ in the following way:
  \begin{align}
    \nu^{\BB}_f &:
    \begin{pmatrix}
      A(f) \\ B(f) \\ C(f) \\ D(f)
    \end{pmatrix}
    \mapsto
    \begin{pmatrix}
      A(f) \\ - B(f) \\ C(f) \\ - D(f)
    \end{pmatrix}
    && \nu^{\WW}_f :
    \begin{pmatrix}
      A(f) \\ B(f) \\ C(f) \\ D(f)
    \end{pmatrix}
    \mapsto
    \begin{pmatrix}
      - A(f) \\ B(f) \\ C(f) \\ - D(f)
    \end{pmatrix}
    \\
    \xi^{\BB}_f &:
    \begin{pmatrix}
      A(f) \\ B(f) \\ C(f) \\ D(f)
    \end{pmatrix}
    \mapsto
    \begin{pmatrix}
      C(f) \\ D(f) \\ A(f) \\ B(f)
    \end{pmatrix}
    && \xi^{\WW}_f :
    \begin{pmatrix}
      A(f) \\ B(f) \\ C(f) \\ D(f)
    \end{pmatrix}
    \mapsto
    \begin{pmatrix}
      D(f) \\ C(f) \\ B(f) \\ A(f)
    \end{pmatrix}
  \end{align}

  Let $B_0,B_1 \subset \BB$ (resp. $W_0,W_1\subset \WW$) be two even
  subsets of black (resp. white) vertices, with simple paths
  $\gamma_{B_0},\gamma_{B_1}$ (resp. $\gamma_{W_0},\gamma_{W_1}$) joining them
  pairwise. As these paths use black (resp. white) diagonals of faces,
  we can identify them with subsets of $\FF$. Let $\gamma =
  (\gamma_B,\gamma_W,\gamma_{B'},\gamma_{W'})$. We define modified
  weights $X'_{\gamma}$ obtained by the following composition of operators:
  \begin{equation}
    \label{eq:operX}
    X'_{\gamma} = \left(\prod_{f\in \gamma_{B_1}}\xi^{\BB}_f
      \prod_{f\in \gamma_{W_1}}\xi^{\WW}_f  \prod_{f\in \gamma_{B_0}}\nu^{\BB}_f \prod_{f\in
        \gamma_{W_0}}\nu^{\WW}_f  \right) X.
  \end{equation}
  We define the \textit{mixed correlator} as:
  \begin{equation}
    \left< \prod_{b\in {B_0}}\sigma_b \prod_{w\in
        {W_0}}\sigma_w  \prod_{b\in B_1}\mu_b  \prod_{w\in W_1}\mu_w\right>_{X,\gamma}^{8V} : =
    \ZZ_{8V}(X'_{\gamma}).
  \end{equation}
  We will also use the shorthand notation
  $\left< \sigma(B_0)\sigma(W_0)\mu(B_1)\mu(W_1) \right>^{8V}_{X,\gamma}$.
\end{Def}

\begin{Rk}\leavevmode
  \begin{itemize}
  \item Mixed correlators may depend on the set of paths $\gamma$, but only
    up to a global sign \cite{Dubedat}.  Also note that in
    \eqref{eq:operX} we always apply order operators before disorder
    ones. This is important because these operators do not always
    commute. The only cases were they do not commute are
    \begin{equation}
      \label{eq:32}
      \begin{split}
        \nu_f^{\BB} \xi_f^{\WW}= -\xi_f^{\WW} \nu_f^{\BB},\\
        \nu_f^{\WW} \xi_f^{\BB}= -\xi_f^{\BB} \nu_f^{\WW},
      \end{split}
    \end{equation}
    but like for path dependence, changing the order might only multiply
    the correlator by a factor $-1$.
  \item Again these correlators can be interpreted as unnormalized
    expectations under the Boltzmann measure. If $b,b'\in B_0$, then
    the couple of order variables $\sigma_b\sigma_{b'}$ represents the
    random variable $(-1)^n$ where $n$ is the number of edges in the
    8V configuration on any path on $\QQ$ between $b$ and $b'$.

    The disorder variables are equivalent to modifying every
    eight-vertex configuration by applying a XOR with the configuration
    of ``half edges'' shown in Figure~\ref{fig:ldis}. The resulting
    ``configuration'' is no longer a subset of edges of $\QQ$, but we
    could still define its weight as in \eqref{eq:1}. The advantage of
    modifying weights with the operators of Definition~\ref{def:corr8}
    is that we never actually have to work with these
    \emph{disordered} configurations.

    \begin{figure}[h]
      \centering
      \def\svgwidth{10cm}
      \import{./}{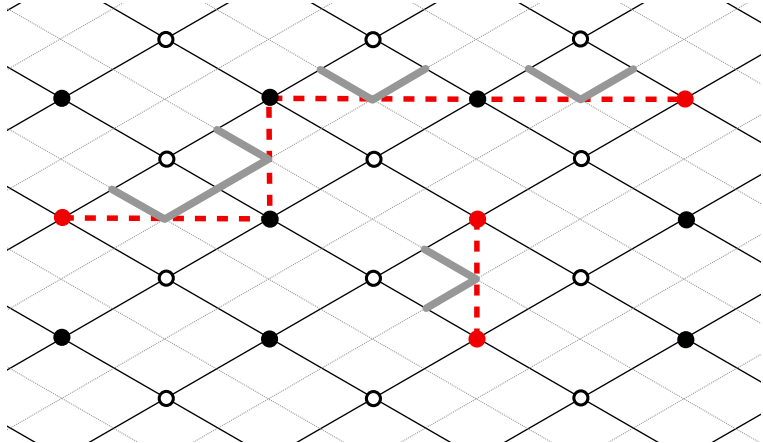}
      \caption{A piece of a quadrangulation, a subset $B_1\subset \BB$
        with paths $\gamma_{B_1}$ joining them pairwise (dashed), and a
        partial configuration (bold grey)}
      \label{fig:ldis}
    \end{figure}

  \end{itemize}
\end{Rk}

\section{Couplings of 8V-models}
\label{sec:coupl}

We review classical results on the Ising-8V correspondence
\cite{Baxter8}, on the 8V duality \cite{Wu:dual}, in terms of order
and disorder variables \cite{Dubedat}. We then apply them to prove couplings results
for different free-fermion 8V-models.

\subsection{Spin-vertex correspondence}
\label{sec:spin-vert-corr}

The spin-vertex correspondence comes from the following simple remark,
that seems to be due to Baxter \cite{Baxter8}. If we superimpose two
spin configurations, one on $G^{\BB}$ and one on $G^{\WW}$, and we
draw the interfaces between $+1$ and $-1$ spins, we get an
8V-configuration. This transformation is two-to-one, and can be made
weight-preserving by choosing the appropriate 8V weights.
\begin{Prop}[\cite{Baxter8,Dubedat}]
  \label{prop:spin_ising}
  Let $\QQ$ be a quadrangulation in the spherical case, and
  $\alpha,\beta : \FF \to (0,\frac{\pi}{2})$. Consider the 8V-weights
  $X: \FF \to \R^4$ given by
  \begin{equation}
    \label{eq:8v_gamma}
    X =
    \begin{pmatrix}
      e^{J^{\alpha}_{\BB}-J^{\beta}_{\WW}} \\
      e^{-J^{\alpha}_{\BB}+J^{\beta}_{\WW}}\\
      e^{J^{\alpha}_{\BB}+J^{\beta}_{\WW}} \\
      e^{-J^{\alpha}_{\BB}-J^{\beta}_{\WW}}
    \end{pmatrix}.
  \end{equation}
  Then for any $B_0,B_1,W_0,W_1$ and paths $\gamma$ as in
  Definition~\ref{def:corr8},
  \begin{equation}
    \label{eq:spin_vertex_corr}
    2 \left< \sigma(B_0)\sigma(W_0)\mu(B_1)\mu(W_1)
    \right>^{8V}_{X,\gamma} =
    \left<\sigma(B_0) \mu(W_1) \right>^{\text{Ising}
      \BB}_{\alpha,\gamma_{B_0},\gamma_{W_1}} \
    \left<\sigma(W_0) \mu(B_1) \right>^{\text{Ising}
      \WW}_{\beta,\gamma_{W_0},\gamma_{B_1}}.
  \end{equation}
  In particular,
  \begin{equation}
    \label{eq:spin_vertex_z}
    2 \ZZ_{8V}(\QQ,X) = \ZZ_{\text{Ising }}^{\BB} (J^{\alpha}_{\BB})
    \times  \ZZ_{\text{Ising}}^{\WW} (J^{\beta}_{\WW}).
  \end{equation}
\end{Prop}

\subsection{Modifications of weights}
One key feature of the 8V-model is its duality relation.
This is an instance of Kramers-Wannier duality \cite{KramersWannier},
and in the case of the eight-vertex model it was discovered by Wu
\cite{Wu:dual} using high-temperature expansion techniques. The
formulation for correlators comes from Dubédat \cite{Dubedat}, and
means that duality exchanges order and disorder. We give an
interpretation in terms of discrete Fourier transform in Appendix~\ref{sec:1form}.

\begin{Prop}[\cite{Wu:dual, Dubedat}]
  \label{prop:8vdu}
  Let $\QQ$ be a quadrangulation in the spherical case, and let $X=(A,B,C,D):\FF
  \to \R^4$ be any 8V-weights. Let
  $\hat{X}=(\hat{A},\hat{B},\hat{C},\hat{D})$ be defined by
  \begin{equation}
    \label{eq:8vdu}
    \forall f \in \FF, \
    \begin{pmatrix}
      \hat{A}(f) \\
      \hat{B}(f) \\
      \hat{C}(f) \\
      \hat{D}(f)
    \end{pmatrix}
    =\frac12
    \begin{pmatrix}
      1 & -1 & 1 & -1 \\
      -1 & 1 & 1 & -1 \\
      1 & 1 & 1 & 1 \\
      -1 & -1 & 1 & 1 \\
    \end{pmatrix}
    \begin{pmatrix}
      A(f) \\
      B(f) \\
      C(f) \\
      D(f)
    \end{pmatrix}.
  \end{equation}
  Then for any $B_0,B_1,W_0,W_1$ and paths $\gamma$ as in
  Definition~\ref{def:corr8}, let
  $\hat{\gamma}=(\gamma_{B_1},\gamma_{W_1},\gamma_{B_0},\gamma_{W_0})$;
  we have
  \begin{equation}
    \label{eq:102}
    \left< \sigma(B_0)\sigma(W_0)\mu(B_1)\mu(W_1) \right>^{8V}_{X,\gamma}
    = \left< \sigma(B_1)\sigma(W_1)\mu(B_0)\mu(W_0)
    \right>^{8V}_{\hat{X},\hat{\gamma}}.
  \end{equation}
  In particular,
  \begin{equation}
    \label{eq:8vdu_Z}
    \ZZ_{8V}(\QQ,X) = \ZZ_{8V}(\QQ,\hat{X}).
  \end{equation}
\end{Prop}

Another transformation of weights consists in multiplying all weights
$D(f)$ by $-1$. As any 8V-configuration contains an even number of
faces of type $D$, this does not change its global weight. However, in
correlators containing disorder operators, the effect is non trivial;
a result of \cite{Dubedat} is that $\mu$ variables become $\sigma
\mu$, while $\sigma$ variables are unchanged, we rephrase it here using
symmetric differences $\triangle$.

\begin{Prop}[\cite{Dubedat}]
  \label{prop:dmd}
  Let $\QQ$ be a quadrangulation in the spherical case, and let
  $X=(A,B,C,D):\FF\to\R^4$ be any 8V-weights. Let $X'=(A,B,C,-D)$.

  Then for any $B_0,B_1,W_0,W_1$ and paths $\gamma$ as in
  Definition~\ref{def:corr8}, let
  $\gamma'=(\gamma_{B_0} \triangle \gamma_{B_1}, \gamma_{W_0}
  \triangle \gamma_{W_1}, \gamma_{B_1},\gamma_{W_1})$; we have
  \begin{equation}
    \left< \sigma(B_0)\sigma(W_0)\mu(B_1)\mu(W_1) \right>^{8V}_{X,\gamma} =
    \left< \sigma(B_0\triangle B_1)\sigma(W_0\triangle W_1)\mu(B_1)\mu(W_1)
    \right>^{8V}_{X',\gamma'}.
  \end{equation}
  In particular,
  \begin{equation}
    \ZZ_{8V}(\QQ,X) = \ZZ_{8V}(\QQ,X').
  \end{equation}
\end{Prop}

\subsection{Free-fermion 8V correlators}
By combining the previous results, we can
relate free-fermion 8V correlations with Ising ones. This has been
done in \cite{Dubedat} when the Ising models are dual of each other
(which in our case corresponds to $\alpha=\beta$), but the proof works
identically when this is not the case.

\begin{Cor}
  \label{cor:8vff_is}
  Let $\QQ$ be a quadrangulation in the spherical case, and let
  $\alpha,\beta:\FF \to (0,\frac{\pi}{2})$.

  For any $B_0,B_1,W_0,W_1$ and paths $\gamma$ as in
  Definition~\ref{def:corr8}, let $\gamma_{\BB} =
  (\gamma_{B_0},\gamma_{W_0}\triangle\gamma_{W_1})$ and $\gamma_{\WW} =
  (\gamma_{W_0},\gamma_{B_0}\triangle\gamma_{B_1})$. Then
  \begin{equation}
    \label{eq:23}
    \left< \sigma(B_0) \sigma(W_0) \mu(B_1) \mu(W_1)
    \right>^{8V}_{X_{\alpha,\beta},\gamma}
    = \ c_0 \
    \left< \sigma(B_0) \mu(W_0\triangle W_1)
    \right>^{\text{Ising} {\BB}}_{\alpha,\gamma_{\BB}}\
    \left< \sigma(W_0) \mu(B_0 \triangle B_1) \right>^{\text{Ising} {\WW}}_{\beta,\gamma_{\WW}}
  \end{equation}
  where
  \begin{equation}
    c_0 =  \frac{1}{2}\prod_{f\in\FF}\sqrt{\cos \alpha(f) \sin \beta(f)
      C_{\alpha,\beta}(f) }.
  \end{equation}
  In particular,
  \begin{equation}
    \ZZ_{8V}(\QQ,X_{\alpha,\beta}) = c_0 \
    \ZZ^{\BB}_{\text{Ising}}(J^\alpha_{\BB}) \
    \ZZ^{\WW}_{\text{Ising}}(J^\beta_{\WW}).
  \end{equation}
\end{Cor}

\begin{proof}
  From Proposition~\ref{prop:spin_ising}, the product of Ising
  correlators on the right-hand side of \eqref{eq:23} is equal
  to
  \begin{equation}
    2 \left< \sigma(B_0)\sigma(W_0)\mu(B_0\triangle B_1)\mu(W_0
      \triangle W_1)
    \right>^{8V}_{X,(\gamma_{\BB},\gamma_{\WW})}
  \end{equation}
  for the weights $X=(A,B,C,D)$ given by
  \begin{equation*}
  \begin{cases}
    A = e^{J^{\alpha}_{\BB}-J^{\beta}_{\WW}} = \sqrt{\frac{1+\sin
        \alpha}{\cos \alpha} \frac{\sin \beta}{1+\cos \beta}},\\
    B = e^{-J^{\alpha}_{\BB}+J^{\beta}_{\WW}} = \sqrt{\frac{\cos \alpha}{1+\sin
        \alpha} \frac{1+\cos \beta}{\sin \beta} },\\
    C = e^{J^{\alpha}_{\BB}+J^{\beta}_{\WW}} = \sqrt{\frac{1+\sin
        \alpha}{\cos \alpha} \frac{1+\cos \beta}{\sin \beta}}, \\
    D = e^{-J^{\alpha}_{\BB}-J^{\beta}_{\WW}} = \sqrt{\frac{\cos \alpha}{1+\sin
        \alpha} \frac{\sin \beta}{1+\cos \beta}}.
  \end{cases}
\end{equation*}
On these weights, we perform the transformation of
Proposition~\ref{prop:dmd}, then of Proposition~\ref{prop:8vdu}, and
again of Proposition~\ref{prop:dmd}. This amounts to defining
$\tilde{X}=(\tilde{A},\tilde{B},\tilde{C},\tilde{D})$ by
\[ \begin{pmatrix}
    \tilde{A} \\
    \tilde{B} \\
    \tilde{C} \\
    \tilde{D}
  \end{pmatrix}
  =\frac12
  \begin{pmatrix}
    1 & -1 & 1 & 1 \\
    -1 & 1 & 1 & 1 \\
    1 & 1 & 1 & -1 \\
    1 & 1 & -1 & 1 \\
  \end{pmatrix}
  \begin{pmatrix}
    A \\
    B \\
    C \\
    D
  \end{pmatrix}.
\]
Following the transformations in the Propositions, we get that the
Ising correlators are equal to
\begin{equation}
    2 \left< \sigma(B_0)\sigma(W_0)\mu(B_1)\mu( W_1)
    \right>^{8V}_{\tilde{X},\gamma}.
  \end{equation}
  Trigonometric computations show that (implicitly at any $f
  \in \FF$):
  \[\begin{pmatrix}
    \tilde{A} \\
    \tilde{B} \\
    \tilde{C} \\
    \tilde{D}
  \end{pmatrix}
  = \frac{1}{\sqrt{\cos \alpha \sin \beta
      \left(1+\cos(\alpha-\beta)\right)}}
  \begin{pmatrix}
    A_{\alpha,\beta} \\
    B_{\alpha,\beta} \\
    C_{\alpha,\beta} \\
    D_{\alpha,\beta}
  \end{pmatrix}
\]
and using the definition of correlators as partition function, we see
that these gauge transformations multiply the correlator by the same factor.
\end{proof}

\subsection{Coupling of free-fermion 8V-models}
With Corollary \ref{cor:8vff_is}, we are able to factor correlators
for the weights $X_{\alpha,\beta}$ into a part that depends on
$\alpha$ and a part that depends on $\beta$. By doing the same for
$X_{\alpha',\beta'}$ and rearranging the Ising correlators, we can get
the correlators of $X_{\alpha,\beta'}$ and $X_{\alpha',\beta}$. This
is expressed in the following ``switching'' result. The constants can
be defined in terms of
\begin{equation}
  \label{eq:defc}
  c_{\alpha,\beta}=\prod_{f\in\FF}C_{\alpha,\beta}(f)
  =\prod_{f\in\FF}\frac{A^2_{\alpha,\beta}(f) + B^2_{\alpha,\beta}(f)}{2}.
\end{equation}
\begin{Theo}
  \label{theo:8vswitch}
  Let $\QQ$ be a quadrangulation in the spherical case, and let
  $\alpha,\beta,\alpha',\beta':\FF \to (0,\frac{\pi}{2})$.
  Let $B_0, B_1, W_0, W_1,\gamma$ (resp. $B'_0, B'_1, W'_0,
  W'_1,\gamma'$) be as in Definition~\ref{def:corr8}.
  Then
  \begin{equation}
    \label{eq:8vswitch_corr}
    \begin{split}
     & \ \ \ \left< \sigma(B_{0}) \sigma(W_{0}) \mu(B_{1}) \mu(W_{1})
    \right>^{8V}_{X_{\alpha,\beta},\gamma} \ \
      \left< \sigma(B'_{0}) \sigma(W'_{0}) \mu(B'_{1}) \mu(W'_{1})
      \right>^{8V}_{X_{\alpha',\beta'},\gamma'} \\
      = &
      c_1 \left< \sigma(B''_{0}) \sigma(W''_{0}) \mu(B''_{1}) \mu(W''_{1})
    \right>^{8V}_{X_{\alpha,\beta'},\gamma''}
      \left< \sigma(B'''_{0}) \sigma(W'''_{0}) \mu(B'''_{1}) \mu(W'''_{1})
    \right>^{8V}_{X_{\alpha',\beta},\gamma'''}
    \end{split}
  \end{equation}
  where
  \begin{equation}
    \label{eq:33}
    \begin{pmatrix}
      B''_0 \\ W''_0 \\ B''_1 \\ W''_1
    \end{pmatrix}
    =
    \begin{pmatrix}
      B_0 \\ W'_0 \\
      B_0 \triangle B'_0 \triangle B'_1 \\
      W_0 \triangle W'_0 \triangle W_1
    \end{pmatrix}
    , \ \ \ \
    \begin{pmatrix}
      B'''_{0} \\ W'''_{0} \\ B'''_{1} \\ W'''_{1}
    \end{pmatrix}
    =
    \begin{pmatrix}
      B'_0 \\ W_{0} \\
      B_0\triangle B'_{0} \triangle B_{1} \\
      W_{0} \triangle W'_{0} \triangle W'_{1}
    \end{pmatrix},
  \end{equation}
  with the same formulas for the paths in $\gamma'',\gamma'''$, and
  \begin{equation}
    c_1 =
    \sqrt{\frac{c_{\alpha,\beta}c_{\alpha',\beta'}}{c_{\alpha,\beta'}c_{\alpha',\beta}}}.
  \end{equation}
  In particular,
  \begin{equation}
    \label{eq:8vswitch_z}
    \ZZ_{8V}(\QQ,X_{\alpha,\beta}) \ \ZZ_{8V}(\QQ,X_{\alpha',\beta'}) =
    \ c_1 \ \ZZ_{8V}(\QQ,X_{\alpha,\beta'}) \ \ZZ_{8V}(\QQ,X_{\alpha',\beta}).
  \end{equation}
\end{Theo}

\begin{proof}
  This immediately comes from writing both the left-hand side and the
  right-hand side in terms of Ising correlators using
  Corollary~\ref{cor:8vff_is}, and checking that they are the same.
\end{proof}

\begin{Ex}
  By taking $B_0=B'_0=B$ and $W_0=W'_0=W$ (i.e. the initial order
  variables being the same), we get the simpler formula
  \begin{equation}
    \label{eq:simple_switch}
    \begin{split}
     & \ \ \  \left< \sigma(B) \sigma(W) \mu(B_{1}) \mu(W_{1})
    \right>^{8V}_{X_{\alpha,\beta},\gamma} \
      \left< \sigma(B) \sigma(W) \mu(B'_{1}) \mu(W'_{1})
      \right>^{8V}_{X_{\alpha',\beta'},\gamma'} \\
      = &
      c_1 \left< \sigma(B) \sigma(W) \mu(B'_{1}) \mu(W_{1})
    \right>^{8V}_{X_{\alpha,\beta'},\gamma''}
      \left< \sigma(B) \sigma(W) \mu(B_{1}) \mu(W'_{1})
    \right>^{8V}_{X_{\alpha',\beta},\gamma'''}
    \end{split}
  \end{equation}
\end{Ex}

This nontrivial equality of correlators (and of partition functions)
suggests that there exists a coupling between pairs of
8V-configurations. Specifically, when $(\alpha,\beta)$ and
$(\alpha',\beta')$ satisfy \eqref{eq:61}, then the 8V-weights define a
Boltzmann probability; if $\tau_{\alpha,\beta},\tau_{\alpha',\beta'}$
are independent and Boltzmann distributed, we want to couple them with
$\tau_{\alpha,\beta'},\tau_{\alpha',\beta}$, while keeping as much
information as possible.

This is the content of Theorem~\ref{theo:xorcoupl}: it is possible to
do so while keeping the XOR of configurations equal (\textit{i.e.} the
XOR of the corresponding sets of dual edges of $\QQ$, which is still
an 8V-configuration). The proof is a direct consequence of
Theorem~\ref{theo:8vswitch}, but requires the introduction of a
discrete Fourier transform on the space of 8V-configurations and is
postponed to the end of Appendix~\ref{sec:1form}. An extended
statement can be formulated for the XOR of configurations with
disorder, see Remark~\ref{rk:xordis}.

\section{Kasteleyn matrices}
\label{sec:dim}
We review the transformation of free-fermion 8V-configurations into
dimers, and we compute special relations for the corresponding
(inverse) Kasteleyn matrices.

\subsection{Free-fermion 8V to dimers}
\label{sec:free-fermion-8v}

\begin{figure}[!ht]
  \centering \begin{tikzpicture}[scale=2]
  \draw [color=gray, dashed] (-2,0) -- (0,-1) -- (2,0) -- (0,1) -- cycle;
  \node [draw=black, fill=black,thick,circle,inner sep=0pt,minimum size=6pt] at (-2,0) {};
  \node [draw=black, fill=white,thick,circle,inner sep=0pt,minimum size=6pt] at (0,1) {};
  \node [draw=black, fill=black,thick,circle,inner sep=0pt,minimum size=6pt] at (2,0) {};
  \node [draw=black, fill=white,thick,circle,inner sep=0pt,minimum size=6pt] at (0,-1) {};
  \draw (-1.3,0.6) -- (-0.8,0.4) -- (-0.8,-0.4) -- (0.8,-0.4) -- (0.8,0.4) -- (-0.8,0.4);
  \draw (0.8,0.4) -- (1.3,0.6);
  \draw (0.8,-0.4) -- (1.3,-0.6);
  \draw (-0.8,-0.4) -- (-1.3,-0.6);
  \draw (0.8,0.4) -- (0.25,0);
  \draw (-0.8,0.4) -- (-0.25,0);
  \draw (0.8,-0.4) -- (0.25,0);
  \draw (-0.8,-0.4) -- (-0.25,0);
  \draw (-0.25,0) -- (0.25,0);

  \draw (0,0.35) node [above] {$(A-D)/C$};
  \draw (0,-0.35) node [below] {$(A-D)/C$};
  \draw (-0.75,0) node [left] {$B/C$};
  \draw (0.75,0) node [right] {$B/C$};
  \draw (0,-0.04) node [above] {$C$};
  \draw (-0.55,0.2) node [] {$\sqrt{D}$};
  \draw (-0.55,-0.2) node [] {$\sqrt{D}$};
  \draw (0.55,0.2) node [] {$\sqrt{D}$};
  \draw (0.55,-0.2) node [] {$\sqrt{D}$};
  \draw (-1,0.5) node [above] {$1$};
  \draw (-1,-0.5) node [below] {$1$};
  \draw (1,0.5) node [above] {$1$};
  \draw (1,-0.5) node [below] {$1$};
\end{tikzpicture}
  \caption{The quadrangulation $\mathcal{Q}$ (dashed) at a face $f$
    and the decorated graph of Fan and Wu \cite{FanWu} (solid lines) with its
    dimer weights. The functions $A,B,C,D$ are implicitly evaluated
    at $f$.}
  \label{fig:dimfw}
\end{figure}
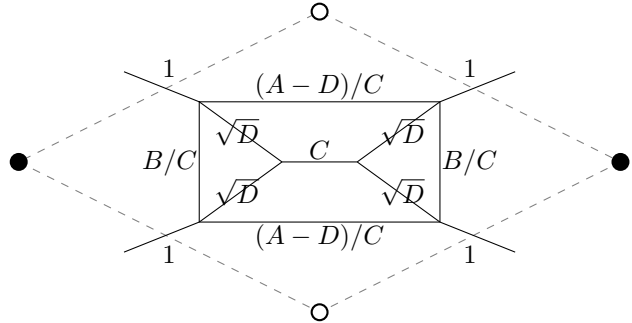

In the case of the square lattice, it has been shown by Fan and Wu
that the 8V-model at its free-fermion point can be transformed into a
dimer model on a planar decorated graph \cite{FanWu}. Their arguments
work identically on any quadrangulation. The corresponding decorated
graph is represented in Figure~\ref{fig:dimfw}.

In the more special case of a free-fermionic 6V model, the graph
becomes bipartite and the dimer model can be studied in more details
\cite{WuLin,Dubedat,BoutillierDeTiliere:xor}. No such bipartite
dimer decoration is known for the 8V-model, and the techniques of
bipartite dimers are unavailable as such.

\begin{figure}[!ht]
  \centering
  \begin{tikzpicture}[scale=2]
  \draw [color=gray, dashed] (-2,0) -- (0,-1) -- (2,0) -- (0,1) -- cycle;
  \node [draw=black, fill=black,thick,circle,inner sep=0pt,minimum size=6pt] at (-2,0) {};
  \node [draw=black, fill=white,thick,circle,inner sep=0pt,minimum size=6pt] at (0,1) {};
  \node [draw=black, fill=black,thick,circle,inner sep=0pt,minimum size=6pt] at (2,0) {};
  \node [draw=black, fill=white,thick,circle,inner sep=0pt,minimum size=6pt] at (0,-1) {};
  \draw (0.8,0.4) -- (1.3,0.6);
  \draw (0.8,-0.4) -- (1.3,-0.6);
  
  \draw (-0.8,0.4) -- (-1.3,0.6);
  
  \draw (-0.8,-0.4) -- (-1.3,-0.6);
  
  \draw (-0.8,0.4) -- (0.8,-0.4);
  \draw (-0.8,-0.4) -- (0.8,0.4);
  \draw (-0.8,0.4) -- (0.8,0.4);
  \draw (-0.8,0.4) -- (-0.8,-0.4);
  \draw (0.8,-0.4) -- (-0.8,-0.4);
  \draw (0.8,-0.4) -- (0.8,0.4);

  \draw (0,0.35) node [above] {$A/C$};
  \draw (0,-0.35) node [below] {$A/C$};
  \draw (-0.75,0) node [left] {$B/C$};
  \draw (0.75,0) node [right] {$B/C$};
  \draw (-0.3,0.2) node [] {$D/C$};
  \draw (-0.3,-0.2) node [] {$D/C$};
  \draw (-1,0.5) node [above] {$1$};

  
  \node [draw=black, fill=white,thick,circle,inner sep=0pt,minimum size=3pt] at (-0.8,0.4) {};
  \node [draw=black, fill=black,thick,circle,inner sep=0pt,minimum size=3pt] at (-0.8,-0.4) {};
  \node [draw=black, fill=white,thick,circle,inner sep=0pt,minimum size=3pt] at (0.8,-0.4) {};
  \node [draw=black, fill=black,thick,circle,inner sep=0pt,minimum size=3pt] at (0.8,0.4) {};
\end{tikzpicture}
  \caption{The decorated graph $G^T$ of Hsue, Lin and Wu
    \cite{HsueLinWu} with its dimer weights.}
  \label{fig:weightshwl}
\end{figure}
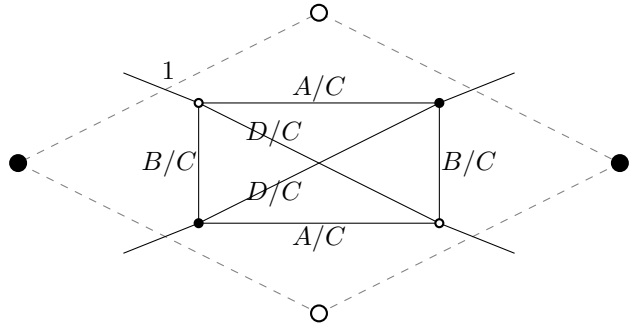

In our setting we will make use of another decorated graph due to
Hsue, Lin and Wu \cite{HsueLinWu}, see
Figure~\ref{fig:weightshwl}. This graph is more symmetric but non
planar, which makes the usual theory of dimers as Pfaffians not
available, but an adapted theory has been developed by Kasteleyn
\cite{Kasteleyn:ising,Kasteleyn:ising2}.

More precisely, let $G^T=(V^T,E^T)$ be a decorated version of $\QQ^*$
obtained by drawing small complete graphs $K_4$ inside faces of $\QQ$
and joining them by ``legs'' that cross the edges of $\QQ$, as
represented in Figure~\ref{fig:weightshwl}. Even if this graph is not
bipartite, we still decompose $V^T$ into a subset of black vertices
$B^T$ and white vertices $W^T$, such that the black (resp. white)
vertices lie on the left (resp. right) of an edge of $\QQ$ oriented
from black to white. For every edge $\mathrm{e}\in E^T$, we define
$\nu_{\mathrm{e}}$ as in Figure~\ref{fig:weightshwl}. We will need a
nonstandard weight for $m\in\mathcal{M}(G^T)$, defined as
\begin{equation}
  \label{eq:49}
  \tilde{w}(m) = (-1)^{N(m)}\prod_{\mathrm{e}\in m} \nu_{\mathrm{e}}
\end{equation}
where $N(m)$ is the number of pairs of intersecting edges in $m$. The
corresponding partition function is
\begin{equation}
  \label{eq:52}
  \tilde{\ZZ}_{\text{dim}}(G^T, \nu) = \sum_{m\in\mathcal{M}(G^T)}\tilde{w}(m).
\end{equation}

Note that at the boundary of every face $f\in\FF$, $m$ uses an even
number of legs. As a result, if we only keep the occupied legs of $m$,
we get an 8V-configuration $\tau \in \Omega(\QQ)$. We denote this by
$m\mapsto \tau$. This mapping is weight-preserving in the
following sense; this was noted when $\QQ$ is the square lattice by
Hsue, Lin and Wu \cite{HsueLinWu} but works in the exact same way for
any quadrangulation:

\begin{Theo}[\cite{HsueLinWu}]
  \label{theo:hlw}
  Let $\QQ$ be a quadrangulation in the spherical or toric case, and
  $X$ a set of standard free-fermionic 8V-weights on $\QQ$. Then for
  every $\tau\in\Omega(\QQ)$,
  \begin{equation}
    \label{eq:50}
    w_{8V}(\tau) = \left( \prod_{f\in\FF}C(f) \right)
    \sum_{\substack{m\in \mathcal{M}(G^T) \\ \text{s.t. } m\mapsto \tau}}\tilde{w}(m).
  \end{equation}
  In particular,
  \begin{equation}
    \label{eq:51}
    \ZZ_{8V}(\QQ,X) = \left( \prod_{f\in\FF}C(f) \right) \tilde{\ZZ}_{\text{dim}}(G^T,\nu).
  \end{equation}
\end{Theo}

We now describe how to compute $\tilde{\ZZ}_{\text{dim}}(G^T,\nu)$ using an
adapted version of Kasteleyn matrices.

\subsection{Skew-symmetric real matrix}

A \emph{Kasteleyn orientation} of a planar or toric graph is an
orientation of the edges such that every face is clockwise-odd,
meaning that it has an odd number of clockwise-oriented edges; by the
planarity condition, such an orientation can always be constructed,
and may be used to identify the partition function of dimers with the
Pfaffian of the corresponding skew-symmetric, weighted adjacency
matrix \cite{Kasteleyn,TemperleyFisher}. Since $G^T$ is non-planar,
there might not exist a usual Kasteleyn orientation, but there still
exists an \emph{admissible} orientation so that the Pfaffian is equal
to $\tilde{\ZZ}_{\text{dim}}(G^T,\nu)$
\cite{Kasteleyn:ising,Kasteleyn:ising2}, which we describe now.

If we remove all edges of $G^T$ that join black vertices (the
\emph{black diagonals} of decorations), we get a planar (or toric)
graph $G^T_B$. Similarly, removing the \emph{white diagonals} gives a
graph $G^T_W$. An orientation of $G^T$ is said to be \emph{admissible}
if its restriction to $G^T_B$ and $G^T_W$ are both Kasteleyn
orientation.  The existence of such an orientation is established in
Section F of \cite{Kasteleyn:ising2}.

\begin{figure}[h]
  \centering
  \def\svgwidth{7cm}
  \input{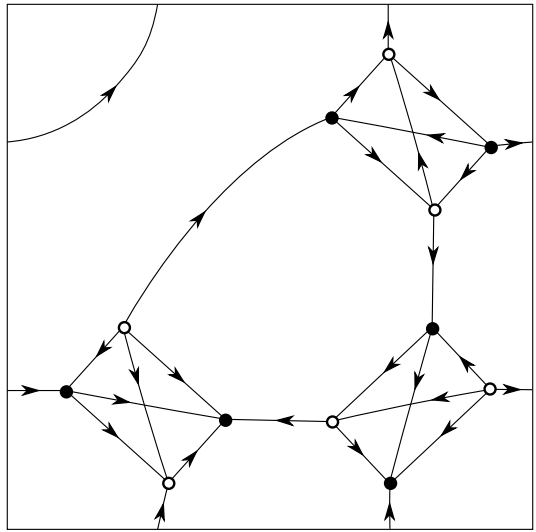}
  \caption{An admissible orientation of $G^T$ in the toric case.}
  \label{fig:or_adm_tor}
\end{figure}

To define a Kasteleyn matrix, we first fix an
admissible orientation of $G^T$. For any standard
$\alpha,\beta : \FF \to \R/2\pi\Z$, the 8V-weights
$X_{\alpha,\beta}$ are standard. Thus we can define dimer weights
$\nu_{\alpha,\beta}$ as in Figure~\ref{fig:weightshwl}.  Let
$\tilde{K}_{\alpha,\beta}$ be the weighted, skew-symmetric adjacency
matrix associated to the oriented weighted graph $G^T$.

In the spherical case, the arguments leading to equation $(79)$ in
\cite{Kasteleyn:ising2} imply the following.
\begin{Prop}[\cite{Kasteleyn:ising,Kasteleyn:ising2}]
  \label{prop:zdk}
  In the spherical case, for any quadrangulation $\QQ$ and any
  standard $\alpha,\beta : \FF \to \R/2\pi\Z$,
  \begin{equation}
    \label{eq:54}
    \tilde{\ZZ}_{\text{dim}}(G^T,\nu_{\alpha,\beta}) = \Pf \tilde{K}_{\alpha,\beta}.
  \end{equation}
\end{Prop}

In the toric case, there also exists an admissible orientation, but
its Pfaffian is no longer equal to
$\tilde{\ZZ}_{\text{dim}}(G^T,\nu_{\alpha,\beta})$. We recall here the
standard way of dealing with this problem; the idea was suggested but
not proved by Kasteleyn \cite{Kasteleyn,Kasteleyn:ising} and was then
proved in various forms of generality in the works of Dolbilin et
al. \cite{Dolbilin_et_al}, Galluccio and Loebl \cite{GalluccioLoebl}, Tesler \cite{Tesler}, Cimasoni and
Reshetikhin \cite{CimasoniReshetikhin}.

Let $m_0$ be the dimer configuration consisting of all dimers in the
decorations that are parallel to the edges of $G^{\BB}$; see the
darker configuration of Figure~\ref{fig:torus}. For any dimer
configuration $m$ on $G^T$, the superimposition of $m$ and $m_0$ is
the disjoint union of alternating loops covering all the
vertices. This union of curves has a well defined homology in
$H_1(\mathbb{T}^2,\Z/2\Z)$, which we denote
$(h^m_x,h^m_y) \in \left(\Z/2\Z\right)^2$.  For any
$\theta,\tau \in \{0,1\}$, let
$\tilde{K}^{\theta,\tau}_{\alpha,\beta}$ be the Kasteleyn matrix where
the weights of edges of $G^T$ crossing $\gamma^{\BB}_x$
(resp. $\gamma^{\BB}_y$) have been multiplied by $(-1)^{\theta}$
(resp. $(-1)^{\tau}$). Then there exists an admissible orientation
such that we have the following.

\begin{Prop}[\cite{Kasteleyn,Kasteleyn:ising,Dolbilin_et_al,
    GalluccioLoebl, Tesler,CimasoniReshetikhin}]
  \label{prop:zdk_torus}
  In the toric case, for any quadrangulation $\QQ$ and any standard
  $\alpha,\beta:\FF \to \R/2\pi\Z$, for any $\theta,\tau \in \{0,1\}$,
  \begin{equation}
    \label{eq:71}
    \Pf \tilde{K}^{\theta,\tau}_{\alpha,\beta} = \sum_{m\in\mathcal{M}(G^T)}
    (-1)^{h^m_x h^m_y + h^m_y + h^m_y + \theta h^m_x + \tau h^m_y} \tilde{w}(m).
  \end{equation}

  Consequently,
  \begin{equation}
    \label{eq:defP}
     \tilde{\ZZ}_{\text{dim}}(G^T,\nu_{\alpha,\beta}) = \frac{1}{2}
     \left( -\Pf \tilde{K}_{\alpha,\beta}^{0,0} + \Pf
      \tilde{K}_{\alpha,\beta}^{0,1} + \Pf \tilde{K}_{\alpha,\beta}^{1,0} +
    \Pf \tilde{K}_{\alpha,\beta}^{1,1}\right).
  \end{equation}
\end{Prop}
For any $(z,w)\in(\C^*)^2$, consider the modified matrix
$\tilde{K}_{\alpha,\beta}(z,w)$, obtained by multiplying the
coefficients $\tilde{K}_{\alpha,\beta}[\mathrm{u},\mathrm{v}]$ by
$z$ (resp $z^{-1}$) when the edge $\textrm{uv}$ crosses
$\gamma^{\BB}_x$ from left to right (resp. right to left), and
similarly for $w$ and $\gamma^{\BB}_y$.  This leads to the definition
of the \emph{characteristic polynomial} of the eight-vertex as
the Laurent polynomial
\begin{equation}
  P^{8V}_{\alpha,\beta}(z,w) = \det \tilde{K}_{\alpha,\beta}(z,w).
\end{equation}
When $(z,w) \notin \{\pm 1\}^2$, this quantity has no reason to factor
as a square product.

\medskip

We conclude this part with a few remarks on the planar case. Then
$\tilde{K}_{\alpha,\beta}$ is an infinite matrix, or equivalently
can be seen as an operator on $\C^{V^T}$:
\begin{equation}
  \label{eq:31}
  \forall f \in \C^{V^T}, \ (\tilde{K}_{\alpha,\beta}f)[\x] = \sum_{\y
    \in V^T} K_{\alpha,\beta}[\x,\y]f[\y].
\end{equation}
This is well defined because for all $\x \in V^T,
\tilde{K}_{\alpha,\beta}[\x,\y]$ is zero for all but a finite number of
$\y$'s.

An \emph{inverse} of $\tilde{K}_{\alpha,\beta}$ is an infinite matrix
$\tilde{K}^{-1}_{\alpha,\beta}$ such that
$\tilde{K}_{\alpha,\beta}\tilde{K}^{-1}_{\alpha,\beta} = \mathrm{Id}$
as a matrix product. This is well defined by the previous remark.

When the graph is $\Z^2$-periodic, let $G^T_1 = G^T / \Z^2$ be the
\emph{fundamental graph}. Note that $G^T_1$ corresponds to the toric
case. For any $(z,w)\in\C^2$ the subspace $V^T_{(z,w)}$ of
$(z,w)$-quasiperiodic functions $f$:
\begin{equation}
  \forall \x \in V^T_1, \ \forall (m,n)\in\Z^2, \ f(\x + (m,n)) =
  z^{-m}w^{-n} f(\x)
\end{equation}
is fixed by $\tilde{K}_{\alpha,\beta}$. The restriction of
$\tilde{K}_{\alpha,\beta}$ to this finite-dimensional subspace is equal to the matrix
$\tilde{K}_{\alpha,\beta}(z,w)$ defined in the toric case for $G^T_1$,
via the identification of $\x \in V^T_1$ with the only
$(z,w)$-quasiperiodic function $\delta_{\x}(z,w)$ that takes value $1$
at $\x$ and $0$ at the other vertices $V^T_1$ of the fundamental domain.

\subsection{Skew-hermitian complex matrix}
\label{sec:skew-herm}

\begin{figure}[!ht]
  \centering \begin{tikzpicture}[scale=2]
  \draw [color=gray, dashed] (-2,0) -- (0,-1) -- (2,0) -- (0,1) -- cycle;
  \node [draw=black, fill=black,thick,circle,inner sep=0pt,minimum size=6pt] at (-2,0) {};
  \node [draw=black, fill=white,thick,circle,inner sep=0pt,minimum size=6pt] at (0,1) {};
  \node [draw=black, fill=black,thick,circle,inner sep=0pt,minimum size=6pt] at (2,0) {};
  \node [draw=black, fill=white,thick,circle,inner sep=0pt,minimum size=6pt] at (0,-1) {};
  \draw (0.8,0.4) -- (1.3,0.6);
  \draw (0.8,-0.4) -- (1.3,-0.6);
  
  \draw [-{Latex[length=2mm,width=2mm]}] (-0.8,0.4) -- (-1.05,0.5);
  \draw (-1.05,0.5) -- (-1.3,0.6);
  
  \draw (-0.8,-0.4) -- (-1.3,-0.6);
  
  \draw [-{Latex[length=2mm,width=2mm]}] (-0.8,0.4) -- (-0.14,0.07);
  \draw (-0.22,0.11) -- (0.8,-0.4);
  \draw [-{Latex[length=2mm,width=2mm]}] (-0.8,-0.4) -- (-0.14,-0.07);
  \draw (-0.22,-0.11) -- (0.8,0.4);
  \draw [-{Latex[length=2mm,width=2mm]}] (-0.8,0.4) -- (-0.2,0.4);
  \draw (-0.22,0.4) -- (0.8,0.4);
  \draw [-{Latex[length=2mm,width=2mm]}] (-0.8,0.4) -- (-0.8,0.1);
  \draw (-0.8,0.12) -- (-0.8,-0.4);
  \draw [-{Latex[length=2mm,width=2mm]}] (0.8,-0.4) -- (0.2,-0.4);
  \draw (0.22,-0.4) -- (-0.8,-0.4);
  \draw [-{Latex[length=2mm,width=2mm]}] (0.8,-0.4) -- (0.8,-0.1);
  \draw (0.8,-0.12) -- (0.8,0.4);

  \draw (0,0.35) node [above] {$iA/C$};
  \draw (0,-0.35) node [below] {$iA/C$};
  \draw (-0.75,0) node [left] {$B/C$};
  \draw (0.75,0) node [right] {$B/C$};
  \draw (-0.3,0.2) node [] {$iD/C$};
  \draw (-0.3,-0.2) node [] {$iD/C$};
  \draw (-1,0.5) node [above] {$-e^{i \phi_e}$};

  \draw [gray] (-1.7,0.25) node {\rotatebox{25}{$e\in \EE$}};
  
  \node [draw=black, fill=white,thick,circle,inner sep=0pt,minimum size=3pt] at (-0.8,0.4) {};
  \node [draw=black, fill=black,thick,circle,inner sep=0pt,minimum size=3pt] at (-0.8,-0.4) {};
  \node [draw=black, fill=white,thick,circle,inner sep=0pt,minimum size=3pt] at (0.8,-0.4) {};
  \node [draw=black, fill=black,thick,circle,inner sep=0pt,minimum size=3pt] at (0.8,0.4) {};
\end{tikzpicture}
  \caption{The skew-hermitian Kasteleyn matrix $K_{\alpha,\beta}$ on
    $G^T$.
  }
  \label{fig:dimhwl}
\end{figure}
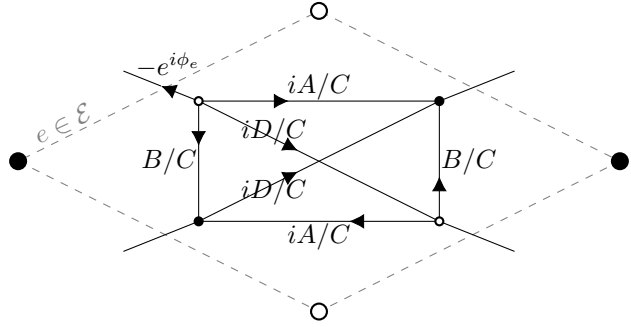

There is another way to define Kasteleyn matrices that is more
intrinsic and does not require fixing an orientation, by using instead
complex arguments on the edges. Let $K_{\alpha,\beta}$ be the
matrix whose entries are indexed by vertices $V^T$ and defined by
Figure~\ref{fig:dimhwl} and by the skew-hermitian condition:
\begin{equation}
  K_{\alpha,\beta} [\u,\v] = - \overline{K_{\alpha,\beta}
    [\v,\u]}.
\end{equation}

The arguments of the entries are inspired by the relation with the
Kac-Ward matrix \cite{KacWard,CimasoniDuminilCopin,ChelkakCimasoniKassel}. The
``angles'' $(\phi_e)_{e\in\EE}$ are defined in the following way:
\begin{itemize}
\item In the spherical and planar cases, we embed the graph $G^{\BB}$
  properly in the plane, with straight edges. Then the white vertices
  of $\QQ$, $\WW$, are in bijection with faces of $G^{\BB}$, and the
  edges $\EE$ of $\QQ$ are in bijection with the \emph{corners} of
  $G^{\BB}$. For every $e\in\EE$, we set $2\phi_e$ to be the direct
  angle at the corner corresponding to $e$, taken in $[0,2\pi)$. See
  Figure~\ref{fig:cycle}.
\item In the toric case, we lift $\QQ$ to a bipartite periodic
  quadrangulation of the plane, and we proceed as in the planar
  case. This yields a periodic choice of angles $\phi$, which can be
  mapped again to the torus.
\end{itemize}

In the toric case, we also define $K_{\alpha,\beta}(z,w)$ just as
before.

\medskip

The following result relates the matrices $\tilde{K}_{\alpha,\beta}$
and $K_{\alpha,\beta}$ by ``gauge equivalence''. In particular, it
shows that all their principal minors are equal.
\begin{Le}
  \label{le:diagconj}
  In the spherical and toric cases, there exists a diagonal
  unitary matrix $\mathcal{D}$, that depends only on the chosen
  admissible orientation of $G^T$, such that
  \begin{equation}
    \label{eq:53}
    K_{\alpha,\beta} = \mathcal{D}^{-1}
    \tilde{K}_{\alpha,\beta} \mathcal{D}
  \end{equation}
\end{Le}

\begin{proof}
  We use Theorem~2.1 of \cite{SaundersSchneider}; see also Appendix~A
  of \cite{DeTiliere}. The existence of such a diagonal (not
  necessarily unitary) matrix is equivalent to having, for every cycle
  $\mathcal{C} = (\x_1,\dots,\x_p,\x_{p+1}=\x_1)$ of adjacent vertices
  on $G^T$,
  \begin{equation}
    \label{eq:67}
    \prod_{i=1}^{p}K_{\alpha,\beta}[\x_i,\x_{i+1}] = \prod_{i=1}^{p}\tilde{K}_{\alpha,\beta}[\x_i,\x_{i+1}].
  \end{equation}
  Since the complex moduli of the entries of $K_{\alpha,\beta}$ and
  $\tilde{K}_{\alpha,\beta}$ are equal, it is sufficient to show that
  the complex arguments in \eqref{eq:67} are the same.
  Notice that for the simple cycles $(\x,\y,\x)$, by the
  skew-symmetric and skew-hermitian properties,
  \begin{equation}
    \label{eq:68}
    \arg(K_{\alpha,\beta}[\x,\y] K_{\alpha,\beta}[\y,\x]) = \pi
    = \arg(\tilde{K}_{\alpha,\beta}[\x,\y] \tilde{K}_{\alpha,\beta}[\y,\x]).
  \end{equation}
  Moreover, if we show that the right-hand side of \eqref{eq:67} is
  real (\textit{i.e.} the argument is $0 [\pi]$), then we only have to
  check one direction for any cycle.

  All in all, by decomposing cycles, it suffices to check that the
  complex arguments in \eqref{eq:67} are equal and real for the
  following cycles:
  \begin{enumerate}
  \item the cycles that winds once in the counter-clockwise direction
    around a vertex of $\BB$, or of $\WW$;
  \item the counter-clockwise $3$-cycles inside decorations that use
    two sides and a diagonal;
  \item in the torus case, two fixed cycles that wind once
    vertically (resp. horizontally) around the torus.
  \end{enumerate}

  \begin{figure}[ht]
    \centering
    \def\svgwidth{6cm}
\begingroup%
  \makeatletter%
  \providecommand\color[2][]{%
    \errmessage{(Inkscape) Color is used for the text in Inkscape, but the package 'color.sty' is not loaded}%
    \renewcommand\color[2][]{}%
  }%
  \providecommand\transparent[1]{%
    \errmessage{(Inkscape) Transparency is used (non-zero) for the text in Inkscape, but the package 'transparent.sty' is not loaded}%
    \renewcommand\transparent[1]{}%
  }%
  \providecommand\rotatebox[2]{#2}%
  \newcommand*\fsize{\dimexpr\f@size pt\relax}%
  \newcommand*\lineheight[1]{\fontsize{\fsize}{#1\fsize}\selectfont}%
  \ifx\svgwidth\undefined%
    \setlength{\unitlength}{197.02679732bp}%
    \ifx\svgscale\undefined%
      \relax%
    \else%
      \setlength{\unitlength}{\unitlength * \real{\svgscale}}%
    \fi%
  \else%
    \setlength{\unitlength}{\svgwidth}%
  \fi%
  \global\let\svgwidth\undefined%
  \global\let\svgscale\undefined%
  \makeatother%
  \begin{picture}(1,0.75665024)%
    \lineheight{1}%
    \setlength\tabcolsep{0pt}%
    \put(0,0){\includegraphics[width=\unitlength,page=1]{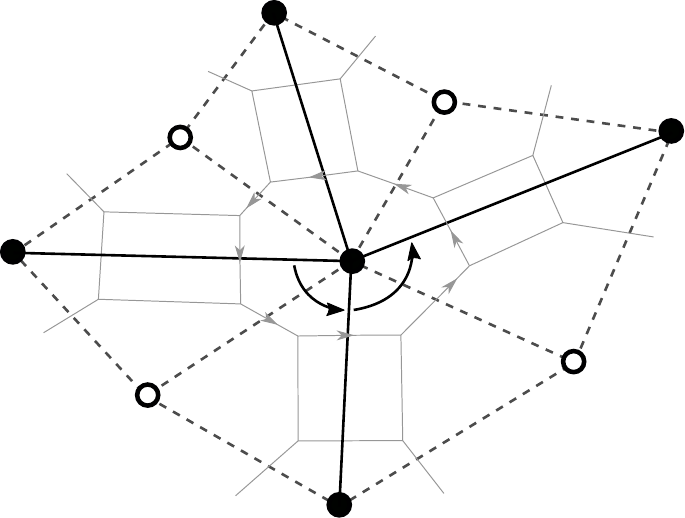}}%
    \put(0.55473489,0.30705977){\color[rgb]{0,0,0}\rotatebox{0.46270714}{\makebox(0,0)[lt]{\lineheight{1.25}\smash{\begin{tabular}[t]{l}$2\phi_1$\end{tabular}}}}}%
    \put(0.3777006,0.30911632){\color[rgb]{0,0,0}\rotatebox{-0.33800911}{\makebox(0,0)[lt]{\lineheight{1.25}\smash{\begin{tabular}[t]{l}$2\phi_p$\end{tabular}}}}}%
    \put(0.67367555,0.42047372){\color[rgb]{0,0,0}\rotatebox{1.0237734}{\makebox(0,0)[lt]{\lineheight{1.25}\smash{\begin{tabular}[t]{l}$\color{gray}{-1}$\end{tabular}}}}}%
    \put(0.66755552,0.30128086){\color[rgb]{0,0,0}\rotatebox{-0.23786291}{\makebox(0,0)[lt]{\lineheight{1.25}\smash{\begin{tabular}[t]{l}$\color{gray}{-e^{i\phi_1}}$\end{tabular}}}}}%
    \put(0.53667463,0.45665256){\color[rgb]{0,0,0}\rotatebox{-0.52688685}{\makebox(0,0)[lt]{\lineheight{1.25}\smash{\begin{tabular}[t]{l}$\dots$\end{tabular}}}}}%
  \end{picture}%
\endgroup%

    \def\svgwidth{5cm}
\begingroup%
  \makeatletter%
  \providecommand\color[2][]{%
    \errmessage{(Inkscape) Color is used for the text in Inkscape, but the package 'color.sty' is not loaded}%
    \renewcommand\color[2][]{}%
  }%
  \providecommand\transparent[1]{%
    \errmessage{(Inkscape) Transparency is used (non-zero) for the text in Inkscape, but the package 'transparent.sty' is not loaded}%
    \renewcommand\transparent[1]{}%
  }%
  \providecommand\rotatebox[2]{#2}%
  \newcommand*\fsize{\dimexpr\f@size pt\relax}%
  \newcommand*\lineheight[1]{\fontsize{\fsize}{#1\fsize}\selectfont}%
  \ifx\svgwidth\undefined%
    \setlength{\unitlength}{172.38394429bp}%
    \ifx\svgscale\undefined%
      \relax%
    \else%
      \setlength{\unitlength}{\unitlength * \real{\svgscale}}%
    \fi%
  \else%
    \setlength{\unitlength}{\svgwidth}%
  \fi%
  \global\let\svgwidth\undefined%
  \global\let\svgscale\undefined%
  \makeatother%
  \begin{picture}(1,0.80732384)%
    \lineheight{1}%
    \setlength\tabcolsep{0pt}%
    \put(0,0){\includegraphics[width=\unitlength,page=1]{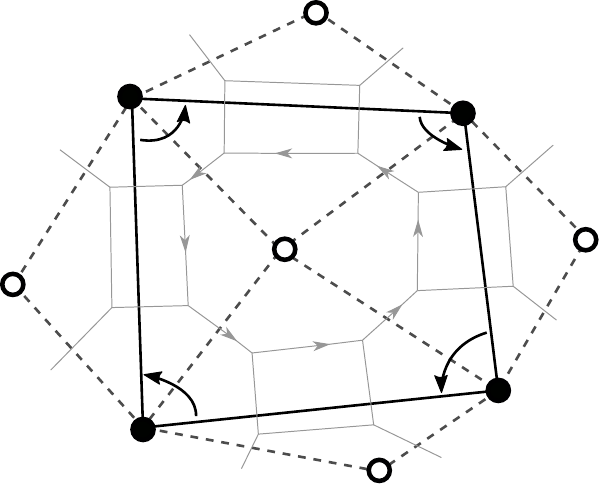}}%
    \put(0.69751923,0.20323501){\color[rgb]{0,0,0}\makebox(0,0)[lt]{\lineheight{1.25}\smash{\begin{tabular}[t]{l}$2\phi_1$\end{tabular}}}}%
    \put(0.2446124,0.19506041){\color[rgb]{0,0,0}\makebox(0,0)[lt]{\lineheight{1.25}\smash{\begin{tabular}[t]{l}$2\phi_p$\end{tabular}}}}%
    \put(0.4633931,0.23452804){\color[rgb]{0,0,0}\makebox(0,0)[lt]{\lineheight{1.25}\smash{\begin{tabular}[t]{l}$\color{gray}{i}$\end{tabular}}}}%
    \put(0.54237084,0.28292251){\color[rgb]{0,0,0}\makebox(0,0)[lt]{\lineheight{1.25}\smash{\begin{tabular}[t]{l}$\color{gray}{e^{-i\phi_1}}$\end{tabular}}}}%
  \end{picture}%
\endgroup%

    \caption{Embedded graph $G^{\BB}$ and unitary part of the entries
      of $K_{\alpha,\beta}$ on simple cycles around black and white
      vertices}
    \label{fig:cycle}
  \end{figure}
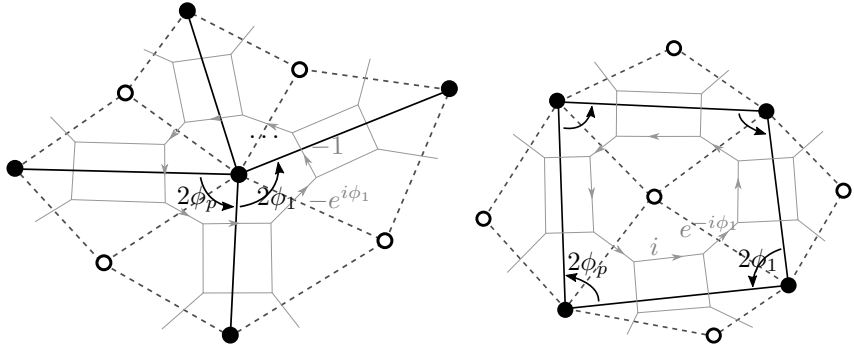

  \textit{Case 1:} let $\mathcal{C}$ be such a cycle corresponding to
  a black vertex $b\in\BB$. Let $\phi_1,\dots,\phi_p$ be the
  successive angles around $b$ as in Figure~\ref{fig:cycle}.  By
  grouping together the successive steps of $\mathcal{C}$ on legs and
  inside decorations, the argument of the left-hand side of
  \eqref{eq:67} is
  \begin{equation}
    \label{eq:69}
    \sum_{i=1}^p \phi_i = \pi.
  \end{equation}
  On the right-hand side, as the cycle is even, the product is also a
  negative real number for an admissible orientation.

  If $\mathcal{C}$ corresponds to a white vertex $w\in\WW$, we also
  set $\phi_1,\dots,\phi_p$ to be the successive angles around $w$ as
  in Figure~\ref{fig:cycle}. Then
  \begin{equation}
    \label{eq:70}
    \sum_{i=1}^p 2\phi_i = (p\pm 2)\pi,
  \end{equation}
  with a $-$ sign when $w$ corresponds to an interior face of
  $G^{\BB}$ when embedded in the plane, and a $+$ sign for the
  exterior face. Again by grouping the steps, the argument for the
  product of the left-hand side of \eqref{eq:67} is
  $\sum_{i=1}^p \left( \frac{\pi}{2} - \phi_i\right) = \pi [2\pi]$,
  and we conclude similarly.

  \textit{Case 2:} In Figure~\ref{fig:dimhwl} we easily check that for
  any of these $3$-cycles, the argument of the product of the elements of
  $K_{\alpha,\beta}$ is $\pi [2\pi]$. By the construction of admissible
  orientations, these are also clockwise-odd so the product for
  $\tilde{K}_{\alpha,\beta}$ is also a negative real number.

  \textit{Case 3:} We show this for a cycle that winds once vertically
  around the torus, the horizontal case being identical. We chose the
  alternating cycle $\mathcal{C}_y$ represented in
  Figure~\ref{fig:torus}. This cycle is obtained by superimposing the
  dimer configuration $m_0$ with a configuration $m_y$ that uses the
  legs that cross edges of $\QQ$ that touch $\gamma^{\BB}_y$ on the
  right, edges parallel to the white diagonal in the decorations of
  faces containing two such edges of $\QQ$, and is equal to $m_0$
  otherwise.

  \begin{figure}[ht]
    \centering
\begingroup%
  \makeatletter%
  \providecommand\color[2][]{%
    \errmessage{(Inkscape) Color is used for the text in Inkscape, but the package 'color.sty' is not loaded}%
    \renewcommand\color[2][]{}%
  }%
  \providecommand\transparent[1]{%
    \errmessage{(Inkscape) Transparency is used (non-zero) for the text in Inkscape, but the package 'transparent.sty' is not loaded}%
    \renewcommand\transparent[1]{}%
  }%
  \providecommand\rotatebox[2]{#2}%
  \newcommand*\fsize{\dimexpr\f@size pt\relax}%
  \newcommand*\lineheight[1]{\fontsize{\fsize}{#1\fsize}\selectfont}%
  \ifx\svgwidth\undefined%
    \setlength{\unitlength}{278.23702114bp}%
    \ifx\svgscale\undefined%
      \relax%
    \else%
      \setlength{\unitlength}{\unitlength * \real{\svgscale}}%
    \fi%
  \else%
    \setlength{\unitlength}{\svgwidth}%
  \fi%
  \global\let\svgwidth\undefined%
  \global\let\svgscale\undefined%
  \makeatother%
  \begin{picture}(1,1)%
    \lineheight{1}%
    \setlength\tabcolsep{0pt}%
    \put(0,0){\includegraphics[width=\unitlength,page=1]{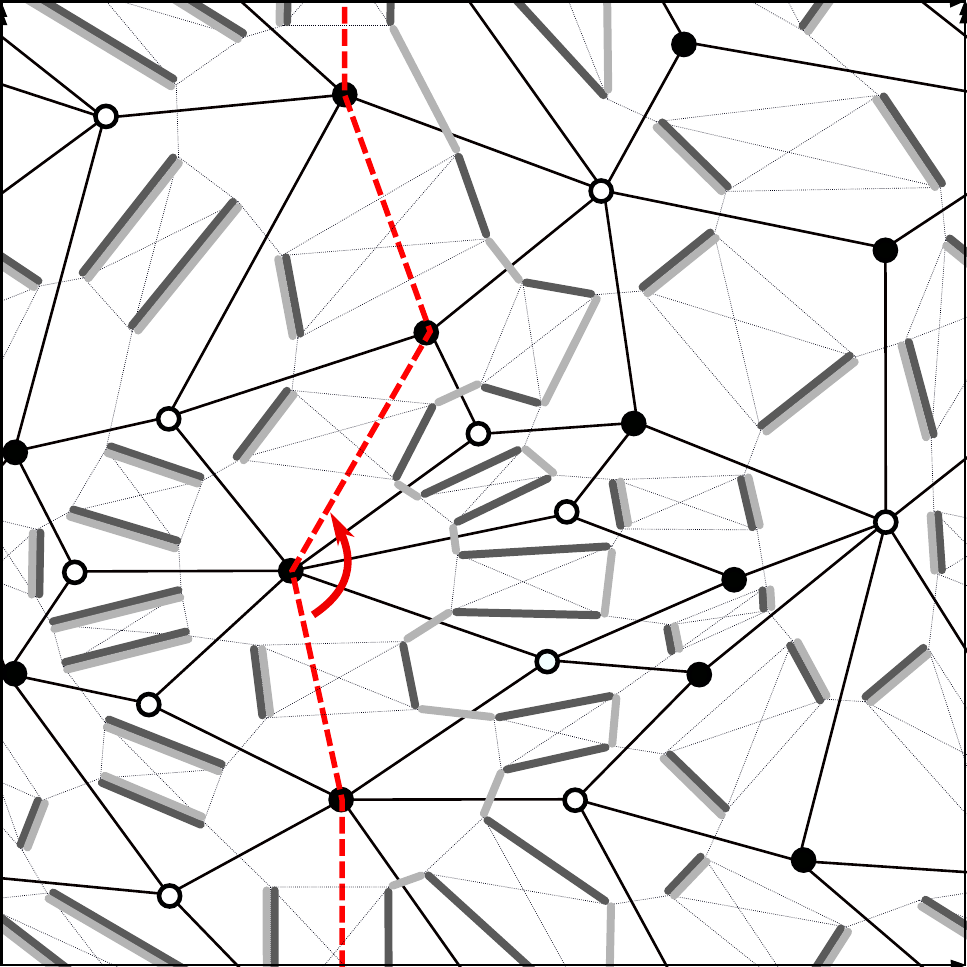}}%
    \put(0.41115944,0.88069725){\color[rgb]{0,0,0}\makebox(0,0)[lt]{\lineheight{1.25}\smash{\begin{tabular}[t]{l}$\textcolor{red}{2\psi_p}$\end{tabular}}}}%
    \put(0,0){\includegraphics[width=\unitlength,page=2]{torusp.pdf}}%
    \put(0.39115926,0.14449522){\color[rgb]{0,0,0}\makebox(0,0)[lt]{\lineheight{1.25}\smash{\begin{tabular}[t]{l}$\textcolor{red}{2\psi_1}$\end{tabular}}}}%
    \put(0.3638754,0.39358654){\color[rgb]{0,0,0}\makebox(0,0)[lt]{\lineheight{1.25}\smash{\begin{tabular}[t]{l}$\textcolor{red}{2\psi_2}$\end{tabular}}}}%
    \put(0.49054551,0.64006507){\color[rgb]{0,0,0}\makebox(0,0)[lt]{\lineheight{1.25}\smash{\begin{tabular}[t]{l}$\textcolor{red}{\dots}$\end{tabular}}}}%
  \end{picture}%
\endgroup%

    \caption{A quadrangulation of the torus with the path $\gamma^{\BB}_y$
      (dashed); the graph $G^T$ equipped with two dimer configurations,
      $m_0$ (dark grey) and $m_y$ (light grey).}
    \label{fig:torus}
  \end{figure}

  Again by decomposing the path, one easily checks that the argument
  on the left-hand side of \eqref{eq:67} is
  \begin{equation}
    \label{eq:72}
    \sum_{i=1}^p \left( \psi_i - \frac{\pi}{2} \right) = 0.
  \end{equation}

  For the right-hand side, we know that $m_0$ has homology $(0,0)$
  while $m_y$ has homology $(0,1)$, so that by
  Proposition~\ref{prop:zdk_torus} the term in $\det
  \tilde{K}_{\alpha,\beta}$ corresponding to the superimposition of
  $m_0$ and $m_y$ must appear with a minus sign. All the double dimers
  in this superimposition give a $+$ sign, because the $-$ sign of the
  product of opposite matrix elements is compensated by the signature
  of a $2$-cycle. Following this reasoning, the product corresponding
  to the alternating cycle $C_y$ must be positive, since the
  signature of the corresponding cycle of the permutation is
  $-1$. This proves that the arguments match.

  \medskip

  The fact that $\mathcal{D}$ does not depend on $\alpha,\beta$ is a
  consequence of the explicit form given in \cite{DeTiliere}. Finally,
  to show that the matrix $\mathcal{D}$ can be taken to be unitary, we
  just have to show that its diagonal elements have the same modulus,
  since multiplying it by a constant leaves relation \eqref{eq:53}
  unchanged. For any two adjacent vertices $\x,\y \in V^T$,
  \begin{equation}
    \label{eq:73}
    K_{\alpha,\beta}[\x,\y] =
    \frac{\mathcal{D}[\y,\y]}{\mathcal{D}[\x,\x]}  \tilde{K}_{\alpha,\beta}[\x,\y]
  \end{equation}
  so that
  $|\mathcal{D}[\x,\x]| =
  |\mathcal{D}[\y,\y]|$. Since the graph is connected, all
  these moduli are equal.
\end{proof}

\subsection{Eight-vertex partition function and correlations}
By injecting the result of Theorem~\ref{theo:hlw} into
Propositions~\ref{prop:zdk}~and~\ref{prop:zdk_torus}, and using
Lemma~\ref{le:diagconj} to transform the determinant of
$\tilde{K}_{\alpha,\beta}$ into that of $K_{\alpha,\beta}$ (we cannot
do the same for Pfaffian, since the latter is only defined for skew-symmetric
matrices \textit{a priori}) we get
\begin{Cor}
  \label{cor:partfunc}
  Let $\QQ$ be a quadrangulation and $\alpha,\beta:\FF \to \R/2\pi\Z$
  be standard. In the spherical case,
  \begin{equation}
    \label{eq:55}
    \begin{split}
    \left(\ZZ_{8V}(\QQ,X_{\alpha,\beta})\right)^2 & = \left(
      \prod_{f\in\FF}C_{\alpha,\beta}(f) \right)^2 \det
    K_{\alpha,\beta}.
    \end{split}
  \end{equation}
  In the toric case,
  \begin{equation}
    \label{eq:66}
    \ZZ_{8V}(\QQ,X_{\alpha,\beta}) = \frac{\prod_{f\in\FF}C_{\alpha,\beta}(f)}{2}
    \left( -\Pf \tilde{K}_{\alpha,\beta}^{0,0} + \Pf
      \tilde{K}_{\alpha,\beta}^{0,1} + \Pf \tilde{K}_{\alpha,\beta}^{1,0} +
      \Pf \tilde{K}_{\alpha,\beta}^{1,1}\right).
  \end{equation}
\end{Cor}

Another standard result is the computation of dimer statistics in
terms of minors of the inverse Kasteleyn matrix; see
\cite{Kenyon:statistics, Kenyon:problems}.  If we adapt this to the 8V
statistics, where we are only interested in the statistics of the legs
dimers, \textit{i.e.} dimers that have weight $1$, we obtain:

\begin{Prop}[\cite{Kenyon:statistics}]
  \label{prop:corrkast}
  Let $\QQ$ be a quadrangulation in the spherical or toric case. Let
  $e_1,\dots,e_p \in \EE$, each $e_i$ corresponding to a leg of $G^T$,
  whose endpoints we denote $\mathrm{b}_i\in B^T$ and
  $\mathrm{w}_i\in W^T$. Let
  $\mathrm{V} =
  \{\mathrm{b}_1,\mathrm{w}_1,\dots,\mathrm{b}_p,\mathrm{w}_p\}$.

  Let $\alpha,\beta:\FF\to\R/2\pi\Z$ satisfy \eqref{eq:61}. Let $\tau$
  be a random 8V-configuration with Boltzmann distribution $\Pr_{8V}$. Then in
  the spherical case,
  \begin{equation}
    \label{eq:4}
    \left(\Pr_{8V}\left(\{e_1,\dots,e_p\} \subset \tau\right)\right)^2
    = \det \left[
        \left(K_{\alpha,\beta}^{-1}\right)_{\mathrm{V}} \right]
  \end{equation}
  where the matrix on the right-hand side is the submatrix of
  $K_{\alpha,\beta}^{-1}$ with rows and columns indexed by
  $\mathrm{V}$.

  In the toric case,
   \begin{equation}
     \label{eq:19}
     \begin{split}
    &\Pr_{8V}\left(\{e_1,\dots,e_p\} \subset \tau\right) = \\
    &\frac{\prod_{f\in\FF}C_{\alpha,\beta}(f)}{2\ZZ_{8V}(\QQ,X_{\alpha,\beta})}
    \left( -\Pf
      \left(\tilde{K}_{\alpha,\beta}^{0,0}\right)_{\mathrm{V}^c} + \Pf
      \left(\tilde{K}_{\alpha,\beta}^{0,1}\right)_{\mathrm{V}^c} + \Pf
      \left(\tilde{K}_{\alpha,\beta}^{1,0}\right)_{\mathrm{V}^c} + \Pf
      \left(\tilde{K}_{\alpha,\beta}^{1,1}\right)_{\mathrm{V}^c}
    \right)
       .
       \end{split}
  \end{equation}
\end{Prop}

\subsection{Relations between matrices $K_{\alpha,\beta}^{-1}$}
\label{sec:invmat}

We now exhibit a symmetry in the 8V-model in the form of a relation
between inverse matrices for different values of $\alpha,\beta$.

\subsubsection{Spherical and planar cases}

Let us define the matrix $T$ with entries indexed by the vertices
$V^T$ of the dimer graph $G^T$, in the
following way: if $\w \in W^T$, then there is a unique ``leg''
adjacent to $\w$. Let us denote $\hat{\w}\in B^T$ the other endpoint of this
leg. Let $e\in \EE$ be the edge of $\QQ$ crossed by
$\{\w\hat{\w}\}$. We define
\begin{equation}
  \label{eq:8}
  \begin{split}
    T(\w,\hat{\w}) &= -e^{i\phi_e},\\
    T(\hat{\w},\w) &= -e^{-i\phi_e},
  \end{split}
\end{equation}
and all the other entries of $T$ are zero. Thus $T$ is a weighted
permutation matrix between vertices $\x$ and their associated
neighbor, which we still denote $\hat{\x}$, $\x$ being black or
white.

\begin{Theo}
  \label{theo:km1sphere}
  In the spherical case or planar case, let
  $(\alpha,\beta)$ and $(\alpha',\beta')$ be standard elements of
  $\left([0,2\pi)^{\FF}\right)^2$. If the
  matrices $K_{\alpha,\beta'}^{-1}, K_{\alpha',\beta}^{-1}$
  are inverses of $K_{\alpha,\beta'}, K_{\alpha',\beta}$, then
  the following are inverses of
  $K_{\alpha,\beta}, K_{\alpha',\beta'}$:

  \begin{equation}
    \label{eq:rel_inv}
    \begin{split}
      K_{\alpha,\beta}^{-1} &= \frac12 \left( (I+T)
        K_{\alpha,\beta'}^{-1} + (I-T) K_{\alpha',\beta}^{-1}\right),\\
      K_{\alpha',\beta'}^{-1} &= \frac12 \left( (I-T)
        K_{\alpha,\beta'}^{-1} + (I+T) K_{\alpha',\beta}^{-1}\right).
    \end{split}
  \end{equation}

\end{Theo}

\begin{proof}
  To simplify notations, we will denote
  \begin{itemize}
  \item
    $(A,B,C,D)$ the weights $X_{\alpha,\beta}$
    ;
  \item
    $(A',B',C',D')$ the weights $X_{\alpha',\beta'}$
    ;
  \item
    $(a,b,c,d)$ the weights $X_{\alpha,\beta'}$, and
    $K^{-1}_{\alpha,\beta'}=\left(u_{\x,\y}\right)_{\x,\y \in V^T}$;
  \item
    $(a',b',c',d')$ the weights $X_{\alpha',\beta}$, and
    $K^{-1}_{\alpha',\beta}=\left(u'_{\x,\y}\right)_{\x,\y \in V^T}$.
  \end{itemize}
  \begin{Le}
    \label{le:lcalc}
    The following relations, implicitly evaluated at any $f\in\FF$,
    hold:
    \begin{equation*}
      \begin{array}{rcl}
        cA+dB-aC-bD=0, &  & c'A-d'B-a'C+b'D=0, \\
        dA-cB+bC-aD=0, &  & d'A+c'B-b'C-a'D=0, \\
        aA+bB-cC-dD=0, &  & a'A+b'B-c'C-d'D=0, \\
        bA-aB+dC-cD=0, &  & b'A-a'B-d'C+c'D=0. \\
      \end{array}
    \end{equation*}
  \end{Le}

  \begin{proof}[Proof of Lemma~\ref{le:lcalc}]
    This is done by direct computations, which are made easier by the
    use of the alternative form of weights \eqref{eq:weight_form2}.
  \end{proof}

  \begin{figure}[!ht]
    \centering \begin{tikzpicture}[scale=2]
  \draw [color=gray, dashed] (-2,0) -- (0,-1) -- (2,0) -- (0,1) -- cycle;
  \node [draw=black, fill=black,thick,circle,inner sep=0pt,minimum size=6pt] at (-2,0) {};
  \node [draw=black, fill=white,thick,circle,inner sep=0pt,minimum size=6pt] at (0,1) {};
  \node [draw=black, fill=black,thick,circle,inner sep=0pt,minimum size=6pt] at (2,0) {};
  \node [draw=black, fill=white,thick,circle,inner sep=0pt,minimum size=6pt] at (0,-1) {};
  \draw (0.8,0.4) -- (1.3,0.6);
  \draw (0.8,-0.4) -- (1.3,-0.6);
  
  \draw (-0.8,0.4) -- (-1.3,0.6);
  
  \draw (-0.8,-0.4) -- (-1.3,-0.6);
  
  \draw (-0.8,0.4) -- (0.8,-0.4);
  \draw (-0.8,-0.4) -- (0.8,0.4);
  \draw (-0.8,0.4) -- (0.8,0.4);
  \draw (-0.8,0.4) -- (-0.8,-0.4);
  \draw (0.8,-0.4) -- (-0.8,-0.4);
  \draw  (0.8,-0.4) -- (0.8,0.4);

  \draw [gray] (-1.7,0.25) node {\rotatebox{25}{$e_1$}};
  \draw [gray] (-1.7,-0.25) node {\rotatebox{-25}{$e_2$}};
  \draw [gray] (1.7,0.25) node {\rotatebox{-25}{$e_4$}};
  \draw [gray] (1.7,-0.25) node {\rotatebox{25}{$e_3$}};
  
  \node [draw=black, fill=white,thick,circle,inner sep=0pt,minimum
  size=3pt] at (-0.8,0.4) {};
  \draw (-0.8,0.4) node [above] {$\x_1$};
  \draw (-1.3,0.6) node [above] {$\hat{\x}_1$};
  \node [draw=black, fill=black,thick,circle,inner sep=0pt,minimum size=3pt] at (-0.8,-0.4) {};
  \draw (-0.8,-0.4) node [below] {$\x_2$};
  \draw (-1.3,-0.6) node [below] {$\hat{\x}_2$};
  \node [draw=black, fill=white,thick,circle,inner sep=0pt,minimum size=3pt] at (0.8,-0.4) {};
  \draw (0.8,-0.4) node [below] {$\x_3$};
  \draw (1.3,-0.6) node [below] {$\hat{\x}_3$};
  \node [draw=black, fill=black,thick,circle,inner sep=0pt,minimum
  size=3pt] at (0.8,0.4) {};
  \draw (0.8,0.4) node [above] {$\x_4$};
  \draw (1.3,0.6) node [above] {$\hat{\x}_4$};
  
  \node [draw=black, fill=black,thick,circle,inner sep=0pt,minimum size=3pt] at (-1.3,0.6) {};
  \node [draw=black, fill=white,thick,circle,inner sep=0pt,minimum size=3pt] at (-1.3,-0.6) {};
  \node [draw=black, fill=black,thick,circle,inner sep=0pt,minimum size=3pt] at (1.3,-0.6) {};
  \node [draw=black, fill=white,thick,circle,inner sep=0pt,minimum size=3pt] at (1.3,0.6) {};
\end{tikzpicture}
    \caption{Notation for $G^T$ around $\x_1\in W^T$.}
    \label{fig:neighb}
  \end{figure}
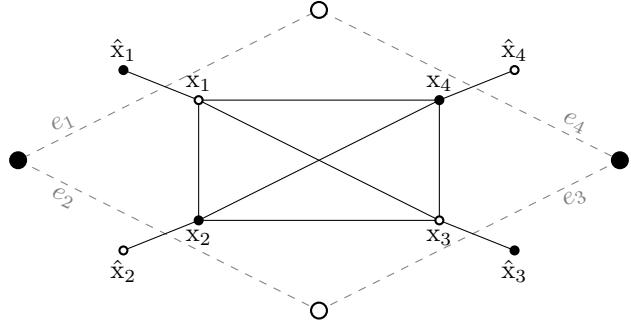
  Let $\x_1\in W^T$. Its neighbors are denoted
  $\hat{\x}_1,\x_2,\x_3,\x_4$ as in Figure~\ref{fig:neighb}. For any
  $\y\in V^T$ and $i\in\{1,\dots,4\}$ we have $\left(K_{\alpha,\beta'}
    K^{-1}_{\alpha,\beta'} \right)[\x_i,\y] = \delta_{\x_i,\y}$,
  which reads:
  \begin{align}
    -e^{i \phi_{e_1}}u_{\hat{\x}_1,\y} + \frac{b}{c} u_{\x_2,\y}
    + i\frac{d}{c} u_{\x_3,\y} + i \frac{a}{c} u_{\x_4,\y} &=
                                                             \delta_{\x_1,\y} \tag{$E_1$}
    \\
    - \frac{b}{c} u_{\x_1,\y} + e^{-i \phi_{e_2}}u_{\hat{\x}_2,\y}
    + i\frac{a}{c} u_{\x_3,\y} + i \frac{d}{c} u_{\x_4,\y} &= \delta_{\x_2,\y}\tag{$E_2$}\\
    i \frac{d}{c} u_{\x_1,\y} + i\frac{a}{c}  u_{\x_2,\y} - e^{i \phi_{e_3}}u_{\hat{\x}_3,\y}
    + \frac{b}{c} u_{\x_4,\y}&= \delta_{\x_3,\y}\tag{$E_3$}\\
    i \frac{a}{c} u_{\x_1,\y} + i\frac{d}{c}  u_{\x_2,\y} - \frac{b}{c}
    u_{\x_3,\y} + e^{-i \phi_{e_4}}u_{\hat{\x}_4,\y}
                                                           &= \delta_{\x_4,\y}\tag{$E_4$}
  \end{align}
  By writing the same equations for $K_{\alpha',\beta}$, we get the same
  relations where $u_{\x,\y}$ is changed into $u'_{\x,\y}$ and
  $(a,b,c,d)$ are changed into $(a',b',c',d')$. We denote these four new
  equations by $(E'_1),(E'_2),(E'_3),(E'_4)$.

  Now we compute
  \begin{align*}
    & C(E_1) - B(E_2) +iD(E_3) - iA(E_4) \\
    + & C(E'_1)+ B(E'_2) - iD(E'_3) + iA(E'_4).
  \end{align*}

  On the right-hand side, this is $2C\delta_{\x_1,\y}$. On the
  left-hand side, we can group the terms corresponding to the same
  $u_{\x,\y}$ (or $u'_{\x,\y}$). For instance, $u_{\x_1,\y}$ will appear with
  coefficient
  \[\frac{b}{c} B - \frac{d}{c}D + \frac{a}{c}A\]
  which is equal to $C$ according to Lemma~\ref{le:lcalc}. All in all,
  using all relations of Lemma~\ref{le:lcalc}, this yields
  \begin{equation}
    \label{eq:13}
    \begin{split}
      -& Ce^{i \phi_{e_1}} \  \left(u_{\hat{\x}_1,\y} + u'_{\hat{\x}_1,\y} -
        e^{-i\phi_{e_1}} \left(u_{\x_1,y} - u'_{\x_1,y}\right) \right) \\
      + & B \  \left(u_{\x_2,\y} + u'_{\x_2,\y} -
        e^{-i\phi_{e_2}} \left(u_{\hat{\x}_2,y} - u'_{\hat{\x}_2,y}\right) \right) \\
      + & iD \  \left(u_{\x_3,\y} + u'_{\x_3,\y} -
        e^{i\phi_{e_3}} \left(u_{\hat{\x}_3,y} - u'_{\hat{\x}_3,y}\right) \right) \\
      + & iA \  \left(u_{\x_4,\y} + u'_{\x_4,\y} -
        e^{-i\phi_{e_4}} \left(u_{\hat{\x}_4,y} - u'_{\hat{\x}_4,y}\right)
      \right) = 2C\delta_{\x1,\y}.
    \end{split}
  \end{equation}
  For $\x,\y \in V^T$, let $e_{\x}\in \EE$ be the edge of the
  quadrangulation closest to $\x$, and let
  $M_{\x,\y}$ be
  \begin{equation}
    \label{eq:14}
    \begin{split}
      &\text{if }\x \in W^T, \ \ \  M_{\x,\y} = \frac12\left(u_{\x,\y} + u'_{\x,\y} - e^{i\phi_{e_{\x}}}
        \left(u_{\hat{\x},\y} - u'_{\hat{\x},\y}\right) \right), \\
      &\text{if }\x \in B^T, \ \ \ \  M_{\x,\y} = \frac12\left(u_{\x,\y} + u'_{\x,\y} - e^{-i\phi_{e_{\x}}}
        \left(u_{\hat{\x},\y} - u'_{\hat{\x},\y}\right) \right).
    \end{split}
  \end{equation}

  Then Equation~\eqref{eq:13} exactly means that the matrix $M=(M_{\x,\y})_{\x,\y
    \in V^T}$ satisfies
  \begin{equation}
    \label{eq:15}
    \left(K_{\alpha,\beta} M\right)[\x_1,\y] = \delta_{\x_1,\y}
  \end{equation}
  when $\x_1\in W^T$.
  A similar computation shows that \eqref{eq:15} also holds when
  $\x_1\in B^T$. As a result, $M$ is an inverse of $K_{\alpha,\beta}$,
  and \eqref{eq:14} is equivalent to
  \begin{equation}
    \label{eq:28}
    M = \frac12 \left[ (I+T)
      K_{\alpha,\beta'}^{-1} + (I-T) K_{\alpha',\beta}^{-1}\right].
  \end{equation}
  The second matrix relation in \eqref{eq:rel_inv} is obtained by
  switching $(\alpha,\beta) \leftrightarrow (\alpha',\beta')$.
\end{proof}

\begin{Rk}
  Theorem~\ref{theo:km1sphere} can be used to give an alternative
  proof of the relation of partition functions
  \eqref{eq:8vswitch_z}. This works exactly as in the forthcoming
  proof of the analogous statement for toric quadrangulations, see
  Theorem~\ref{theo:pol8vswitch}.
\end{Rk}

\subsubsection{Toric case}

\begin{Theo}
  \label{theo:km1torus}
  Let $\QQ$ be a quadrangulation in the toric case. Let
  $(\alpha,\beta)$ and $(\alpha',\beta')$ be two standard elements
  of $\left([0,2\pi)^{\FF}\right)^2$.

  Let $(z,w) \in (\C^*)^2$ be such that $K_{\alpha,\beta'}(z,w)$
  and $K_{\alpha',\beta}(z,w)$ are invertible. Then
  $K_{\alpha,\beta}(z,w)$ and $K_{\alpha',\beta'}(z,w)$ are
  invertible and their inverses are given by
  \begin{equation}
    \begin{split}
      K_{\alpha,\beta}^{-1}(z,w) &= \frac12 \left[ (I+T)
        K_{\alpha,\beta'}^{-1}(z,w) + (I-T) K^{-1}_{\alpha',\beta}(z,w)\right],\\
      K_{\alpha',\beta'}^{-1}(z,w) &= \frac12 \left[ (I-T)
        K_{\alpha,\beta'}^{-1}(z,w) + (I+T) K_{\alpha',\beta}^{-1}(z,w)\right].
    \end{split}
  \end{equation}
\end{Theo}

\begin{proof}
  The proof, being based on a local computation of matrix products, is
  identical to that of Theorem~\ref{theo:km1sphere}. One simply has to
  take into account the possible multiplication by $z^{\pm 1}$ and
  $w^{\pm 1}$ in the weights when the face considered is crossed by
  $\gamma^{\BB}_x,\gamma^{\WW}_y$, or both. For instance, if it is
  crossed by $\gamma^{\BB}_x$, in the notation of the proof of
  Theorem~\ref{theo:km1sphere}, one has to compute
  \begin{align*}
  & C(E_1) - zB(E_2) +izD(E_3) - iA(E_4) \\
   + & C(E'_1)+ zB(E'_2) - izD(E'_3) + iA(E'_4).
  \end{align*}
  to get the correct matrix relation. The other cases are similar.
\end{proof}

\begin{Theo}
  \label{theo:pol8vswitch}
  Let $\QQ$ be a quadrangulation in the toric case. Let
  $(\alpha,\beta)$ and $(\alpha',\beta')$ be two standard elements
  of $\left([0,2\pi)^{\FF}\right)^2$. Then the characteristic
  polynomials satisfy
  \begin{equation}
    \label{eq:24}
    P^{8V}_{\alpha,\beta} P^{8V}_{\alpha',\beta'} = c_2
    P^{8V}_{\alpha,\beta'} P^{8V}_{\alpha',\beta}
  \end{equation}
  where
  \begin{equation}
    \label{eq:7}
    c_2
    =  \frac{c_{\alpha,\beta'} c_{\alpha',\beta}}{c_{\alpha,\beta}
     c_{\alpha',\beta'}}
    = \prod_{f\in\FF}
    \frac{C_{\alpha,\beta'}(f)C_{\alpha',\beta}(f)} {C_{\alpha,\beta}(f)C_{\alpha',\beta'}(f)}.
  \end{equation}
\end{Theo}

To prove Theorem~\ref{theo:pol8vswitch}, we also
need the following diagonal matrix $D$, whose rows and columns are
indexed by the vertices of $G^T$:
\begin{equation}
  \label{eq:9}
    D_{\x\x} =
    \begin{cases}-1 & \text{if} \ \x \in W^T, \\
      1 & \text{if} \ \x \in B^T.
  \end{cases}
\end{equation}

\begin{Le}
  \label{le:commut}
  Let $\QQ$ be a quadrangulation in the toric case, and let
  $\alpha,\beta:\FF \to [0,2\pi[$ be standard, and
  $(z,w)\in \left(\C^*\right)^2$. The commutator of
  $K_{\alpha,\beta}(z,w)$ with $T$ is
  \begin{equation}
    \label{eq:commK}
    \left[K_{\alpha,\beta}(z,w),T\right] = -K_{\alpha,\beta}(z,w) D
    K_{\alpha,\beta}(z,w).
  \end{equation}
  If $P_{\alpha,\beta}(z,w) \neq 0$, the commutator of
  $K_{\alpha,\beta}^{-1}(z,w)$ with $T$ is
  \begin{equation}
    \label{eq:commKi}
    \left[K^{-1}_{\alpha,\beta}(z,w),T\right] = D.
  \end{equation}
\end{Le}

\begin{proof}
  Equality \eqref{eq:commK} can be verified by a straightforward
  computation of the matrix elements. For instance, in the
  notations of Figure~\ref{fig:neighb} (we drop the $(z,w)$ in the
  computation to simplify notations) the matrix element
  $[\x_1,\hat{\x}_2 ]$ are
  \begin{align*}
    \left(K_{\alpha,\beta}T \right)[\x_1,\hat{\x}_2] -
    \left(TK_{\alpha,\beta} \right)[\x_1,\hat{\x}_2] =
    & K_{\alpha,\beta}[\x_1,\x_2] T[\x_2,\hat{\x}_2]- 0 \\
    = & \frac{B}{C} \left(-e^{-i\phi_{e_2}}\right); \\
    - \left(K_{\alpha,\beta}DK_{\alpha,\beta} \right)[\x_1,\hat{\x}_2]
    = & - K_{\alpha,\beta}[\x_1,\x_2] D[\x_2,\x_2]
        K_{\alpha,\beta}[\x_2,\hat{\x}_2] \\
    = & - \frac{B}{C} \left(e^{-i\phi_{e_2}}\right).
  \end{align*}
  Another important case is the matrix element $[\x_1,\x_1]$:
  \begin{align*}
    \left(K_{\alpha,\beta}T \right)[\x_1,\x_1] -
    \left(TK_{\alpha,\beta} \right)[\x_1,\x_1] =
    & K_{\alpha,\beta}[\x_1,\hat{\x_1}] T[\hat{\x_1},\x_1]-
      T[\x_1,\hat{\x_1}] K_{\alpha,\beta}[\hat{\x_1},\x_1] \\
    = & \left(-e^{i \phi_{e_1}}\right) \left(-e^{-i \phi_{e_1}}\right)
        - \left(-e^{i \phi_{e_1}}\right) e^{-i \phi_{e_1}}; \\
    = & 2; \\
    - \left(K_{\alpha,\beta}DK_{\alpha,\beta} \right)[\x_1,\hat{\x}_2]
    = & - \left( \left(-e^{i \phi_{e_1}}\right) e^{-i \phi_{e_1}} +
        \frac{B}{C} \left(-\frac{B}{C}\right) + i\frac{D}{C} (-1)
        i\frac{D}{C} + i\frac{A}{C} i\frac{A}{C} \right)\\
    = & - \left(-1 + \frac{-B^2-A^2+D^2}{C^2}\right) \\
    = & 2.
  \end{align*}
  All other cases are similar.

  For the second point,
  $P(z,w)= \det\left(\tilde{K}_{\alpha,\beta}(z,w)\right) = \det
  \left(K_{\alpha,\beta}(z,w)\right)$ Lemma~\ref{le:diagconj}, so the
  matrix $K_{\alpha,\beta}(z,w)$ is invertible. Then \eqref{eq:commKi}
  is simply obtained by multiplying \eqref{eq:commK} by
  $K_{\alpha,\beta}^{-1}(z,w)$ on both sides.
\end{proof}

\begin{proof}[Proof of Theorem~\ref{theo:pol8vswitch}]
  We prove the polynomials relation for $(z,w)$ such that none of the
  four polynomials is zero at $(z,w)$ (\textit{i.e.} the four
  Kasteleyn matrices are invertible); the relation is then obtained by
  analytic continuation. By noting that $T^2 = I$, we can rewrite
  Theorem~\ref{theo:km1torus} as a block-matrix relation:
  \begin{equation}
    \label{eq:11}
    \begin{pmatrix}
      I & I \\
      I & -I
    \end{pmatrix}
    \begin{pmatrix}
      K_{\alpha,\beta}^{-1}(z,w) & 0 \\
      K_{\alpha',\beta'}^{-1}(z,w) & K_{\alpha',\beta'}^{-1}(z,w)
    \end{pmatrix}
    =
    \begin{pmatrix}
      I & I \\
      T & -T
    \end{pmatrix}
    \begin{pmatrix}
      K_{\alpha,\beta'}^{-1}(z,w) &  \frac12 (I-T)
      K_{\alpha,\beta'}^{-1}(z,w) \\
      K_{\alpha',\beta}^{-1}(z,w) &  \frac12 (I+T)
      K_{\alpha',\beta}^{-1}(z,w)
    \end{pmatrix}
  \end{equation}
  We take the determinant of these. The matrices $ \begin{pmatrix}
    I & I \\
    I & -I
  \end{pmatrix}$ and $ \begin{pmatrix}
    I & I \\
    T & -T
  \end{pmatrix}$ can be written in block-diagonal form, with blocks
  corresponding to the two copies of the pair $(\x,\hat{\x})$ for
  $\x\in W^T$. For the two matrices, the blocks have determinant $4$
  and there are $2\FF$ blocks, so both their
  determinants are equal to $2^{4\FF}$.

  The determinant of both sides of \eqref{eq:11} can now be computed;
  we successively use the formula for determinants of block matrices,
  Lemma~\ref{le:commut} to exchange $T$ and the matrices
  $K^{-1}_{\cdot,\cdot}(z,w)$, and $\det(D) = (-1)^{2\FF}=1$; we drop
  the $(z,w)$ in the notations to make the computation clearer:
  \begin{equation}
    \label{eq:dets}
    \begin{split}
      \left|K_{\alpha,\beta}^{-1}\right|\left|K_{\alpha',\beta'}^{-1}\right|
      &= \left|K_{\alpha',\beta}^{-1}\right| \left|\frac12 (I-T)
        K_{\alpha,\beta'}^{-1} - K_{\alpha,\beta'}^{-1} K_{\alpha',\beta}\frac12 (I+T)
        K_{\alpha',\beta}^{-1}\right| \\
      &= \left|K_{\alpha',\beta}^{-1}\right|
      \left|K_{\alpha,\beta'}^{-1}\right| \left| -
        \frac12\left(K_{\alpha,\beta'}TK_{\alpha,\beta'}^{-1}
          + K_{\alpha',\beta}TK_{\alpha',\beta}^{-1}\right)\right| \\
      &= \left|K_{\alpha',\beta}^{-1}\right|
      \left|K_{\alpha,\beta'}^{-1}\right| \left|-T +\frac12\left(K_{\alpha,\beta'}+K_{\alpha',\beta}\right)D\right|\\
      &= \left|K_{\alpha',\beta}^{-1}\right|
      \left|K_{\alpha,\beta'}^{-1}\right| \left|-TD +\frac12\left(K_{\alpha,\beta'}+K_{\alpha',\beta}\right)\right|.
    \end{split}
  \end{equation}
  Notice that $TD$ is exactly equal the part of $K_{\alpha,\beta}$
  that corresponds to legs of $G^T$. Thus
  $-TD +\frac12\left(K_{\alpha,\beta'} + K_{\alpha',\beta}\right)$ is
  a block-diagonal matrix, where blocks correspond to decorations
  inside of the faces $\FF$ of $\QQ$. When the face is not crossed by
  $\gamma^{\BB}_x$ nor $\gamma^{\BB}_y$, the block we get is
  represented in Figure~\ref{fig:locdeco}, where (in the notation of
  the proof of Theorem~\ref{theo:km1sphere}):
  \begin{align*}
    \tilde{a} &= \frac{1}{2}\left(\frac{a}{c} + \frac{a'}{c'}\right) \\
    \tilde{b} &= \frac{1}{2}\left(\frac{b}{c} + \frac{b'}{c'}\right) \\
    \tilde{d} &= \frac{1}{2}\left(\frac{d}{c} + \frac{d'}{c'}\right)
  \end{align*}

  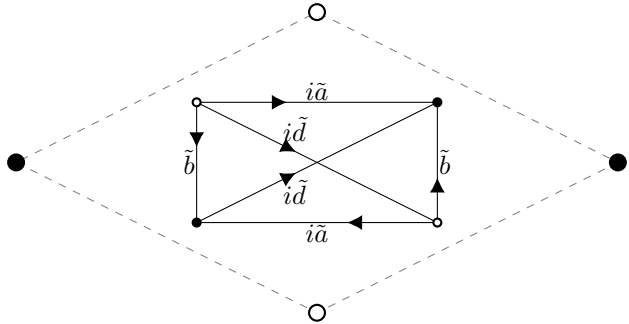
\begin{figure}[h!]
    \centering \begin{tikzpicture}[scale=2]
  \draw [color=gray, dashed] (-2,0) -- (0,-1) -- (2,0) -- (0,1) -- cycle;
  \node [draw=black, fill=black,thick,circle,inner sep=0pt,minimum size=6pt] at (-2,0) {};
  \node [draw=black, fill=white,thick,circle,inner sep=0pt,minimum size=6pt] at (0,1) {};
  \node [draw=black, fill=black,thick,circle,inner sep=0pt,minimum size=6pt] at (2,0) {};
  \node [draw=black, fill=white,thick,circle,inner sep=0pt,minimum size=6pt] at (0,-1) {};
  
  
  
  \draw [-{Latex[length=2mm,width=2mm]}] (-0.8,0.4) -- (-0.14,0.07);
  \draw (-0.22,0.11) -- (0.8,-0.4);
  \draw [-{Latex[length=2mm,width=2mm]}] (-0.8,-0.4) -- (-0.14,-0.07);
  \draw (-0.22,-0.11) -- (0.8,0.4);
  \draw [-{Latex[length=2mm,width=2mm]}] (-0.8,0.4) -- (-0.2,0.4);
  \draw (-0.22,0.4) -- (0.8,0.4);
  \draw [-{Latex[length=2mm,width=2mm]}] (-0.8,0.4) -- (-0.8,0.1);
  \draw (-0.8,0.12) -- (-0.8,-0.4);
  \draw [-{Latex[length=2mm,width=2mm]}] (0.8,-0.4) -- (0.2,-0.4);
  \draw (0.22,-0.4) -- (-0.8,-0.4);
  \draw [-{Latex[length=2mm,width=2mm]}] (0.8,-0.4) -- (0.8,-0.1);
  \draw (0.8,-0.12) -- (0.8,0.4);

  \draw (0,0.35) node [above] {$i\tilde{a}$};
  \draw (0,-0.35) node [below] {$i\tilde{a}$};
  \draw (-0.75,0) node [left] {$\tilde{b}$};
  \draw (0.75,0) node [right] {$\tilde{b}$};
  \draw (-0.15,0.22) node [] {$i\tilde{d}$};
  \draw (-0.15,-0.2) node [] {$i\tilde{d}$};

  
  \node [draw=black, fill=white,thick,circle,inner sep=0pt,minimum size=3pt] at (-0.8,0.4) {};
  \node [draw=black, fill=black,thick,circle,inner sep=0pt,minimum size=3pt] at (-0.8,-0.4) {};
  \node [draw=black, fill=white,thick,circle,inner sep=0pt,minimum size=3pt] at (0.8,-0.4) {};
  \node [draw=black, fill=black,thick,circle,inner sep=0pt,minimum size=3pt] at (0.8,0.4) {};
\end{tikzpicture}
    \caption{Coefficients of the matrix $-TD +\frac12\left(K_{\alpha,\beta'} +
        K_{\alpha',\beta}\right)$ at a face.}
    \label{fig:locdeco}
  \end{figure}
  The determinant of this block can be easily computed using \eqref{eq:weight_form2}, giving
  \begin{align*}
    \left(\tilde{a}^2 + \tilde{b}^2 - \tilde{d}^2\right)^2 &=
                                                             \frac14\left(1+\frac{a}{c}\frac{a'}{c'} + \frac{b}{c}\frac{b'}{c'} -
                                                             \frac{d}{c}\frac{d'}{c'} \right)^2 \\
                                                           &= \frac{\frac{a^2+b^2}{c^2}\frac{a'^2+b'^2}{c'^2}}{\frac{A^2+B^2}{C^2}\frac{A'^2+B'^2}{C'^2}}
  \end{align*}
  When the face is crossed by $\gamma^{\BB}_x$ or $\gamma^{\BB}_y$,
  some weights are multiplied by $z^{\pm 1}, w^{\pm 1}$ but the
  determinant is the same.

  All in all, \eqref{eq:dets} becomes
  \begin{equation}
    \label{eq:rel_det_i}
    \begin{split}
      & \left( \prod_{f\in\FF}\frac{A(f)^2+B(f)^2}{C(f)^2}\frac{A'(f)^2+B'(f)^2}{C'(f)^2} \right)
      \left|K_{\alpha,\beta}^{-1}(z,w)\right|\left|K_{\alpha',\beta'}^{-1}(z,w)\right|
      \\
      =& \left( \prod_{f\in\FF}\frac{a(f)^2+b(f)^2}{c(f)^2}\frac{a'(f)^2+b'(f)^2}{c'(f)^2} \right)
      \left|K_{\alpha,\beta'}^{-1}(z,w)\right|\left|K_{\alpha',\beta}^{-1}(z,w)\right|.
    \end{split}
  \end{equation}
  Using relation \eqref{eq:relc}
  finishes the proof.
\end{proof}

An important case appears when we set $\alpha=\beta'$ and
$\beta=\alpha'$. The model with weights $X_{\alpha,\alpha}$ is
actually a 6V model, and the weights of diagonals in $G^T$ become
null. This gives the bipartite decorated graph of Wu and Lin
\cite{WuLin}, see also \cite{Dubedat,BoutillierDeTiliere:xor} which we
denote $G^Q$.

More precisely, our Kasteleyn skew-hermitian matrix
$K_{\alpha,\alpha}(z,w)$ can be related to Boutillier, de Tilière
and Raschel's $\mathcal{K}$ matrix from Section $5$ of
\cite{BoutillierDeTiliereRaschel} - whose rows are indexed by white
vertices and columns by black vertices of $G^Q$ - via
\begin{equation}
  \label{eq:26}
  K_{\alpha,\alpha}(z,w) =
  i\begin{blockarray}{ccc}
    {\color{gray} W^T} & {\color{gray} B^T} \\
    \begin{block}{(cc)c}
      0 & \overline{\mathcal{K}}_{\alpha}(z,w) &  {\color{gray} W^T} \\
       ^t\mathcal{K}_{\alpha}(z^{-1},w^{-1}) & 0 &  {\color{gray} B^T} \\
    \end{block}
  \end{blockarray}.
\end{equation}
The determinant of $\mathcal{K}(z,w)$ is the characteristic polynomial
of a bipartite dimer model; we denote it by
$P^{6V}_{\alpha}(z,w)$. Thus $P^{6V}_{\alpha}(z,w)$ is the determinant of a
matrix twice as small as $K_{\alpha,\beta}(z,w)$.

\begin{Cor}
  \label{cor:pol8v6v}
  Let $\QQ$ be a quadrangulation in the toric case. Let
  $(\alpha,\beta)$ and $(\alpha',\beta')$ be two standard elements
  of $\left([0,2\pi)^{\FF}\right)^2$. Then the characteristic
  polynomial of the 8V-model satisfies
  \begin{equation}
    P^{8V}_{\alpha,\beta} = \tilde{c} P^{6V}_{\alpha} P^{6V}_{\beta}
  \end{equation}
  for some constant $\tilde{c}$ satisfying
  \begin{equation}
    |\tilde{c}| = \prod_{f\in\FF} \frac{2}{|C_{\alpha,\beta}(f)|}.
  \end{equation}
\end{Cor}

\begin{proof}
  Equation \eqref{eq:26} yields
  \begin{equation}
    \label{eq:20}
    P^{8V}_{\alpha,\alpha}(z,w) = \overline{P^{6V}_{\alpha}}(z,w)\
    P^{6V}_{\alpha}(z^{-1},w^{-1}).
  \end{equation}

  However, $P^{6V}_{\alpha}$ has extra symmetries. First, as it is the
  characteristic polynomial of a (bipartite) dimer model, up to a global
  factor its entries are real (see for instance Proposition 3.1 in
  \cite{kos}), so that
  $\overline{P^{6V}_{\alpha}} = c_3 P^{6V}_{\alpha}$ for some constant
  $c_3\in S^1$. It also corresponds to the dimer model on the decorated
  graph $G^Q$, and the characteristic polynomial in that case is
  proportional to that of Fisher's decorated graph \cite{Fisher} (see
  Section 4 of \cite{Dubedat}). By Corollary 16 of
  \cite{BoutillierDeTiliereRaschel},the characteristic polynomial on
  Fisher's graph has a symmetry $(z,w)\leftrightarrow
  (z^{-1},w^{-1})$. This gives
  $P^{6V}_{\alpha}(z,w) = P^{6V}_{\alpha}(z^{-1},w^{-1})$.
  As a result \eqref{eq:20} becomes
  \begin{equation}
    \label{eq:29}
    P^{8V}_{\alpha,\alpha} = c_3 \left(P^{6V}_{\alpha}\right)^2
  \end{equation}
  We can now apply Theorem \ref{theo:pol8vswitch} with $\alpha=\beta'$
  and $\beta=\alpha'$. By the same argument as for
  \eqref{eq:switch_a_b},
  $P^{8V}_{\alpha,\beta} = P^{8V}_{\beta,\alpha}$. Thus
  Theorem~\ref{theo:pol8vswitch} becomes
  \begin{equation}
    \label{eq:30}
    \left(P^{8V}_{\alpha,\beta}\right)^2 = c_2c_3^2  \left(P^{6V}_{\alpha} P^{6V}_{\beta}\right)^2.
  \end{equation}
  By analytic continuation and by computing the constant we get the
  desired relation.
\end{proof}

\section{$Z$-invariant regime}
\label{sec:zinv}

In this section we restrict to the planar case. The graph may be
periodic (in which case we will still make use of the toric case) or
not. We study the $Z$-invariant regime of the model, which is a regime
where the star-triangle relations are satisfied.

\subsection{Checkerboard Yang-Baxter equations}

Here we generalize Baxter's star-triangle relations
\cite{Baxter8,Baxter:exactly} in our ``checkerboard'' setting, and we
find free-fermion solutions.

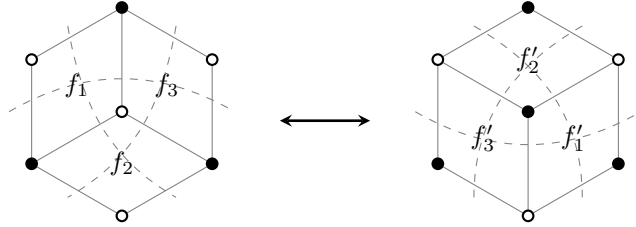
\begin{figure}[h]
  \centering
  \begin{tikzpicture}[scale=0.6]

\begin{scope}[yshift=0cm] 
  \begin{scope}[xshift=9cm]
  \coordinate (hh) at (0,2.31) ;
  \coordinate (hd) at (2,1.155) ;
  \coordinate (hg) at (-2,1.155) ;
  \coordinate (bd) at (2,-1.155) ;
  \coordinate (bg) at (-2,-1.155) ;
  \coordinate (bb) at (0,-2.31) ;
  \coordinate (x) at (0,1.155) ;
  \coordinate (y) at (1,-0.577) ;
  \coordinate (z) at (-1,-0.577) ;
  \draw [gray] (hh) -- (hd) -- (bd) -- (bb) -- (bg) -- (hg) -- cycle;
  \draw [gray] (hd) -- (0,0);
  \draw [gray] (hg) -- (0,0);
  \draw [gray] (bb) -- (0,0);
  \node [draw=black, fill=black,thick,circle,inner sep=0pt,minimum size=4pt] at (hh) {};
  \node [draw=black, fill=white,thick,circle,inner sep=0pt,minimum size=4pt] at (hd) {};
  \node [draw=black, fill=white,thick,circle,inner sep=0pt,minimum size=4pt] at (hg) {};
  \node [draw=black, fill=black,thick,circle,inner sep=0pt,minimum size=4pt] at (bd) {};
  \node [draw=black, fill=black,thick,circle,inner sep=0pt,minimum size=4pt] at (bg) {};
  \node [draw=black, fill=white,thick,circle,inner sep=0pt,minimum size=4pt] at (bb) {};
  \node [draw=black, fill=black,thick,circle,inner sep=0pt,minimum
  size=4pt] at (0,0) {};
  \draw [draw=gray,dashed] (-2.5,0) to[out=-30,in=210] (2.5,0);
  \draw [draw=gray,dashed] (-1.2,1.9) to[out=-30,in=90] (1.2,-1.9);
  \draw [draw=gray,dashed] (1.2,1.9) to[out=210,in=90] (-1.2,-1.9);
  \draw (x) node [] {$f'_2$};
  \draw (z) node [] {$f'_3$};
  \draw (y) node [] {$f'_1$};
\end{scope}
\begin{scope}[xshift=3.5cm, yshift=-0.2cm]
  \draw[>=stealth,<->, line width = 1pt] (0,0) -- (2,0);
\end{scope}
\begin{scope}[xshift=0cm]
  \coordinate (hh) at (0,2.31) ;
  \coordinate (hd) at (2,1.155) ;
  \coordinate (hg) at (-2,1.155) ;
  \coordinate (bd) at (2,-1.155) ;
  \coordinate (bg) at (-2,-1.155) ;
  \coordinate (bb) at (0,-2.31) ;
  \coordinate (x) at (0,-1.155) ;
  \coordinate (y) at (1,0.577) ;
  \coordinate (z) at (-1,0.577) ;
  \draw [gray] (hh) -- (hd) -- (bd) -- (bb) -- (bg) -- (hg) -- cycle;
  \draw [gray] (bd) -- (0,0);
  \draw [gray] (bg) -- (0,0);
  \draw [gray] (hh) -- (0,0);
  \node [draw=black, fill=black,thick,circle,inner sep=0pt,minimum size=4pt] at (hh) {};
  \node [draw=black, fill=white,thick,circle,inner sep=0pt,minimum size=4pt] at (hd) {};
  \node [draw=black, fill=white,thick,circle,inner sep=0pt,minimum size=4pt] at (hg) {};
  \node [draw=black, fill=black,thick,circle,inner sep=0pt,minimum size=4pt] at (bd) {};
  \node [draw=black, fill=black,thick,circle,inner sep=0pt,minimum size=4pt] at (bg) {};
  \node [draw=black, fill=white,thick,circle,inner sep=0pt,minimum size=4pt] at (bb) {};
  \node [draw=black, fill=white,thick,circle,inner sep=0pt,minimum size=4pt] at (0,0) {};
  \draw [draw=gray,dashed] (-2.5,0) to[out=30,in=-210] (2.5,0);
  \draw [draw=gray,dashed] (-1.2,1.9) to[out=-80,in=150] (1.2,-1.9);
  \draw [draw=gray,dashed] (1.2,1.9) to[out=-100,in=30] (-1.2,-1.9);
  \draw (x) node [] {$f_2$};
  \draw (y) node [] {$f_3$};
  \draw (z) node [] {$f_1$};
\end{scope}
\end{scope}

\end{tikzpicture}

  \caption{``Star-triangle'' move on the quadrangulation (solid lines)
  and its dual on which the 8V-configurations are defined (dashed lines).}
  \label{fig:startri}
\end{figure}

Let us suppose that the quadrangulation $\QQ$ contains three adjacent
faces in the configuration on the left of
Figure~\ref{fig:startri}. Then we can transform it locally into the
configuration on the right. We need to update the weights of the
eight-vertex model at the same time. This can be done in such a way
that there exists a coupling of the configurations on the right and of
the left quadrangulations, such that they agree everywhere except at
the central dashed ``triangles''.

Specifically, let us denote $(a_i,b_i,c_i,d_i)$ the 8V-weights at
$f_i$, and $(a'_i,b'_i,c'_i,d'_i)$ those at $f'_i$. By conditioning on
every possible boundary condition, we get the following equations for
the existence of a coupling: for every $i,j,k$ with
$\{i,j,k\}=\{1,2,3\}$,

\begin{equation}
  \label{eq:37}
  \begin{split}
    c_ic_jc_k + a_ia_ja_k &\propto c'_ic'_jc'_k + b'_ib'_jb'_k \\
    a_ic_jc_k + c_ia_ja_k &\propto c'_ia'_ja'_k + b'_id'_jd'_k \\
    c_ib_jb_k + a_id_jd_k &\propto b'_ic'_jc'_k + c'_ib'_jb'_k \\
    c_id_jd_k + a_ib_jb_k &\propto c'_id'_jd'_k + b'_ia'_ja'_k \\
    c_ib_jd_k + a_id_jb_k &\propto d'_ia'_jc'_k + a'_id'_jb'_k
  \end{split}
\end{equation}
where the proportionality constants are all the same. We call
equations \eqref{eq:37} the \emph{Yang-Baxter equations} of our model.

\begin{Rk} \noindent
  \begin{itemize}
  \item Most of the equations (all but the last one) are invariant
    under some nontrivial subgroup of the permutation of indices
    $\{i,j,k\}$. All in all \eqref{eq:37} contains $16$ distinct
    equations.
  \item We presented the ``star-triangle'' move as going from the left
    configuration to the right one, but it can of course be done in
    both ways, giving the same set of equations.
  \end{itemize}
\end{Rk}

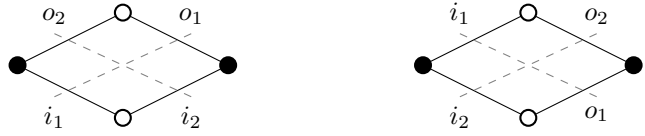
\begin{figure}[h]
  \centering
  \begin{tikzpicture}[scale=0.7]
  \draw [] (-2,0) -- (0,-1) -- (2,0) -- (0,1) -- cycle;
  \node [draw=black, fill=black,thick,circle,inner sep=0pt,minimum size=6pt] at (-2,0) {};
  \node [draw=black, fill=white,thick,circle,inner sep=0pt,minimum size=6pt] at (0,1) {};
  \node [draw=black, fill=black,thick,circle,inner sep=0pt,minimum size=6pt] at (2,0) {};
  \node [draw=black, fill=white,thick,circle,inner sep=0pt,minimum size=6pt] at (0,-1) {};
  \draw [color=gray, dashed] (-1.3,0.6) -- (1.3,-0.6);
  \draw [color=gray, dashed] (-1.3,-0.6) -- (1.3,0.6);

  \draw (-1.3,0.6) node [above] {$o_2$};
  \draw (-1.3,-0.6) node [below] {$i_1$};
  \draw (1.3,-0.6) node [below] {$i_2$};
  \draw (1.3,0.6) node [above] {$o_1$};
\end{tikzpicture} \hspace{2cm} \begin{tikzpicture}[scale=0.7]
  \draw [] (-2,0) -- (0,-1) -- (2,0) -- (0,1) -- cycle;
  \node [draw=black, fill=black,thick,circle,inner sep=0pt,minimum size=6pt] at (-2,0) {};
  \node [draw=black, fill=white,thick,circle,inner sep=0pt,minimum size=6pt] at (0,1) {};
  \node [draw=black, fill=black,thick,circle,inner sep=0pt,minimum size=6pt] at (2,0) {};
  \node [draw=black, fill=white,thick,circle,inner sep=0pt,minimum size=6pt] at (0,-1) {};
  \draw [color=gray, dashed] (-1.3,0.6) -- (1.3,-0.6);
  \draw [color=gray, dashed] (-1.3,-0.6) -- (1.3,0.6);

  \draw (-1.3,0.6) node [above] {$i_1$};
  \draw (-1.3,-0.6) node [below] {$i_2$};
  \draw (1.3,-0.6) node [below] {$o_1$};
  \draw (1.3,0.6) node [above] {$o_2$};
\end{tikzpicture}
  \caption{Entries of $R(f)$ (left) and $\bar{R}(f)$ (right) are indexed by the
    occupation state of $(i_1,i_2)$ and $(o_1,o_2)$, in the order
    $(\text{absent}, \text{absent}), (\text{absent},\text{present}),
    (\text{present},\text{absent}), (\text{present},\text{present})$.}
  \label{fig:Rmat}
\end{figure}

Equations \eqref{eq:37} are often written in matrix form. For the
checkerboard setting, we define $R$ and $\bar{R}$ matrices containing
the weights at every face, with the indexing of Figure~\ref{fig:Rmat}:
\begin{equation}
  R(f) =
  \begin{blockarray}{ccccc}
     & \begin{tikzpicture}[scale=0.2]
  \draw [color=gray, dashed] (-1.3,0.6) -- (0,0);
  \draw [color=gray, dashed] (0,0) -- (1.3,0.6);
\end{tikzpicture} & \begin{tikzpicture}[scale=0.2]
  \draw [color=gray, thick] (-1.3,0.6) -- (0,0);
  \draw [color=gray, dashed] (0,0) -- (1.3,0.6);
\end{tikzpicture} &
     \begin{tikzpicture}[scale=0.2]
  \draw [color=gray, dashed] (-1.3,0.6) -- (0,0);
  \draw [color=gray, thick] (0,0) -- (1.3,0.6);
\end{tikzpicture} & \begin{tikzpicture}[scale=0.2]
  \draw [color=gray, thick] (-1.3,0.6) -- (0,0);
  \draw [color=gray, thick] (0,0) -- (1.3,0.6);
\end{tikzpicture} \\
    \begin{block}{c(cccc)}
     \begin{tikzpicture}[scale=0.2]
  \draw [color=gray, dashed] (-1.3,-0.6) -- (0,0);
  \draw [color=gray, dashed] (0,0) -- (1.3,-0.6);
\end{tikzpicture} & C(f) & 0 & 0 & A(f) \\
     \begin{tikzpicture}[scale=0.2]
  \draw [color=gray, dashed] (-1.3,-0.6) -- (0,0);
  \draw [color=gray, thick] (0,0) -- (1.3,-0.6);
\end{tikzpicture} & 0 & D(f) & B(f) & 0 \\
     \begin{tikzpicture}[scale=0.2]
  \draw [color=gray, thick] (-1.3,-0.6) -- (0,0);
  \draw [color=gray, dashed] (0,0) -- (1.3,-0.6);
\end{tikzpicture} & 0 & B(f) & D(f) & 0 \\
     \begin{tikzpicture}[scale=0.2]
  \draw [color=gray, thick] (-1.3,-0.6) -- (0,0);
  \draw [color=gray, thick] (0,0) -- (1.3,-0.6);
\end{tikzpicture} & A(f) & 0 & 0 & C(f) \\
    \end{block}
  \end{blockarray}\ , \ \ \ \ \ \ \
  \bar{R}(f) =
  \begin{blockarray}{ccccc}
     & \begin{tikzpicture}[scale=0.2]
  \draw [color=gray, dashed] (1.3,-0.6) -- (0,0);
  \draw [color=gray, dashed] (1.3,0.6) -- (0,0);
\end{tikzpicture} & \begin{tikzpicture}[scale=0.2]
  \draw [color=gray, dashed] (1.3,-0.6) -- (0,0);
  \draw [color=gray, thick] (1.3,0.6) -- (0,0);
\end{tikzpicture} &
     \begin{tikzpicture}[scale=0.2]
  \draw [color=gray, thick] (1.3,-0.6) -- (0,0);
  \draw [color=gray, dashed] (1.3,0.6) -- (0,0);
\end{tikzpicture} & \begin{tikzpicture}[scale=0.2]
  \draw [color=gray, thick] (1.3,-0.6) -- (0,0);
  \draw [color=gray, thick] (1.3,0.6) -- (0,0);
\end{tikzpicture} \\
    \begin{block}{c(cccc)}
     \begin{tikzpicture}[scale=0.2]
  \draw [color=gray, dashed] (-1.3,0.6) -- (0,0);
  \draw [color=gray, dashed] (-1.3,-0.6) -- (0,0);
\end{tikzpicture} & C(f) & 0 & 0 & B(f) \\
     \begin{tikzpicture}[scale=0.2]
  \draw [color=gray, dashed] (-1.3,0.6) -- (0,0);
  \draw [color=gray, thick] (-1.3,-0.6) -- (0,0);
\end{tikzpicture} & 0 & D(f) & A(f) & 0 \\
     \begin{tikzpicture}[scale=0.2]
  \draw [color=gray, thick] (-1.3,0.6) -- (0,0);
  \draw [color=gray, dashed] (-1.3,-0.6) -- (0,0);
\end{tikzpicture} & 0 & A(f) & D(f) & 0 \\
     \begin{tikzpicture}[scale=0.2]
  \draw [color=gray, thick] (-1.3,0.6) -- (0,0);
  \draw [color=gray, thick] (-1.3,-0.6) -- (0,0);
\end{tikzpicture} & B(f) & 0 & 0 & C(f) \\
    \end{block}
  \end{blockarray}\ .
\end{equation}

These matrices are elements of $\mathrm{End}(V\otimes V)$,
where $V$ is a complex vector space of dimension $2$. For
$i,j\in\{1,2,3\}$, $i<j$, we define
$\mathcal{R}_{i,j}(f) \in \mathrm{End}(V\otimes V\otimes V)$ that acts
as $R(f)$ on the components $i$ and $j$, and as the identity on the
other component. We similarly define
$\bar{\mathcal{R}}_{i,j}(f)$. Then equations \eqref{eq:37} are
equivalent to (see for instance \cite{PerkAuYang})
\begin{equation}
  \mathcal{R}_{1,2}(f_1) \bar{\mathcal{R}}_{1,3}(f_2)
  \mathcal{R}_{2,3}(f_3) \propto
  \bar{\mathcal{R}}_{2,3}(f'_3)
  \mathcal{R}_{1,3}(f'_2) \bar{\mathcal{R}}_{1,2}(f'_1).
\end{equation}

\subsection{Lozenge graphs}
One way to make sure that \eqref{eq:37} always hold is to make the 8V
weights depend on the geometry of the embedding. This has been
done for several models on special embedded graphs called
\emph{isoradial}; see for instance \cite{Kenyon:lap}. In our context
it is more natural to talk only about lozenge graphs.
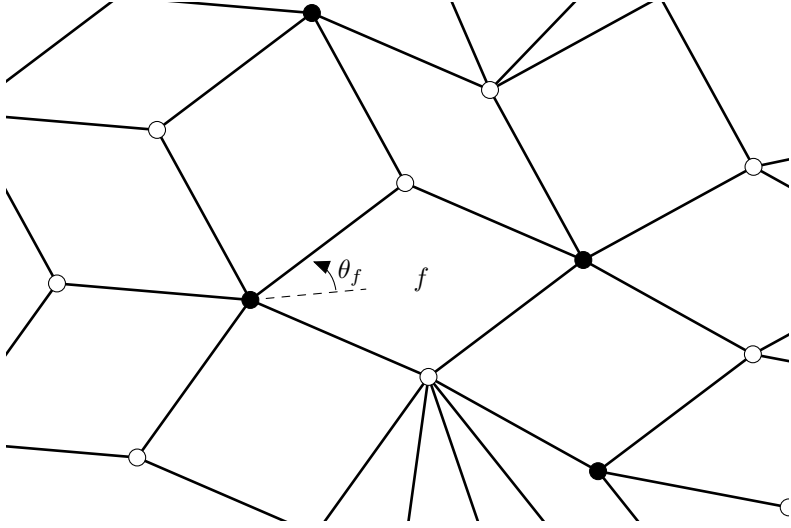
\begin{figure}[!h]
  \centering
  \begin{tikzpicture}[line cap=round,line join=round,>=triangle 45,x=1cm,y=1cm,scale=0.8]
\clip(-8.5,-5.1) rectangle (4.5,3.5);
\draw [line width=1pt] (-4.44,-1.44)-- (-1.48,-2.72);
\draw [line width=1pt] (-1.48,-2.72)-- (1.0930448057625224,-0.7759217023127627);
\draw [line width=1pt] (1.0930448057625224,-0.7759217023127627)-- (-1.8669551942374785,0.5040782976872384);
\draw [line width=1pt] (-1.8669551942374785,0.5040782976872384)-- (-4.44,-1.44);
\draw [line width=1pt] (1.0930448057625224,-0.7759217023127627)-- (-0.45743625558097656,2.0517997264024395);
\draw [line width=1pt] (-0.45743625558097656,2.0517997264024395)-- (-3.417436255580978,3.3317997264024406);
\draw [line width=1pt] (-1.8669551942374785,0.5040782976872384)-- (-3.417436255580978,3.3317997264024406);
\draw [line width=1pt] (-5.990481061343502,1.3877214287152015)-- (-4.44,-1.44);
\draw [line width=1pt] (-5.990481061343502,1.3877214287152015)-- (-3.417436255580978,3.3317997264024406);
\draw [line width=1pt] (-5.990481061343502,1.3877214287152015)-- (-9.20369753036258,1.6620203955826813);
\draw [line width=1pt] (-9.20369753036258,1.6620203955826813)-- (-6.630652724600058,3.606098693269918);
\draw [line width=1pt] (-6.630652724600058,3.606098693269918)-- (-3.417436255580978,3.3317997264024406);
\draw [line width=1pt] (-7.653216469019079,-1.165701033132518)-- (-9.20369753036258,1.6620203955826813);
\draw [line width=1pt] (-7.653216469019079,-1.165701033132518)-- (-4.44,-1.44);
\draw [line width=1pt] (-4.44,-1.44)-- (-6.32010847092546,-4.0601511669280335);
\draw [line width=1pt] (-9.533324939944535,-3.7858522000605497)-- (-6.32010847092546,-4.0601511669280335);
\draw [line width=1pt] (-9.533324939944535,-3.7858522000605497)-- (-7.653216469019079,-1.165701033132518);
\draw [line width=1pt] (-6.32010847092546,-4.0601511669280335)-- (-3.360108470925459,-5.340151166928035);
\draw [line width=1pt] (-3.360108470925459,-5.340151166928035)-- (-1.48,-2.72);
\draw [line width=1pt] (-1.48,-2.72)-- (1.3380480363429872,-4.287994025774154);
\draw [line width=1pt] (1.3380480363429872,-4.287994025774154)-- (3.9110928421055102,-2.3439157280869165);
\draw [line width=1pt] (3.9110928421055102,-2.3439157280869165)-- (1.0930448057625224,-0.7759217023127627);
\draw [line width=1pt] (1.0930448057625224,-0.7759217023127627)-- (3.921300542318484,0.7735845056897428);
\draw [line width=1pt] (3.921300542318484,0.7735845056897428)-- (6.739348578661471,-0.7944095200844115);
\draw [line width=1pt] (6.739348578661471,-0.7944095200844115)-- (3.9110928421055102,-2.3439157280869165);
\draw [line width=1pt] (-0.45743625558097656,2.0517997264024395)-- (2.3708194809749847,3.6013059344049454);
\draw [line width=1pt] (3.921300542318484,0.7735845056897428)-- (2.3708194809749847,3.6013059344049454);
\draw [line width=1pt] (-3.8021105142764045,-8.534620480237146)-- (-3.360108470925459,-5.340151166928035);
\draw [line width=1pt] (-3.8021105142764045,-8.534620480237146)-- (-1.9220020433509475,-5.914469313309112);
\draw [line width=1pt] (-1.9220020433509475,-5.914469313309112)-- (-1.48,-2.72);
\draw [line width=1pt] (-0.43660240457052013,-5.771445797954142)-- (-1.48,-2.72);
\draw [line width=1pt] (-0.8786044479214705,-8.965915111263255)-- (-1.9220020433509475,-5.914469313309112);
\draw [line width=1pt] (-0.8786044479214705,-8.965915111263255)-- (-0.43660240457052013,-5.771445797954142);
\draw [line width=1pt] (-0.43660240457052013,-5.771445797954142)-- (1.5731691553734533,-8.29351206925638);
\draw [line width=1pt] (0.5297715599439718,-5.24206627130224)-- (1.5731691553734533,-8.29351206925638);
\draw [line width=1pt] (0.5297715599439718,-5.24206627130224)-- (-1.48,-2.72);
\draw [line width=1pt] (0.5297715599439718,-5.24206627130224)-- (3.34781959628696,-6.810060297076392);
\draw [line width=1pt] (1.3380480363429872,-4.287994025774154)-- (3.34781959628696,-6.810060297076392);
\draw [line width=1pt] (1.3380480363429872,-4.287994025774154)-- (4.506301426788489,-4.889800019372249);
\draw [line width=1pt] (4.506301426788489,-4.889800019372249)-- (7.079346232551011,-2.945721721685012);
\draw [line width=1pt] (3.9110928421055102,-2.3439157280869165)-- (7.079346232551011,-2.945721721685012);
\draw [line width=1pt] (-0.45743625558097656,2.0517997264024395)-- (-1.7388619432577967,5.011182805181191);
\draw [line width=1pt] (-0.45743625558097656,2.0517997264024395)-- (1.7771401207285915,4.377024106637434);
\draw [line width=1pt] (3.921300542318484,0.7735845056897428) -- (5,1);

\draw [fill=black] (-3.417436255580978,3.3317997264024406) circle (4pt);
\draw [fill=black] (-4.44,-1.44) circle (4pt);
\draw  (-1.6,-1.1) node [] {$f$};
\draw [dashed] (-4.44,-1.44) -- (-2.52,-1.26);
\path [->] (-3.02,-1.27) edge[bend right] node[right] {$\theta_f$} (-3.4,-0.8);
\draw [fill=white] (-1.48,-2.72) circle (4pt);
\draw [fill=white] (-1.8669551942374785,0.5040782976872384) circle (4pt);
\draw [fill=black] (1.0930448057625224,-0.7759217023127627) circle (4pt);
\draw [fill=white] (-0.45743625558097656,2.0517997264024395) circle (4pt);
\draw [fill=white] (-5.990481061343502,1.3877214287152015) circle (4pt);
\draw [fill=white] (-7.653216469019079,-1.165701033132518) circle (4pt);
\draw [fill=white] (-6.32010847092546,-4.0601511669280335) circle (4pt);
\draw [fill=white] (3.921300542318484,0.7735845056897428) circle (4pt);
\draw [fill=white] (3.9110928421055102,-2.3439157280869165) circle (4pt);
\draw [fill=black] (-3.360108470925459,-5.340151166928035) circle (4pt);
\draw [fill=black] (1.3380480363429872,-4.287994025774154) circle (4pt);
\draw [fill=white] (4.506301426788489,-4.889800019372249) circle (4pt);

\end{tikzpicture}
  \label{fig:los}
  \caption{A portion of a lozenge graph.}
\end{figure}

We say that the planar quadrangulation $\QQ$ is a \emph{lozenge graph}
if it is embedded in such a way that all faces are nondegenerate
rhombi, with edge length equal to $1$. Then for every $f\in\FF$, there
is a natural parameter $\theta(f)\in (0,\pi/2)$, which is the
half-angle of the black corners of the rhombus.
For a vertex $\x \in V^T$, we also denote $\theta(\x) = \theta(f)$
where $f$ is the face containing $\x$.

A lozenge graph $\QQ$ is said to be \emph{quasicrystalline} if the
number $l$ of possible directions $\pm e^{i\alpha}$ of the edges of
the rhombi is finite. In that case there exists an $\epsilon > 0$ such
that for all faces $f$, $\theta(f) \in (\epsilon, \frac{\pi}{2}-\epsilon)$.

Let $k$ be a complex number such that $k^2\in (-\infty,1)$, which will
serve as an \emph{elliptic modulus}. We denote by $\mathrm{K}(k)$ (or simply
$\mathrm{K}$) the complete elliptic integral of the first kind associated to
$k$. We denote by $\am(\cdot|k)$ the Jacobi amplitude with modulus
$k$. For every complex number $\theta \in \C$, we define $\theta_k$ as
\begin{equation}
  \label{eq:10}
  \theta_k = \frac{2\mathrm{K}(k)}{\pi} \theta.
\end{equation}

\begin{Prop}
  \label{prop:pdszinv}
  Let $\QQ$ be a lozenge graph. Let $k,l$ be two elliptic
  moduli, with $k^2\leq l^2$. For every $f\in\FF$, let
  \begin{equation}
    \label{eq:38}
     \begin{split}
       \alpha(f) & = \am\left(\left. \theta(f)_k \right| k\right), \\
       \beta(f) & = \am\left(\left. \theta(f)_l \right|
         l\right).
     \end{split}
 \end{equation}
 Then $\alpha,\beta$ satisfy \eqref{eq:61}, and the weights
 $X_{\alpha,\beta}$ satisfy the Yang-Baxter equations
 \eqref{eq:37}.
\end{Prop}
\begin{proof}
  The rhombi are supposed to be nondegenerate so that
  $\theta_f\in(0,\frac{\pi}{2})$, and for $u\in (0,\mathrm{K}(k))$ one has
  $\am(u|k)\in(0,\frac{\pi}{2})$ (see for instance
  \cite{AbramowitzStegun}). To show that \eqref{eq:61} holds, it
  suffices to show that for all $\lambda \in (0,1)$,
  \begin{equation}
    \label{eq:45}
    f_{\lambda}(k) = \am\left(\lambda \mathrm{K}(k)|k\right)
  \end{equation}
  is an increasing function of $k^2\in(-\infty,1)$.
  This has been shown in \cite{Jordan} (see also \cite{CarlsonTodd}
  for a reference in English) on the domain $k^2\in [0,1)$, but the
  proof works identically for $k^2\in (-\infty,1)$.

  We now prove \eqref{eq:37}.
  It is easy to check that these equations are unchanged if we
  multiply the weights $d$ at every face by $-1$. Simple but lengthy
  computations also show that they are still satisfied if we apply the
  duality of Proposition~\ref{prop:8vdu} at every face. As a result,
  we just have to check equations~\ref{eq:37} for the weights
  \eqref{eq:8v_gamma} (we show how these weights can be transformed
  into $X_{\alpha,\beta}$ in the proof of
  Corollary~\ref{cor:8vff_is}). This holds \textit{iff} the Ising
  models defined by $\alpha$ and $\beta$ of \eqref{eq:38} satisfy the
  star-triangle relations on lozenge graphs, which is the case as
  shown in \cite{BoutillierDeTiliere:per}.
\end{proof}

\begin{Rk}
  The weights of this $Z$-invariant 8V model are
  \begin{equation}
    \label{eq:18}
    \begin{split}
      A(f) &= \sn(\theta(f)_{k}|k) + \sn(\theta(f)_{l}|l) \\
      B(f) &= \cn(\theta(f)_{k}|k) + \cn(\theta(f)_{l}|l) \\
      C(f) &= 1 +  \sn(\theta(f)_{k}|k) \sn(\theta(f)_{l}|l) +
      \cn(\theta(f)_{k}|k) \cn(\theta(f)_{l}|l) \\
      D(f) &= \cn(\theta(f)_{k}|k) \sn(\theta(f)_{l}|l) -
      \sn(\theta(f)_{k}|k) \cn(\theta(f)_{l}|l)
    \end{split}
  \end{equation}
  The \emph{dual} modulus of $k$ is defined as $k'=\sqrt{1-k^2}$. When
  $k' = \frac{1}{l'}$, (or $l=k^*$ in the notations of
  \cite{BoutillierDeTiliereRaschel}), the bipartite coloring no longer
  matters and we recover the $Z$-invariant weights of Baxter
  \cite{Baxter8,Baxter:exactly} at the free-fermion point.

  When $k=l$, we get a $Z$-invariant 6V model whose corresponding
  bipartite dimer model has been studied in
  \cite{BoutillierDeTiliereRaschel}.
\end{Rk}

From now on, we suppose that $\QQ$ is a lozenge graph, and that two
elliptic moduli $k^2\leq l^2$ are chosen. \emph{We replace the indices
$\alpha,\beta$ by $k,l$}, meaning that they correspond to the
$\alpha,\beta$ of \eqref{eq:38}. We also slightly modify our Kasteleyn
matrices by setting $\phi_e = \theta_e$ in the notations of
Figure~\ref{fig:dimhwl}. These angles also satisfy \eqref{eq:69},
\eqref{eq:70}, \eqref{eq:72} so the results of Section \ref{sec:dim}
still hold.

\subsection{Local expression for $K_{k,l}^{-1}$}
\label{sec:local}

In the case where $k=l$, we have $\alpha=\beta$ and we already know
that this corresponds to a free-fermionic \emph{six}-vertex model --
or equivalently to dimers on a bipartite decorated graph $G^Q$. The
operator $K_{k,k}$ can be written as
\begin{equation}
  \label{eq:Kaablock}
  K_{k,k} =
  i
  \begin{blockarray}{ccc}
    {\color{gray} W^Q} & {\color{gray} B^Q} \\
    \begin{block}{(cc)c}
      0 & \overline{\mathcal{K}_k} &  {\color{gray} W^Q} \\
      ^t\mathcal{K}_k & 0 &  {\color{gray} B^Q} \\
    \end{block}
  \end{blockarray}
\end{equation}
where $\mathcal{K}_k$ is the operator from black to white vertices,
associated to the elliptic modulus $k$, defined in Section~5 of
\cite{BoutillierDeTiliereRaschel}; we only change notation slightly to
emphasize the dependence on $k$. In the following subsection we recall
the tools of \cite{BoutillierDeTiliereRaschel:lap,
  BoutillierDeTiliereRaschel} that are required to compute a local
formula for $\mathcal{K}_k^{-1}$. Our definitions differ from those of \cite{BoutillierDeTiliereRaschel:lap,
  BoutillierDeTiliereRaschel} by the multiplication of the arguments
by a factor $\frac{\pi}{2\mathrm{K}(k)}$, which is aimed at making the
dependence in $k$ more apparent.

\subsubsection{Inverse of $\mathcal{K}_k$}
\label{sec:inverse-mathcalk_k}

Let $b\in B^T$ and $w\in W^T$. We chose a path on $\QQ$ going from the
half-edge closest to $\b$ to the half-edge closest to $\w$, which we
denote
$\frac12 e^{i\alpha_1}, e^{i\alpha_2}, \dots, e^{i\alpha_{n-1}},
\frac12 e^{i\alpha_n}$. We also set $a_1, \dots, a_{n-1}$ to be the
successive vertices of $\QQ$ in that path. See
Figure~\ref{fig:pathQ}. The following definitions do not depend on the
choice of this path.

\begin{figure}[h]
  \centering
  \def\svgwidth{10cm}
  \import{./}{pathQ.tex}
  \caption{A path on $\QQ$ from $\b$ to $\w$.}
  \label{fig:pathQ}
\end{figure}

The \emph{discrete $k$-massive exponential function} is defined in
\cite{BoutillierDeTiliereRaschel:lap} as
\begin{equation}
  \label{eq:21}
  \mathrm{e}_{a_1,a_{n-1}}(u|k) = \prod_{j=2}^{n-1} i \sqrt{k'} \
  \sc\left(\left. \left(\frac{u-\alpha_j}{2}\right)_k \right| k \right).
\end{equation}
This is a well-defined function of the complex argument $u$. It is
moreover $2\pi$-periodic, and $2i\pi \frac{\mathrm{K}'}{\mathrm{K}}$-periodic when $a_1$
and $a_{n-1}$ are the same color (\textit{i.e.} the product contains
an even number of terms), $2i\pi \frac{\mathrm{K}'}{\mathrm{K}}$-antiperiodic otherwise.

Let also
\begin{equation}
  \label{eq:27}
  h(u\mid k) =
  \begin{cases}
    \nc\left(\left. \left(\frac{u-\alpha_1}{2}\right)_k \right| k
    \right) \
    \nc\left(\left. \left(\frac{u-\alpha_n}{2}\right)_k \right| k
    \right) \ (-k')
    &\text{if } a_1\in \BB, a_{n-1} \in \BB \\
    \nc\left(\left. \left(\frac{u-\alpha_1}{2}\right)_k \right| k
    \right) \
    \dc\left(\left. \left(\frac{u-\alpha_n}{2}\right)_k \right| k
    \right) \ (-\sqrt{k'})
    &\text{if } a_1\in \BB, a_{n-1} \in \WW \\
    \dc\left(\left. \left(\frac{u-\alpha_1}{2}\right)_k \right| k
    \right) \
    \nc\left(\left. \left(\frac{u-\alpha_n}{2}\right)_k \right| k
    \right) \ \sqrt{k'}
    &\text{if } a_1\in \WW, a_{n-1} \in \BB \\
    \dc\left(\left. \left(\frac{u-\alpha_1}{2}\right)_k \right| k
    \right) \
    \dc\left(\left. \left(\frac{u-\alpha_n}{2}\right)_k \right| k
    \right)
    &\text{if } a_1\in \WW, a_{n-1} \in \WW.
  \end{cases}
\end{equation}
Then the function $e^{-\frac{i}{2} (\alpha_n-\alpha_1)}h(u|k)$ is
well defined and has the same (anti)periodicity as
$e_{a_1,a_{n-1}}(u|k)$. As a result the following function is
meromorphic on the torus $\T(k) = \C / \left(2\pi \Z + 2i\pi
  \frac{\mathrm{K}'}{\mathrm{K}}\Z\right)$:
\begin{equation}
  \label{eq:39}
  f_{\b,\w}(u|k) = e^{i\theta(\w)}e^{-\frac{i}{2}
    (\alpha_n-\alpha_1)} \ h(u|k) \ e_{a_1,a_{n-1}}(u|k).
\end{equation}
Its only possible poles are the $\alpha_i + \pi$. On can chose the
paths joining $\b$ and $\w$ such that the angles $\alpha_i$ all lie in
an open interval of length $\pi$. Let $\Gamma_{\b,\w|k}$ be a vertical
contour on $\T(k)$ avoiding this sector.

\begin{Theo*}[\cite{BoutillierDeTiliereRaschel}, Theorem 37]
  For $\b\in B^T,\w \in W^T$, let
  \begin{equation}
    \label{eq:40}
    \mathcal{K}_k^{-1}[\b,\w] =
    \frac{\mathrm{K}(k)}{2i\pi^2}\int_{\Gamma_{\b,\w | k}}f_{\b,\w|k}(u)\  \mathrm{d}u.
  \end{equation}
  Then $\mathcal{K}_k^{-1}$ is an inverse of the operator
  $\mathcal{K}_k$. For $k\neq 0$, it is the only inverse with bounded
  coefficients.
\end{Theo*}

\subsubsection{Inverse of $K_{k,l}$}
\label{sec:inverse-k_alpha-beta}

By \eqref{eq:Kaablock}, the following are inverses of
$K_{k,k}$ and $K_{l,l}$:
\begin{equation}
  \label{eq:Km1block}
  K^{-1}_{k,k} =
  -i
  \begin{blockarray}{ccc}
    {\color{gray} W^T} & {\color{gray} B^T} \\
    \begin{block}{(cc)c}
      0 & ^t\mathcal{K}^{-1}_k &  {\color{gray} W^T} \\
     \overline{\mathcal{K}^{-1}_k} & 0 &  {\color{gray} B^T} \\
    \end{block}
  \end{blockarray}, \ \ \ \
  K^{-1}_{l,l} =
  -i
  \begin{blockarray}{ccc}
    {\color{gray} W^T} & {\color{gray} B^T} \\
    \begin{block}{(cc)c}
      0 & ^t\mathcal{K}^{-1}_l &  {\color{gray} W^T} \\
       \overline{\mathcal{K}^{-1}_l} & 0 &  {\color{gray} B^T} \\
    \end{block}
  \end{blockarray}.
\end{equation}

\begin{Cor}
  \label{cor:inv_inf}
  The operator
  \begin{equation}
    \label{eq:41}
    K_{k,l}^{-1} = \frac12 \left( (I+T)
        K_{k,k}^{-1} + (I-T) K_{l,l}^{-1}\right)
  \end{equation}
  is an inverse of $K_{k,l}$. It is the only inverse with
  bounded coefficients.

  Its coefficients read, for $\w,\w' \in W^T, \b,\b' \in B^T$,
  \begin{align}
    \label{eq:22}
    K_{k,l}^{-1} [\w,\b] &= \frac{-i}{2}
                                    \left(\mathcal{K}_k^{-1}[\b,\w] +
                                    \mathcal{K}_l^{-1}[\b,\w] \right),
    \\
    K_{k,l}^{-1} [\b,\w] &= \frac{-i}{2}
                                    \left(\overline{\mathcal{K}_k^{-1}[\b,\w]} +
                                    \overline{\mathcal{K}_l^{-1}[\b,\w]}
                                    \right),
    \\
    K_{k,l}^{-1} [\w,\w'] &= \frac{ie^{i\theta(\w)}}{2}
                                    \left(\overline{\mathcal{K}_k^{-1}[\hat{\w},\w']} -
                                    \overline{\mathcal{K}_l^{-1}[\hat{\w},\w']} \right),
    \\
    K_{k,l}^{-1} [\b,\b'] &= \frac{ie^{-i\theta(\b)}}{2}
                                    \left(\mathcal{K}_k^{-1}[\b',\hat{\b}] -
                                    \mathcal{K}_l^{-1}[\b',\hat{\b}] \right).
  \end{align}
\end{Cor}
\begin{proof}
  This is a direct consequence of \eqref{eq:Kaablock}, \eqref{eq:40}
  and Theorem~\ref{theo:km1sphere}.
\end{proof}

\subsubsection{Asymptotics of coefficients}
\label{sec:asympt-coeff}

The asymptotics of the coefficients of $\mathcal{K}^{-1}$ for points
$\b$ and $\w$ far away is also computed in
\cite{BoutillierDeTiliereRaschel}. To state the result, using the
notations of Section~\ref{sec:inverse-mathcalk_k}, we also introduce
the following real function:
\begin{equation}
  \label{eq:75}
  \chi(u,k) = \frac{1}{|a_1-a_{n-1}|}
  \log\left[\mathrm{e}_{a_1,a_{n-1}}\left( \left. u+ i\pi\frac{\mathrm{K}'}{\mathrm{K}}
      \right| k \right) \right].
\end{equation}
As stated before, the $\alpha_i$ can be taken in an interval of length
$\pi$; let $\alpha$ be the center of this interval. According to
\cite{BoutillierDeTiliereRaschel:lap}, for any $k\in(0,1)$ the
equation $\frac{\partial \chi}{\partial u} (u,k) = 0$ has a unique
solution in $\alpha + (-\frac{\pi}{2},\frac{\pi}{2})$. Let $u_0(k)$ be
this solution, then $u_0(k)$ corresponds to a local minimum of
$\chi(\cdot|k)$, and $\chi(u_0(k),k) < 0$.

\begin{Theo}[\cite{BoutillierDeTiliereRaschel}, Theorem~38]
  \label{theo:asympt_Kk}
  Let $\QQ$ be a quasicrystalline planar lozenge graph, and $k\in(0,1)$. Then
  when $|\b - \w| \to \infty$,
  \begin{equation}
    \label{eq:76}
    \mathcal{K}_k^{-1}[\b,\w] = \frac{ \mathrm{K}(k) \ e^{i\theta(\w)}e^{-\frac{i}{2}
    (\alpha_n-\alpha_1)} \ h\left( \left. u_0(k)+i\pi\frac{\mathrm{K}'}{\mathrm{K}}
    \right| k\right) +
  o(1)} {\sqrt{2\pi^2 |a_1 - a_{n-1}| \frac{\partial^2 \chi}{\partial u
      ^2}(u_0(k),k)}}
e^{|a_1-a_{n-1}|\chi\left(u_0(k),k\right)}.
  \end{equation}
\end{Theo}

The case $k=0$ can be deduced from Theorem~4.3 of \cite{Kenyon:lap}
and corresponds to a polynomial decay of the coefficients of the
inverse matrix.

To get precise asymptotics for $K_{k,l}^{-1}$, we need to
compare two terms coming from $\mathcal{K}_k^{-1}$ and
$\mathcal{K}_l^{-1}$. The following Lemma lets us compare the main
term, $e^{|a_1-a_{n-1}|\chi\left(u_0(k),k\right)}$. The conclusion is
natural, since the case $k=0$ corresponds to critical models (where
the decay of correlations is polynomial), while as $k$ gets bigger the
decay is exponential and should have a faster rate. Thus for
$0<k<l<1$, only the term corresponding to $k$ remains in the
asymptotics.

\begin{Le}
  \label{le:monot}
  The function $k\mapsto \left| \chi(u_0(k),k) \right|$ is increasing
  in $(0,1)$.
\end{Le}
The proof can be found in Appendix~\ref{sec:proof-lemma}.

\begin{Rk}
  In the case of the $Z$-invariant elliptic Laplacian
  \cite{BoutillierDeTiliereRaschel:lap}, Ising or free-fermion 6V
  model \cite{BoutillierDeTiliereRaschel}, the characteristic
  polynomial defines a Harnack curve of genus $1$; in fact every
  Harnack curve of genus $1$ with a central symmetry can be obtained
  in this way. This means that its amoeba's complement has only one
  bounded component, or ``oval'' (see Figure~\ref{fig:amoebas}). The
  boundary of this convex oval is parametrized by functions
  $\chi(\cdot,k)$ for appropriate paths, and the value of
  $\chi(u_0(k),k)$ corresponds to the position of an extremal point of
  the oval in the path direction. Thus Lemma~\ref{le:monot} shows that
  as $k$ goes from $0$ to $1$, these ovals are actually included into
  each other. In \cite{BoutillierDeTiliereRaschel:lap} the authors
  show that the area of the oval grows from $0$ to $\infty$, but the
  monotonic inclusion is new.
\end{Rk}

We can now deduce the asymptotics of coefficients for
$K_{k,l}^{-1}$. There is a technical difficulty due to the
fact that the prefactor $h\left( \left. u_0(k)+i\pi\frac{\mathrm{K}'}{\mathrm{K}}
  \right| k\right)$ in Theorem~\ref{theo:asympt_Kk} can be zero. This
may happen when $u_0(k)$ is equal to $\alpha_1$ or to $\alpha_n$, in
the notation of Figure~\ref{fig:pathQ}. We do not expect this to
happen except for a finite number of moduli $k$, but we could not get
rid of this hypothesis.

\begin{Cor}
  \label{cor:asymp}
  Let $\QQ$ be a quasicrystalline planar lozenge graph, and $0 \leq k <
  1$. We let $|\b - \w| \to \infty$; suppose that there is an
  $\epsilon>0$ such that $|u_0(k)-\alpha_1|>\epsilon,
  |u_0(k)-\alpha_n|>\epsilon$ for all $\b,\w$. Then
    \begin{align}
      K_{k,l}^{-1} [\w,\b] &=
                                      \frac{ -i \mathrm{K}(k) \ e^{i\theta(\w)}e^{-\frac{i}{2}
    (\alpha_n-\alpha_1)} \ h\left( \left. u_0(k)+i\pi\frac{\mathrm{K}'}{\mathrm{K}}
    \right| k\right) +
  o(1)} {2\sqrt{2\pi^2 |a_1 - a_{n-1}| \frac{\partial^2 \chi}{\partial u
      ^2}(u_0(k),k)}}
e^{|a_1-a_{n-1}|\chi\left(u_0(k),k\right)}.
  \end{align}
\end{Cor}
\begin{proof}
  This comes immediately from Corollary~\ref{cor:inv_inf},
  Theorem~\ref{theo:asympt_Kk} and Lemma~\ref{le:monot}. The fact that $h$
  is bounded away from zero is a consequence of the technical
  hypothesis, and the fact that
  $\frac{\partial^2 \chi}{\partial u^2}(u_0(k),k)$ is bounded and
  bounded away from zero is proven in
  \cite{BoutillierDeTiliereRaschel:lap}.
\end{proof}
The other coefficients of $K_{k,l}^{-1}$ can be computed in a
similar way using Corollary~\ref{cor:inv_inf}, giving the same
exponential behavior. When $k=0 < l < 1$, the decay is polynomial, so
that all these models can be considered as ``critical''.

We conclude this part on asymptotics with the computation of a
critical parameter.
\begin{Prop}
  \label{prop:exp_crit}
  Let $\QQ$ be a quasicrystalline planar lozenge graph. For any $\b,\w$, as
  $k\to 0$, there exists positive constants $c,C>0$ such that the
  exponential rate of decay $\chi\left(u_0(k),k\right)$ satisfies
  \begin{equation}
    -C k^2 \leq \chi\left(u_0(k),k\right) \leq -c k^2.
  \end{equation}
\end{Prop}
\begin{proof}
  In the notations of Appendix~\ref{sec:proof-lemma}, we showed that the
  minimum of $g$ is $\frac12 \log(k')$, so that
  \begin{equation}
    \chi\left(u_0(k),k\right) \geq \frac{n-2}{2r} \log(k') =
    \frac{n-2}{2r} \log\left((1-k^2)^{1/2}\right) \sim - \frac{n-2}{4r} k^2.
  \end{equation}
  On the other hand, by Lemma~16 of
  \cite{BoutillierDeTiliereRaschel:lap}, there exists an $\epsilon>0$
  such that
  $\chi\left(u_0(k),k\right)<
  \log(\sqrt{k'}\nd\left(\epsilon|k\right))$. As
  $\nd\left(\epsilon|k\right) \to 1$ when $k\to 0$, this is equivalent
  to $-\frac14 k^2$.
\end{proof}

\subsection{Free energy and Gibbs measure}
\label{sec:gibbs}
A Gibbs measure can be constructed by taking the limit of Boltzmann
measures on toric graphs, \textit{i.e.} to consider periodic boundary
conditions. When $\QQ$ is a $\Z^2$-periodic quadrangulation, we can
define a toric exhaustion by $\QQ_n = \QQ / n\Z^2$. Using this idea
and the machinery developed for finite graphs in
Section~\ref{sec:dim}, we can prove Theorem~\ref{theo:gibbs}.

\begin{proof}[Proof of Theorem~\ref{theo:gibbs}]
  The proof follows closely from the arguments of \cite{ckp}, see also
  Theorem~6 of \cite{BoutillierDeTiliere:per}. We sketch the main
  steps here.

  First consider the case where $\QQ$ is $\Z^2$-periodic. We denote
  $\Pr_{8V}^n$ the Boltzmann probability on $\QQ_n$. We use
  Kolmogorov's extension theorem; to do so, it is sufficient to show
  that the right-hand side of \eqref{eq:42} is the limit as
  $n\to \infty$ of the probability that $e_1,\dots,e_m\in\tau$ in the
  toric graph $\QQ_n$. This probability is given by
  \eqref{eq:19}. Thus we want to compute the limit of
  $\Pf\left(\tilde{K}_{n;k,l}^{s,t}\right)_{V^c}$ (where the $n$ means
  that the matrix is defined on $\QQ_n$) for any $s,t\in\{0,1\}$. When
  $k\neq 0$ or $(s,t) \neq (0,0)$, the matrix
  $\tilde{K}_{n;k,l}^{s,t}$ is invertible and by Jacobi's identity,
  \begin{equation}
    \label{eq:100}
    \Pf\left(\tilde{K}_{n;k,l}^{s,t}\right)_{V^c} = \Pf
    \left[\left(\tilde{K}_{n;k,l}^{s,t}\right)^{-1}\right]_{V} \Pf\left(\tilde{K}_{n;k,l}^{s,t}\right).
  \end{equation}
  By
  Lemma~\ref{le:diagconj} and Theorem~\ref{theo:km1torus},
  \begin{equation}
    \label{eq:98}
    \left(\tilde{K}_{n;k,l}^{s,t}\right)^{-1} = \mathcal{D}^{-1}
    \frac12 \left[ (I+T)
      \left(K^{s,t}_{n;k,k}\right)^{-1} +
      (I-T) \left(K^{s,t}_{n;l,l}\right)^{-1} \right]
    \mathcal{D}.
  \end{equation}
  As $n\to\infty$, the coefficients of
  $\left(K^{s,t}_{n;k,k}\right)^{-1}$ tend to that of the infinite
  matrix $K_{k,k}^{-1}$ by the 6V case
  \cite{BoutillierDeTiliereRaschel}. Using Corollary~\ref{cor:inv_inf}
  we get that the coefficients of
  $\left(\tilde{K}_{n;k,l}^{s,t}\right)^{-1}$ converge to that of the
  infinite matrix $\tilde{K}_{k,l}^{-1}$ (where the orientation of the
  infinite graph is obtained by periodizing the orientation of
  $\QQ_1$).

  As a result, \eqref{eq:100} implies that for $(s,t)\neq(0,0)$ or
  $k\neq 0$,
  \begin{equation}
    \Pf\left(\tilde{K}_{n;k,l}^{s,t}\right)_{V^c} \sim_{n\to \infty} \Pf
    \left[\tilde{K}_{k,l}^{-1}\right]_{V} \Pf\left(\tilde{K}_{n;k,l}^{s,t}\right).
  \end{equation}
  When $k=0$ and $(s,t)=(0,0)$, the generic arguments in
  \cite{BoutillierDeTiliere:per} imply that
  \begin{equation}
    \frac{\Pf\left(\tilde{K}_{n;0,l}^{0,0}\right)_{V^c}}{\ZZ_{8V}(\QQ_n,X_{0,l})}
    = O\left(\frac{1}{n}\right).
  \end{equation}
  Using these estimates, Proposition~\ref{prop:corrkast} implies
  \begin{equation}
    \label{eq:103}
    \Pr_{8V}^n(e_1,\dots,e_m\in\tau) \sim \Pf
    \left[\tilde{K}_{k,l}^{-1}\right]_{V} \left(
      \frac{\prod_{f\in\FF_n}C_{k,l}(f)}{2\ZZ_{8V}(\QQ_n,X_{k,l})}
      \sum_{s,t}\epsilon_{s,t} \Pf\left(\tilde{K}_{n;k,l}^{s,t}\right)
    \right)
  \end{equation}
  where $\epsilon_{0,0}=-1$ and the others are $1$. By
  Corollary~\ref{cor:partfunc}, the right-hand side is simply $\Pf
  \left[ \tilde{K}_{k,l}^{-1}\right]_{V} = \sqrt{\det
    \left[\left(K_{k,l}^{-1}\right)_{V}\right]}$.

  The non-periodic case can be deduced from the periodic case by the
  generic arguments of \cite{DeTiliere:quadri}. It comes from the
  uniqueness of an inverse of the infinite Kasteleyn matrix with
  bounded coefficients and the locality property of
  Corollary~\ref{cor:inv_inf}.
\end{proof}

When $\QQ$ is $\Z^2$-periodic, the \emph{free energy} is defined as
\begin{equation}
  \label{eq:43}
  F^{k,l}_{8V} = - \lim_{n\to\infty} \frac{1}{n^2} \log
  \ZZ_{8V}(\QQ_n, X_{k,l}).
\end{equation}
Its existence and exact value can be deduced from that of dimers
\cite{ckp,kos}, giving the following:

\begin{Prop}
  Let $\QQ$ be a periodic lozenge graph, and $0\leq k < l < 1$. Let
  $P^{8V}_{k,l}$ be the characteristic polynomial of the 8V-model on
  the toric graph $\QQ_1$. Then
  \begin{equation}
    \label{eq:freee}
    F^{k,l}_{8V} = -\sum_{f \in \FF_1}\log C_{k,l}(f) \ - \frac12 \iint_{\T^2}
    \left|\log P^{8V}_{k,l}(z,w)\right|
    \frac{\mathrm{d}z}{2i\pi z}\frac{\mathrm{d}w}{2i\pi w}.
  \end{equation}
\end{Prop}

\appendix

\section{8V-configurations as $1$-forms}
\label{sec:1form}
This section aims at providing a simple algebraic framework to
understand 8V duality and order-disorder variables. Specifically, we
write configurations as elements of certain $\Z_2$-modules, so we use
additive notations; similar definitions can be found for various
models, in multiplicative notation, in \cite{DubedatD}. We do this for
a quadrangulation $\QQ$ only in the spherical case.

\subsection{Setup}
\label{sec:setup}

A spin configuration on the vertices $\VV$ of $\QQ$ can be seen as an
element $\bm{\sigma} \in \Z_2^{\VV}$ (we will use bold notations to
represent objects defined in $\Z_2$-modules). Then the spin-vertex
correspondence sketched at the beginning of
Section~\ref{sec:spin-vert-corr} can be seen as a linear map
$\bm{\Phi}: \Z_2^{\VV} \to \left(\Z_2^2\right)^{\FF}$, such that for a
spin configuration
$\bm{\sigma} = \left(\bm{\sigma}_v\right)_{v\in\VV}$ and a face
$f\in \FF$ with boundary vertices $b,b'\in\BB$ and $w,w'\in\WW$,
\begin{equation}
  \label{eq:phi_def}
  \bm{\Phi}\left(\bm{\sigma}\right)_f =
  (\bm{\sigma}_b \bm{+} \bm{\sigma}_{b'}, \bm{\sigma}_w \bm{+} \bm{\sigma}_{w'}).
\end{equation}
Thus an 8V-configuration can be represented as an element $\t =
(\a_f,\bm{\beta}_f)_{f\in\FF} \in \left(\Z_2^2\right)^{\FF}$, with $\a_f$
taking the value $\0$ when $\t_f$ is of type $A$ or $C$
(\textit{i.e.} the black spins are equal) and $\1$ when it is of type
$B$ or $D$, and similarly for $\bm{\beta}_f$ and white spins. Note
that this characterizes the 8V configuration up to global complement,
so that this setup is only relevant in the zero field case.

Thus we define the set of 8V-configurations as
$\H = \text{Im } \bm{\Phi}$, which is a strict subset of
$\left(\Z_2^2\right)^{\FF}$. Also note that if
$\t = (\a_f,\bm{\beta}_f)_{f\in\FF} \in \H$, then for any $b\in\BB$,
$\sum_{f \sim b} \bm{\beta}_f = \bm{0}$ where the sum is over all
faces adjacent to $b$. Similarly, if $w\in \WW$, $\sum_{f \sim w}
\bm{\alpha}_f = \bm{0}$. This motivates the definition of $\bm{\Psi}:
\left(\Z_2^2\right)^{\FF} \to \Z_2^\VV$, such that if $\t =
(\bm{\alpha}_f,\bm{\beta}_f)_{f\in\FF} \in \left(\Z_2^2\right)^{\FF}$,
\begin{equation}
  \label{eq:psidef}
  \bm{\Psi}(\t)_x =
  \begin{cases}
    \sum_{f \sim x} \bm{\beta}_f & \mbox{if } x \in \BB, \\
    \sum_{f \sim x} \bm{\alpha}_f & \mbox{if } x \in \WW.
  \end{cases}
\end{equation}

The applications $\bm{\Phi}$ and $\bm{\Psi}$ can be considered as dual
of each other. To do so, we equip $\left(\Z_2^2\right)^{\FF}$ with the
symplectic form $\braket{\cdot | \cdot}$
\begin{equation}
  \braket{\bm{\tau} | \bm{\tau'}} = \sum_{f\in F}\bm{\alpha}_f \bm{\beta'}_f \bm{+} \bm{\alpha'}_f \bm{\beta}_f,
\end{equation}
and $\Z_2^\VV$ with the canonical bilinear symmetric form
$(\cdot,\cdot)$
\begin{equation}
  (\bm{\sigma},\bm{\sigma'}) = \sum_{v\in V}\bm{\sigma}_v \bm{\sigma'}_v.
\end{equation}

\begin{Prop}
  \label{prop:exact}
  \leavevmode
  \begin{enumerate}
  \item The applications $\bm{\Psi}$ and $\bm{\Phi}$ are dual of each
    other, meaning that for any $\bm{\sigma}\in \Z_2^\VV$ and
    $\bm{\tau} \in \left(\Z_2^2\right)^{\FF}$,
    \begin{equation}
      \label{eq:dualps}
      \braket{\bm{\Phi \sigma} | \bm{\tau}} = (\bm{\sigma}, \bm{\Psi \tau}).
    \end{equation}

  \item $\bm{H} = \text{\emph{Im }} \bm{\Phi} = \ker{\bm{\Psi}}.$
    In other words, the following sequence is exact
    \begin{equation}
      \label{eq:dia}
      \Z_2^\VV \xrightarrow[]{\bm{\Phi}} \left(\Z_2^2\right)^{\FF}  \xrightarrow[]{\bm{\Psi}} \Z_2^\VV.
    \end{equation}
  \item $\bm{H} = \bm{H}^\bot.$
  \end{enumerate}
\end{Prop}

\begin{proof}
  Let $\bm{\sigma} = (\bm{\sigma}_v)_{v\in \VV}$ and
  $\bm{\tau} = \left( \bm{\alpha}_f,\bm{\beta}_f \right)_{f\in
    \FF}$. By linearity, it is enough to prove \eqref{eq:dualps} when
  $\bm{\sigma}, \bm{\tau}$ are elements of the canonical basis,
  \textit{i.e.} when $\bm{\sigma}_v$ is $\bm{0}$ for all vertices but
  one, and $\bm{\alpha}_f,\bm{\beta}_f$ are all $\bm{0}$ except one.

  If $\s_b$ is $\bm{1}$ for one black vertex $b\in \BB$ and $\bm{0}$ for
  all other vertices, then $\bm{\Phi}\s$ is $(\1,\0)$ on faces
  adjacent to $b$ and $(\0,\0)$ otherwise. Two cases may appear:
  \begin{itemize}
  \item if $\a _f = \bm{1}$ at some face $f$ and all the other
    components of $\t$ are $\bm{0}$, then $\bm{\Psi}\t$ is $\1$ on the
    white vertices of $f$ and $\0$ everywhere else, and we have
    $$\braket{\bm{\Phi}\s | \t} = (\s, \bm{\Psi}\t) = \0.$$
  \item if $\bm{\beta} _f = \bm{1}$ at some face $f$ and all the other
    components of $\t$ are $\bm{0}$, then $\bm{\Psi}\t$ is $\1$ on the
    black vertices of $f$ and $\0$ everywhere else, and we have
    \begin{equation*}
      \braket{\bm{\Phi}\s | \t} = (\s, \bm{\Psi}\t) =
      \begin{cases}
        \1 \ \ \text{if $u$ is a vertex of $f$},\\
        \0 \ \ \text{otherwise}.
      \end{cases}
    \end{equation*}
  \end{itemize}
  The case where $\s_w$ is $\1$ at a white vertex $w\in W$ and $\0$
  elsewhere is similar. This proves \textit{1}.

  We now prove \textit{2}. We already know that
  $\text{Im}\ \bm{\Phi} \subset \ker{\bm{\Psi}}$. Let us show that they
  have the same dimension.
  \begin{itemize}
  \item The kernel of $\bm{\Phi}$ is clearly composed of elements of
    $\Z_2^\VV$ constant on $\BB$ and constant on $\WW$, so it has dimension
    $2$. By the rank-nullity theorem, $\text{Im} \ \bm{\Phi}$ has
    dimension $|\VV|-2$.
  \item The applications $\bm{\Phi}$ and $\bm{\Psi}$ are dual of each
    other so they have the same rank. By the rank-nullity theorem,
    $\ker \bm{\Psi}$ has dimension $2|\FF| - |\VV| + 2$.
  \item We have Euler's formula
    $|\VV|-|\EE|+|\FF| = 2$,
    and the graph is a quadrangulation so
    $4|\FF| = 2|\EE|$.
    Combining these gives
    $|\VV|-2 = 2|\FF|-|\VV|+2$
    as needed.
  \end{itemize}

  Since $\bm{\Phi}$ and $\bm{\Psi}$ are dual of each other,
  $\text{Im}\ \bm{\Phi} = \left(\ker{\bm{\Psi}}\right)^\bot$
  and \textit{3} is obvious from \textit{2}.
\end{proof}

\begin{Rk}\leavevmode
  \begin{itemize}
  \item It is clear now that we are working with an avatar of discrete Hodge
    theory.  The applications $\bm{\Phi}$ and $\bm{\Psi}$ are in fact the
    $d$ applications defined by Mercat for the \textit{double} of a
    chain complex \cite{Mercat}. For that reason, we will now simply
    denote the sequence \eqref{eq:dia} as
    \begin{equation}
      \label{eq:dia2}
      \Z_2^\VV \xrightarrow[]{\bm{d}} \left(\Z_2^2\right)^{\FF}  \xrightarrow[]{\bm{d}} \Z_2^\VV
    \end{equation}
    so, for instance, an 8V configuration is a closed $1$-form (i.e. a
    $\t \in \left(\Z_2^2\right)^{\FF}$ s.t. $\bm{d}\t = \0$).
  \item The elements of $\left(\Z_2^2\right)^{\FF} \setminus \H$ do
    not correspond to 8V-configurations, but can be thought of as
    configurations with defects. More precisely, if
    $\bm{d}\t = \1_{B_1\cup W_1}$, with $B_1\subset \BB$ and
    $W_1 \subset \WW$, then $B_1$ and $W_1$ have to be of even
    cardinality, and $\t$ corresponds to the disordered configurations
    of \cite{Dubedat} mentioned in \ref{sec:8v-corr}. We will
    alternatively denote $\bm{d}\t = B_1\cup W_1$.
  \item Properties similar to Proposition~\ref{prop:exact} might hold
    when $\QQ$ is not a quadrangulation of the sphere but of the
    torus, or other surfaces. These are beyond the scope of
    the present paper.
  \end{itemize}
\end{Rk}

\subsection{Fourier transform}
\label{sec:fourier-transform}
Let $g:\left(\Z_2^2\right)^{\FF} \rightarrow \C$. We define its Fourier transform
$\hat{g}: \left(\Z_2^2\right)^{\FF} \rightarrow \C$ by
\begin{equation}
  \label{eq:deffour}
  \hat{g}(\t) = 2^{-|\FF|} \sum_{\t'\in \left(\Z_2^2\right)^{\FF}}(-1)^{\braket{\t | \t'}}g(\t').
\end{equation}
The normalization is such that we have the Inverse Fourier transform
formula is
$\hat{\hat{g}} = g$.

Another important formula is Poisson's summation identity. For any
subspace $\bm{F}\subset \left(\Z_2^2\right)^{\FF}$,
\begin{equation}
  \sum_{\t \in \bm{F}}g(\t) = \sum_{\t \in \bm{F}^\bot}\hat{g}(\t).
\end{equation}

\begin{Ex}
  For 8V weights $X:\FF\to\R^4$, the weight function $w_{8V}$ that we
  defined for 8V-configuration \eqref{eq:1} can be extended to a
  function on $\left(\Z_2^2\right)^{\FF}$. Then it is easy to check
  that its Fourier transform is actually the weight function for the
  dual weights $\hat{X}$ \eqref{eq:8vdu}. Then Poisson's summation
  identity applied to $\H$, given that $\H=\H^{\bot}$, becomes
  \begin{equation}
    \sum_{\t \in \bm{H}}w_{8V}(\t) = \sum_{\t \in \H}\hat{w}_{8V}(\t)
  \end{equation}
  which is the duality relation for partition functions
  \eqref{eq:8vdu_Z}.
\end{Ex}

\subsection{Correlators}
\label{sec:correlators}

We now describe how correlators of Definition~\ref{def:corr8} fit into
this description. In the absence of disorder, the order variables
$\sigma(B_0)\sigma(W_0)$ correspond to a random variable taking value
$1$ (resp. $-1$) when there is an even (resp. odd) number of edges in
$\t$ between the $B_0,W_0$ joined pairwise. If we fix paths
$\gamma_{B_0},\gamma_{W_0}$, and if $\t =
(\bm{\alpha}_f,\bm{\beta}_f)_{f\in\FF}$, this is equivalent to
considering
\begin{equation}
  \prod_{f\in \gamma_{B_0}}(-1)^{\bm{\alpha}_f} \ \prod_{f\in
    \gamma_{W_0}}(-1)^{\bm{\beta}_f}.
\end{equation}
If we define $\t_{\gamma} = (\1_{\gamma_{W_0}}, \1_{\gamma_{B_0}})$
(where the paths are identified with subsets of $\FF$),
then this quantity is exactly $(-1)^{\braket{\t_{\gamma} | \t }}$. On
the other hand, disorder variables at $B_1,W_1$ correspond to
configurations $\t$ with $\bm{d}\t = B_1\cup W_1$. Thus we have:
\begin{equation}
  \left<\sigma(B_0)\sigma(W_0)\mu(B_1)\mu(W_1) \right>^{8V}_{X,\gamma}
  = 2 \sum_{\substack{\t \text{ s.t.} \\ \bm{d}\t = B_1\cup
      W_1}}(-1)^{\braket{\t_{\gamma}|\t}}w_{8V}(\t).
\end{equation}
The factor $2$ comes from the fact that the representation of
8V-configurations in $\left(\Z_2^2\right)^{\FF}$ is two-to-one.

\begin{proof}[Proof of Theorem~\ref{theo:xorcoupl}]
  Our goal is to prove that for any 8V-configuration $\tau\in\Omega(\QQ)$,
  \begin{equation}
    \Pr_{8V}\left(\tau_{\alpha,\beta} \oplus \tau_{\alpha',\beta'} =
      \tau\right) = \Pr_{8V}\left(\tau_{\alpha,\beta'} \oplus
      \tau_{\alpha',\beta} = \tau\right).
  \end{equation}
  By definition of Boltzmann probabilities, this is equivalent to (we
  indicate the dependence of $w_{8V}$ in the $\alpha,\beta$
  variables):
  \begin{equation}
    \sum_{\substack{\tau_{\alpha,\beta}, \tau_{\alpha',\beta'} \text{ s.t.} \\
        \tau_{\alpha,\beta} \oplus \tau_{\alpha',\beta'} = \tau}}
    \frac{w^{\alpha,\beta}_{8V}(\tau_{\alpha,\beta})}{\ZZ_{8V}(\QQ,X_{\alpha,\beta})}
    \frac{w^{\alpha',\beta'}_{8V}(\tau_{\alpha',\beta'})}{\ZZ_{8V}(\QQ,X_{\alpha',\beta'})}
    \ =
    \sum_{\substack{\tau_{\alpha,\beta'}, \tau_{\alpha',\beta} \text{ s.t.} \\
        \tau_{\alpha,\beta'} \oplus \tau_{\alpha',\beta} = \tau}}
    \frac{w^{\alpha,\beta'}_{8V}(\tau_{\alpha,\beta'})}{\ZZ_{8V}(\QQ,X_{\alpha,\beta'})}
    \frac{w^{\alpha',\beta}_{8V}(\tau_{\alpha',\beta})}{\ZZ_{8V}(\QQ,X_{\alpha',\beta})}.
  \end{equation}
  We already know that the product of partition functions are
  proportional with a factor $c_1$ \eqref{eq:8vswitch_z}, so we just
  have to show that
  \begin{equation}
    \label{eq:xor_prop}
    \sum_{\substack{\tau_{\alpha,\beta}, \tau_{\alpha',\beta'} \text{ s.t.} \\
        \tau_{\alpha,\beta} \oplus \tau_{\alpha',\beta'} = \tau}}
    w^{\alpha,\beta}_{8V}(\tau_{\alpha,\beta})
    w^{\alpha',\beta'}_{8V}(\tau_{\alpha',\beta'})
    \ = \
    c_1 \sum_{\substack{\tau_{\alpha,\beta'}, \tau_{\alpha',\beta} \text{ s.t.} \\
        \tau_{\alpha,\beta'} \oplus \tau_{\alpha',\beta} = \tau}}
    w^{\alpha,\beta'}_{8V}(\tau_{\alpha,\beta'})
    w^{\alpha',\beta}_{8V}(\tau_{\alpha',\beta}).
  \end{equation}

  \medskip

  To prove \eqref{eq:xor_prop}, we first rewrite the correlators of
  Theorem~\ref{theo:8vswitch} in the formalism of $1$-forms. In the
  particular case $B=B_0=B'_0$, $W=W_0=W'_0$ and
  $B_1=B'_1=W_1=W'_1=\emptyset$, let
  $\t_{\gamma}=(\1_{\gamma_W},\1_{\gamma_B})$, then
  \eqref{eq:8vswitch_corr} reads
  \begin{equation}
    \sum_{\t', \t'' \in \H}
    (-1)^{\braket{\t'+\t''|\t_{\gamma}}}w^{\alpha,\beta}_{8V}(\t')
    w^{\alpha',\beta'}_{8V}(\t'')
    = c_1
    \sum_{ \t', \t'' \in \H}
    (-1)^{\braket{\t'+\t''|\t_{\gamma}}}w^{\alpha,\beta'}_{8V}(\t')
    \bm{w}^{\alpha',\beta}_{8V}(\t'').
  \end{equation}
  Reordering these sums according to $\t=\t'+\t''$ gives
  \begin{equation}
    \sum_{\t \in \left(\Z_2^2\right)^{\FF}} (-1)^{\braket{\t|\t_{\gamma}}}
    \sum_{\substack{\t', \t'' \in \H \ \text{\textit{s.t}} \\ \t'
        \bm{+} \t'' = \t}} w^{\alpha,\beta}_{8V}(\t')
    \bm{w}^{\alpha',\beta'}_{8V}(\t'')
    = c_1
    \sum_{\t \in \left(\Z_2^2\right)^{\FF}} (-1)^{\braket{\t|\t_{\gamma}}}
    \sum_{\substack{\t', \t'' \in \H \ \text{\textit{s.t}} \\ \t'
        \bm{+} \t'' = \t}} w^{\alpha,\beta'}_{8V}(\t')
    \bm{w}^{\alpha',\beta}_{8V}(\t'').
  \end{equation}
  Note that we always have $\t=\t'+\t'' \in \H$, so when $\t \notin
  \H$ the inner sum is empty. We rewrite this as
  \begin{equation}
    \sum_{\t \in \left(\Z_2^2\right)^{\FF}}
    (-1)^{\braket{\t|\t_{\gamma}}} f(\t) = 0
  \end{equation}
  where
  \begin{equation}
    f(\t) = \sum_{\substack{\t', \t'' \in \H \ \text{\textit{s.t}} \\ \t'
        \bm{+} \t'' = \t }} w^{\alpha,\beta}_{8V}(\t') w^{\alpha',\beta'}_{8V}(\t'') \ - c_1
    \sum_{\substack{\t', \t'' \in \H \ \text{\textit{s.t}} \\ \t' \bm{+}
        \t'' = \t }} w^{\alpha,\beta'}_{8V}(\t')
    w^{\alpha',\beta}_{8V}(\t'').
  \end{equation}

  In other words, we have $\hat{f}(\t_{\gamma}) = 0.$ This is true for
  any $B,W$ and paths $\gamma$ joining them pairwise. Conversely, any
  element $\t \in \left(\Z_2^2\right)^{\FF}$ can be considered as such
  a $\t_{\gamma}$ -- namely, if $\bm{d} \t = B \cup W$, then
  $\t = (\1_{\gamma_W},\1_{\gamma_B})$ for some paths
  $\gamma_B,\gamma_W$ that satisfy the hypothesis of
  Theorem~\ref{theo:8vswitch}. This means that $\hat{f}$ is actually
  the null function, and by injectivity of the Fourier transform, so
  is $f$. This proves \eqref{eq:xor_prop}.
\end{proof}

\begin{Rk}
  \label{rk:xordis}
  In the previous proof, if we let $B_1,B'_1,W_1,W'_1$ be any even
  subsets of black and white vertices of $\QQ$, we get
  \begin{equation}
    \sum_{\substack{\t', \t'' \in \left(\Z_2^2\right)^{\FF} \ \text{\textit{s.t}} \\ \t'
        \bm{+} \t'' = \t \\ \bm{d}\t' = B_1\cup W_1 \\ \bm{d}\t'' =
        B'_1 \cup W'_1 }}
    \frac{w^{\alpha,\beta}_{8V}(\t')}{\ZZ_{8V}(\QQ,X_{\alpha,\beta})}
    \frac{w^{\alpha',\beta'}_{8V}(\t'')}{\ZZ_{8V}(\QQ,X_{\alpha',\beta'})}
    =
    \sum_{\substack{\t', \t'' \in \left(\Z_2^2\right)^{\FF} \ \text{\textit{s.t}} \\ \t'
        \bm{+} \t'' = \t \\ \bm{d}\t' = B'_1\cup W_1 \\ \bm{d}\t'' =
        B_1 \cup W'_1 }}
    \frac{w^{\alpha,\beta'}_{8V}(\t')}{\ZZ_{8V}(\QQ,X_{\alpha,\beta'})}
    \frac{w^{\alpha',\beta}_{8V}(\t'')}{\ZZ_{8V}(\QQ,X_{\alpha',\beta})}
  \end{equation}
  which expresses a coupling for the XOR of 8V-configurations \emph{with disorder}.
\end{Rk}

\section{Proof of Lemma~\ref{le:monot}}
\label{sec:proof-lemma}

  By rotating the graph, we can suppose that $\alpha=0$, \textit{i.e.}
  the angles $\alpha_i$ and $u_0(k)$ all lie in
  $(-\frac{\pi}{2},\frac{\pi}{2})$. We also fix a $k\in(0,1)$ and
  suppose that $u_0(k)\geq 0$, the other case being symmetric.

  Using the chain rule we have
  \begin{equation}
    \label{eq:77}
    \frac{\mathrm{d}}{\mathrm{d}k}\chi(u_0(k),k) =
    \frac{\mathrm{d}}{\mathrm{d}k} u_0(k) \frac{\partial
      \chi}{\partial u} (u_0(k),k) + \frac{\partial \chi}{\partial k} (u_0(k),k).
  \end{equation}
  By
  definition of $u_0(k)$ the first term of the sum is null so we just
  have to show that $\frac{\partial \chi}{\partial k}$ is negative at
  $(u_0(k),k)$.

  We denote $r=|a_1-a_{n-1}|$; this does not depend on $k$. By using
  the change of arguments in Jacobi elliptic functions (see Table 16.8
  in \cite{AbramowitzStegun}),
  \begin{equation}
    \label{eq:78}
    \chi(u,k) = \frac{1}{r} \sum_{j=2}^{n-1}
    \log\left[\sqrt{k'}\nd\left(\left. \left(\frac{u-\alpha_j}{2}\right)_k \right|
      k \right)
    \right].
  \end{equation}
  Let
  \begin{equation}
    \label{eq:79}
    g(u,k) = \log\left[\sqrt{k'}\nd\left( \left. \left( \frac{u}{2} \right)_k
        \right| k \right)
    \right].
  \end{equation}
  By the properties of the function $\nd(\cdot,k)$ (see 16.2 in
  \cite{AbramowitzStegun}), for any $k\in (0,1)$, $g(\cdot,k)$ is
  decreasing on $[-\pi,0]$ and increasing on $[0,\pi]$. Its minimum is
  $g(0,k) = \frac12\log(k')<0$. As a result, if all the angles
  $\alpha_j$ are equal, then $u_0(k)$ has the same value and
  $\chi(u_0(k),k) = \frac{n-2}{2r} \log(k')$, which is indeed a
  decreasing function of $k$. We now suppose that the $\alpha_j$ are
  not all equal. We need some extra properties on $g$.
  \begin{Le}
    \label{le:funh}
    For all $k\in(0,1)$,
    \begin{enumerate}
    \item $g(-u,k) = g(u,k)$ and $g(\pi-u,k) = -g(u,k)$.
    \item $\frac{\partial g}{\partial k} (u,k)$ is
      a strictly decreasing function of $u$ on $[-\pi,0]$, and
      strictly increasing on $[0,\pi]$. It
      is zero at $u=\pm \frac{\pi}{2}$.
    \end{enumerate}
  \end{Le}

  \begin{Le}
    \label{le:card}
    We have the following inequality of cardinals:
    \begin{equation}
      \label{eq:80}
      \#\left\{ j \in [2,n-1] \mid \alpha_j < u_0(k) - \frac{\pi}{2} \right\} <
      \#\left\{ j \in [2,n-1] \mid \alpha_j > u_0(k) \right\}.
    \end{equation}
  \end{Le}

  We prove these two Lemmas later, and first show how they imply
  Lemma~\ref{le:monot}.  By differentiation of \eqref{eq:78}, for
  $u\in[0,\frac{\pi}{2})$ we have (using Lemma~\ref{le:funh} to remove
  possible terms equal to zero):
  \begin{equation}
    \label{eq:81}
    \begin{split}
      r\ \frac{\partial \chi}{\partial k} (u,k) &=
      \sum_{j=2}^{n-1}\frac{\partial g}{\partial k} (u-\alpha_j,k) \\
      &= \sum_{j \mid \alpha_j < u - \frac{\pi}{2}}\frac{\partial
        g}{\partial k} (u-\alpha_j,k) \ +
      \sum_{j \mid u - \frac{\pi}{2} < \alpha_j \leq u}\frac{\partial
        g}{\partial k} (u-\alpha_j,k) \ +
      \sum_{j \mid u < \alpha_j }\frac{\partial
        g}{\partial k} (u-\alpha_j,k)
    \end{split}
  \end{equation}
  By Lemma~\ref{le:funh}, the terms in the first sum are positive
  while those in the second an third sums are negative. We show that
  for $u=u_0(k)$, the first sum is, in absolute value, smaller than
  the third one, which is enough to conclude.

  For the first sum, if $-\frac{\pi}{2} < \alpha_j < u -
  \frac{\pi}{2}$ then $\frac{\pi}{2} < u-\alpha_j < u + \frac{\pi}{2}$
  and by Lemma~\ref{le:funh},
  \begin{equation}
    \label{eq:82}
    0 < \frac{\partial g}{\partial k} (u-\alpha_j,k)  < \frac{\partial g}{\partial k} \left(u+\frac{\pi}{2},k\right).
  \end{equation}
  Thus the first sum $S_1$ in \eqref{eq:81} satisfies
  \begin{equation}
    \label{eq:83}
    0 \leq S_1 \leq \left(\frac{\partial g}{\partial k}
      \left(u+\frac{\pi}{2},k\right)\right) \ \#\left\{ j \in [2,n-1] \mid \alpha_j < u - \frac{\pi}{2} \right\}.
  \end{equation}

  Similarly, for the third sum $S_3$, we have
  \begin{equation}
    \label{eq:84}
    S_3 \leq \left(\frac{\partial g}{\partial k}
      \left(u-\frac{\pi}{2},k\right)\right) \ \#\left\{ j \in [2,n-1]
      \mid  \alpha_j > u \right\}  < 0.
  \end{equation}
  By Lemma~\ref{le:funh},
  $g\left(u+\frac{\pi}{2},k\right) = - g\left(u-\frac{\pi}{2},k\right)
  > 0$, and by differentiating the same symmetry holds for
  $\frac{\partial g}{\partial k}$. Hence \eqref{eq:84} becomes
  \begin{equation}
    \label{eq:85}
    |S_3| \geq \left(\frac{\partial g}{\partial k}
      \left(u+\frac{\pi}{2},k\right)\right) \ \#\left\{ j \in [2,n-1]
      \mid \alpha_j > u \right\}.
  \end{equation}
  Using \eqref{eq:85}, \eqref{eq:83} and Lemma~\ref{le:card} we see
  that for $u=u_0(k),\ |S_3| > S_1$ as needed.
\qed

\begin{proof}[Proof of Lemma~\ref{le:funh}]
  The first point is a direct consequence of the change of arguments
  in elliptic functions, see Table 16.8 in \cite{AbramowitzStegun}.

  For the second point, first notice that for all $k$, using Table
  16.5 in \cite{AbramowitzStegun}, $g\left(\frac{\pi}{2},k\right) = 1$
  so $\frac{\partial g}{\partial k} \left(\frac{\pi}{2},k\right) =
  0$. Using the symmetries of the first point of the Lemma, it remains
  to check that $\frac{\partial g}{\partial k} (u,k)$ is a strictly
  increasing function of $u$ on $\left[ 0,\frac{\pi}{2} \right]$.

  Using the derivatives of elliptic functions with respect to $u$ and
  $k$ (see Sections 2.5 and 3.10 in \cite{Lawden}), and setting
  $v=\left(\frac{u}{2}\right)_k$, we get
  \begin{equation}
    \label{eq:86}
    \frac{\partial g}{\partial k} (u,k) = - \frac{k}{2k'^2}  + \frac{k}{k'^2} \left(
      \frac{v}{\mathrm{K}(k)}\mathrm{E}(k) - \mathrm{E}(v,k) + \frac{\sn\dn}{\cn}
      \left(v|k \right) \right) \frac{\sn\cn}{\dn} \left(v|k \right)
  \end{equation}
  where $\mathrm{E}$ is the elliptic integral of the second kind:
  \begin{align}
    \label{eq:87}
    \mathrm{E}(v,k) & = \int_{0}^v \dn^2(t|k) \mathrm{d}t, \\
    \mathrm{E}(k) & = \mathrm{E}(\mathrm{K}(k),k).
  \end{align}
  As $v = \frac{\mathrm{K}(k)}{\pi}u$, it is sufficient to prove that the
  right-hand side of \eqref{eq:86} is a strictly increasing function of $v$ on
  $\left[0,\frac{\mathrm{K}(k)}{2}\right]$. On that interval,
  \begin{itemize}
  \item
    $v \mapsto \frac{v}{\mathrm{K}(k)}\mathrm{E}(k) - \mathrm{E}(v,k) + \frac{\sn\dn}{\cn} \left(v|k
    \right)$ is strictly increasing because its derivative in $v$ is
    (using Section 2.5 in \cite{Lawden})
    \begin{equation}
      \label{eq:88}
      \frac{\mathrm{E}(k)}{\mathrm{K}(k)} + k'^2\sc(v|k) > 0
    \end{equation}
  \item $v \mapsto \frac{\sn\cn}{\dn} \left(v|k \right)$ is strictly
    increasing because, using the ascending Landen transform
    $\tilde{k} = \frac{1-k'}{1+k'}$ (see 16.14.1 in
    \cite{AbramowitzStegun}), this is equal to
    \begin{equation}
      \label{eq:89}
      \frac{1+\tilde{k}}{2} \sn\left( \left. 2\frac{\mathrm{K}(\tilde{k})}{\mathrm{K}(k)} v \right| \tilde{k}\right)
    \end{equation}
    and $\sn(\cdot|\tilde{k})$ is strictly increasing on $[0,\mathrm{K}(\tilde{k})]$.
  \end{itemize}
  As a result, \eqref{eq:86} is a strictly increasing function of $v$
  on $\left[0,\frac{\mathrm{K}(k)}{2}\right]$.
\end{proof}

\begin{proof}[Proof of Lemma~\ref{le:card}]
  We take again $\tilde{k} = \frac{1-k'}{1+k'}$. By equation $(26)$ in
  \cite{BoutillierDeTiliereRaschel:lap}, $u_0(k)$ is also the unique
  element of $\left(-\frac{\pi}{2},\frac{\pi}{2}\right)$ such that
  \begin{equation}
    \label{eq:90}
    \sum_{j=2}^{n-1} \sn\left(
      \left. (u_0(k)-\alpha_j)_{\tilde{k}}\right| \tilde{k} \right)
    = 0.
  \end{equation}
  Let
  $s_j = \sn\left( \left. (u_0(k)-\alpha_j)_{\tilde{k}}\right|
    \tilde{k} \right)$.  We fix an $\epsilon>0$ such that the angles
  $\alpha_i$ and $u_0(k)$ all lie in
  $\left(-\frac{\pi}{2} + \epsilon,\frac{\pi}{2} -
    \epsilon\right)$. Since we supposed that $u_0(k) \geq 0$, we have
  $u_0(k)-\alpha_j \in \left[-\frac{\pi}{2} + \epsilon, \pi -
    \epsilon\right]$. As a result,
  $\left( u_0(k)-\alpha_j\right)_{\tilde{k}} \in
  \left[-\mathrm{K}(\tilde{k}) + \epsilon_{\tilde{k}},
    2\mathrm{K}(\tilde{k}) - \epsilon_{\tilde{k}}\right]$. By the
  properties of the $\sn$ function, this implies that $s_j< 0$ when
  $\alpha_j > u_0(k)$, that $s_j > 0$ when $\alpha_j < u_0(k)$, and
  that $s_j=0$ when $\alpha_j=u_0(k)$. As a result,
  \begin{equation}
    \label{eq:91}
    \sum_{j \mid \alpha_j < u_0(k)} s_j = \sum_{j \mid \alpha_j > u_0(k)}(-s_j)
  \end{equation}
  where all the terms in the sums are positive. In particular,
  \begin{equation}
    \label{eq:92}
    \sum_{j \mid \alpha_j < u_0(k) - \frac{\pi}{2}} s_j \leq \sum_{j \mid \alpha_j > u_0(k)}(-s_j).
  \end{equation}

  When $ \alpha_j < u_0(k) - \frac{\pi}{2}$, then
  $\left( u_0(k)-\alpha_j\right)_{\tilde{k}} \in \left[\mathrm{K}(\tilde{k}),
    \mathrm{K}(\tilde{k}) + \left(u_0(k) -
      \epsilon\right)_{\tilde{k}}\right)$. Since
  $\sn(\cdot,\tilde{k})$ is decreasing on $[ \mathrm{K}(\tilde{k}), 2 \mathrm{K}(\tilde{k})]$, in that case
  \begin{equation}
    \label{eq:93}
    0 < \sn\left(\left. \mathrm{K}(\tilde{k}) + \left(u_0(k) -
          \epsilon\right)_{\tilde{k}} \right| \tilde{k} \right) < s_j \leq 1.
  \end{equation}
  When $\alpha_j > u_0(k)$, then
  $\left( u_0(k)-\alpha_j\right)_{\tilde{k}} \in \left( - \mathrm{K}(\tilde{k})
    + \left(u_0(k) + \epsilon\right)_{\tilde{k}}, 0\right)$. Since
  $\sn(\cdot,\tilde{k})$ is increasing on $[-\mathrm{K}(\tilde{k}),0]$ and odd,
  in that case
  \begin{equation}
    \label{eq:94}
    0 < -s_j < \sn\left(\left. \mathrm{K}(\tilde{k}) - \left(u_0(k) +
          \epsilon\right)_{\tilde{k}}\right| \tilde{k} \right).
  \end{equation}
  Moreover, using again the symmetry and monotonicity of the $\sn$ function,
  \begin{equation}
    \label{eq:95}
    \begin{split}
      \sn\left(\left. \mathrm{K}(\tilde{k}) - \left(u_0(k) +
            \epsilon\right)_{\tilde{k}}\right| \tilde{k} \right) & =
      \sn\left(\left. 2\mathrm{K}(\tilde{k}) - \left(\mathrm{K}(\tilde{k}) - \left(u_0(k) +
              \epsilon\right)_{\tilde{k}}\right)\right| \tilde{k}
      \right) \\
      & =  \sn\left(\left. \mathrm{K}(\tilde{k}) + \left(u_0(k) +
            \epsilon\right)_{\tilde{k}}\right| \tilde{k}
      \right) \\
      & < \sn\left(\left. \mathrm{K}(\tilde{k}) + \left(u_0(k) -
            \epsilon\right)_{\tilde{k}}\right| \tilde{k}
      \right).
    \end{split}
  \end{equation}
  As a result, we get the following inequalities:
  \begin{equation}
    \label{eq:96}
    \begin{split}
      &\sn\left(\left. \mathrm{K}(\tilde{k}) + \left(u_0(k) -
            \epsilon\right)_{\tilde{k}} \right| \tilde{k} \right) \ \
      \#\left\{ j \in [2,n-1] \mid \alpha_j < u_0(k) - \frac{\pi}{2}
      \right\} \\
      \leq &\sum_{j \mid \alpha_j < u_0(k) - \frac{\pi}{2}} s_j \\
      \leq &\sum_{j \mid \alpha_j > u_0(k)}(-s_j) \\
      \leq &\sn\left(\left. \mathrm{K}(\tilde{k}) - \left(u_0(k) +
            \epsilon\right)_{\tilde{k}} \right| \tilde{k} \right) \ \
      \#\left\{ j \in [2,n-1] \mid \alpha_j > u_0(k)
      \right)\\
      < &\sn\left(\left. \mathrm{K}(\tilde{k}) + \left(u_0(k) -
            \epsilon\right)_{\tilde{k}} \right| \tilde{k} \right) \ \
      \#\left\{ j \in [2,n-1] \mid \alpha_j > u_0(k) \right).
    \end{split}
  \end{equation}
  In the last inequality, we used the fact that the cardinal is not
  zero since these $j$ are exactly those that give a negative term in
  in \eqref{eq:90}; those negative terms have to exist because the
  $\alpha_j$ are not all equal. Dividing by
  $\sn\left(\left. \mathrm{K}(\tilde{k}) + \left(u_0(k) -
        \epsilon\right)_{\tilde{k}} \right| \tilde{k} \right) >0$, we
  get the claim of Lemma~\ref{le:card}.

\end{proof}

\bibliographystyle{abbrv} \bibliography{ff}

\end{document}